\documentclass[12pt]{article}
\usepackage{authblk}
\usepackage{geometry}
\usepackage{amsthm}
\usepackage{mathrsfs}
\usepackage{subfigure}
\usepackage{customcommands}
\usepackage{comment}
\usepackage{bm}
\usepackage{Chrisstyle}

\usepackage{amsmath}
\usepackage{amssymb}
\usepackage{amsfonts}
\usepackage{braket}
\usepackage[normalem]{ulem}
\usepackage{color}
\usepackage{bbold}
\usepackage{ulem}
\usepackage{mathtools}
\usepackage{tensor} 
\usepackage{thmtools} 
\usepackage{thm-restate} 

\newcommand{\diff}{\mathrm{d}}
\newcommand{\cl}{\text{cl}}
\newcommand{\qtheta}{\tilde{\theta}}
\newcommand{\qvartheta}{\tilde{\vartheta}}
\newcommand{\htime}{{\rm H}}
\newcommand{\proj}{\Pi}
\newcommand{\mc}{\text{mc}}
\newcommand{\clas}[1]{{#1}_{\text{cl}}}

\DeclareMathOperator{\sinc}{sinc}

\DeclareMathOperator{\Si}{Si}
\DeclareMathOperator{\erfc}{erfc}

\newcommand{\edit}[1]{#1}

\title{On the reconstruction map in JT gravity}

\author{Chris Akers, Andrew Lucas, and Amit Vikram}

\affiliation{Department of Physics and Center for Theory of Quantum Matter, University of Colorado, Boulder CO 80309 USA}

\emailAdd{chris.akers@colorado.edu}

\emailAdd{andrew.j.lucas@colorado.edu}

\emailAdd{amitvikram.anand@colorado.edu}

\abstract{An open question in AdS/CFT is how to reconstruct semiclassical bulk operators precisely enough that non-perturbative quantum effects can be computed. We propose a set of physically-motivated requirements for such a reconstruction map, and explicitly construct a map satisfying these requirements in Jackiw-Teitelboim (JT) gravity. Our map is found by canonically quantizing ``action-angle" variables for JT gravity, which are chosen to ensure that the spectrum of the fundamental quantum theory matches known results from the gravitational path integral. We then study unitary quantum dynamics in this theory, and obtain analytical predictions for the dynamics of the wormhole length, including its quantum fluctuations, leveraging techniques from quantum ergodicity theory.  Level repulsion in the non-perturbative JT spectrum implies that the average wormhole length is non-monotonic in time, that  fluctuations in wormhole length are non-perturbatively suppressed until nearly the Heisenberg time, and that the late-time-evolved Hartle-Hawking state has a heavy-tailed distribution of lengths. We discuss the implications of our results for the ``complexity = volume" conjecture.
}

\begin{document}
\maketitle

\section{Introduction}

Consider semiclassical gravity, which we define as a quantum field theory on a curved background coupled to gravity in a perturbative $G_N$ expansion.
We can regard this as an effective theory of an underlying, fundamental, quantum gravity theory.
As an effective theory, its predictions cannot always be trusted because sometimes they receive large corrections from non-perturbative effects.
The anti de Sitter / conformal field theory (AdS/CFT) correspondence offers a path towards computing these effects: given an operator in the semiclassical AdS gravity theory, we \emph{reconstruct} it in the CFT and evaluate it there.
Unfortunately, doing this with the necessary precision is not yet possible in general, because we do not understand these reconstruction maps precisely enough. 

Recently, a precise reconstruction map \cite{Iliesiu:2024cnh} was proposed in Jackiw-Teitelboim (JT) gravity \cite{Jackiw:1984je, Teitelboim:1983ux}.
Our goal is to review this reconstruction map, describe some of its features -- both desirable and (from our perspective) perhaps undesirable -- and then propose a modified version without the `undesirable' features.
We then use this new reconstruction map to learn  things about non-perturbative quantum gravity and the dynamics of quantum wormholes on long time scales, using techniques from quantum ergodicity theory \cite{dynamicalqergodicity}.

\subsection{Summary of previous work}

JT gravity \cite{Jackiw:1984je, Teitelboim:1983ux} is a 1+1 dimensional theory of gravity defined by the action
\begin{equation}
	S_\mathrm{JT} = S_0 \int \mathrm{d}^2x \sqrt{-g} R + \int \mathrm{d}^2x \sqrt{-g} \Phi(R + 2) + \text{(boundary terms)}
\end{equation}
In this number of dimensions, the first term is topological, and a larger parameter $S_0$ serves to suppress contributions from non-trivial spacetime topologies to a path integral computation.
Consider the theory at $S_0 \to \infty$, which we will call $\mathrm{JT}_\infty$.
All solutions in this theory have the same topology of two asymptotic boundaries connected by a wormhole behind the horizon of two black holes.  The only dynamical degree of freedom is the length of this wormhole \cite{Harlow:2018tqv}.
Quantized $\mathrm{JT}_\infty$ is equivalent to a particle moving in one dimension with an exponential potential: it has a Hilbert space $\mathcal{H}_0$ and Hamiltonian $H_0$ given by
\begin{equation}
    \mathcal{H}_0 = \mathrm{L}^2(\mathbb{R})~,~~~~ H_0 = -\frac{1}{2}\partial_x^2 + 2 \mathrm{e}^{-x}
\end{equation}
where $x$ is the wormhole length \cite{Harlow:2018tqv}.
This Hamiltonian has eigenstates $\ket{E}_0$ for $E > 0$, with wavefunctions
\begin{equation}
    \phi_E(x) := \braket{x|E}_0 = 4 \mathrm{K}_{\mathrm{i}\sqrt{8E}}(4 \mathrm{e}^{-x/2})
\end{equation} 
where these $\mathrm{K}$ are modified Bessel functions, with inner product
\begin{equation}
    \braket{E|E'}_0 = \frac{\delta(E - E')}{\rho_0(E)}~,~~~~\int\limits_0^\infty \mathrm{d}E\; \rho_0(E) \ket{E}\bra{E}_0 = \mathbb{1}_0~,
\end{equation}
where we have defined the density of states
\begin{equation}
    \rho_0(E) = \frac{1}{4 \pi^2} \sinh(2 \pi \sqrt{2 E}). \label{eq:JTinftyDOS}
\end{equation}
We regard $\mathrm{JT}_\infty$ as our semiclassical gravity theory, with $S_0 \to \infty$ analogous to $G_N \to 0$ in higher dimensions. 
Note in particular the continuous-energy spectrum.

At finite $S_0$, the topology can fluctuate.
When computing with the path integral, we must sum over different topologies \cite{Saad:2018bqo, Saad:2019lba}. 
This theory has a known dual description as a random matrix model \cite{Saad:2019lba}, where each matrix can be regarded as a Hamiltonian in a Hilbert space $\mathcal{H}$, with discrete spectrum $E_1, E_2,... $ 
with typical spacing $\Delta E \simeq \mathrm{e}^{-S_0}$ and leading order density of states $\mathrm{e}^{S_0} \rho_0(E)$. 
The gravitational path integral calculates averaged quantities over this ensemble of random matrix theories.

Now consider a single draw $H$ from this ensemble.
We will follow \cite{Iliesiu:2024cnh} and consider $\mathrm{JT}_\infty$ as an effective description of that fundamental theory.
We can then ask how the predictions of $\mathrm{JT}_\infty$ are corrected in the fundamental theory.
In particular, let us consider the wormhole length starting in the Hartle-Hawking state
\begin{equation}
    |\beta\rangle = \frac{1}{\sqrt{Z(\beta)}} \sum_{n=1}^\infty \mathrm{e}^{-\beta E_n/2} |E_n\rangle   \label{eq:HHstate}
\end{equation}
where $Z(\beta)$ is the partition function.  
$\mathrm{JT}_\infty$ predicts $\braket{x(t)}$ grows linearly forever in $t$. 
What does the fundamental theory predict?
This question has a long history \cite{Susskind:2014rva, Iliesiu:2021ari, Stanford:2022fdt, Iliesiu:2024cnh}, and some general comments can be made.
First, we should expect that the true length matches the prediction of the effective theory for a while -- any short experiment shouldn't be able to resolve the energy differences.
However, eternal growth can't happen for any reasonable choice of $\hat x$: in the fundamental theory, $|\beta\rangle$ has non-negligible support on a finite $\mathrm{e}^S$ number of energy eigenstates, where $S$ is the thermal entropy of the state.  
Assuming the matrix elements $\langle E_n | \hat x|E_m\rangle$ are finite, the length must eventually stop growing.
In fact, it is expected to \emph{plateau} (rather than, say, to coherently decrease in length back to its starting size), with small fluctuations around the plateau value for extremely long times after the Heisenberg time, $t_{\mathrm{H}} \sim \mathrm{e}^S$.

That said, it is one thing to know $\mathrm{JT}_\infty$ is wrong at late times, and another thing to actually compute the finite $S_0$ corrections.
Explicit computation of these non-perturbative corrections was a major goal of \cite{Iliesiu:2024cnh}.
The first hurdle they encountered is to find an answer to the following question:
\begin{center}
    \emph{Given a $\mathrm{JT}_\infty$ state $\ket{\psi}_0$ and operator $\hat{x}_0$, what are the corresponding state and operator in the fundamental theory?}
\end{center}
The starting point of \cite{Iliesiu:2024cnh} is to define a \emph{holographic map}\footnote{The presentation of these ideas in \cite{Iliesiu:2024cnh} is slightly different, but the physics is equivalent.}
\begin{equation}
    V: \mathcal{H}_0 \to \mathcal{H}
\end{equation}
which takes as input a state in $\mathrm{JT}_\infty$ and outputs the ``corresponding'' state in the fundamental theory.
This map is deduced by studying the non-perturbative corrections to the inner product that appear when $S_0$ is finite.
At infinite $S_0$, two length eigenstates have inner product
\begin{equation}
    \braket{x|x'}_0 = \delta(x - x')~.
\end{equation}
At finite $S_0$, \cite{Iliesiu:2024cnh} argues that the inner product is modified to
\begin{equation}
    \braket{x|x'} = \mathrm{e}^{-S_0} \sum_{i=1}^\infty \phi_{E_i}(x) \phi_{E_i}(x')~.
\end{equation}
This modification can be understood as follows.
Starting with the $\mathrm{JT}_\infty$ inner product, insert the identity in the energy basis, and then simply restrict the integral over all energies to a sum over the energies in the fundamental theory,
\begin{equation}
    \braket{x|x'}_0 = \int \mathrm{d}E\; \rho_0(E) \phi_E(x)\phi_E(x') \longrightarrow \mathrm{e}^{-S_0} \sum_{i = 1}^\infty \phi_{E_i}(x) \phi_{E_i}(x') =: \braket{x|x'}~.
\end{equation}
This modified inner product can be quite different, and in general is non-zero when $x \neq x'$.

This defines the holographic map $V$ as the linear map satisfying
\begin{equation}\label{eq:V_def_inner_prod}
    \braket{x|x'} = \braket{x|V^\dagger V|x'}_0~.
\end{equation}
Technically, this $V$ is still ambiguous under $V \to U V$ for any unitary $U$.
We follow \cite{Iliesiu:2024cnh} and fix this ambiguity by choosing $V$ to be the one satisfying both \eqref{eq:V_def_inner_prod} and which is equivariant under time evolution: \begin{equation}
    V H_0 = H V.
\end{equation}
We then write the holographic map as
\begin{equation}\label{eq:V_JT}
    V \ket{x}_0 = \mathrm{e}^{-S_0/2} \sum_{i = 1}^\infty \phi_{E_i}(x) \ket{E_i}~.
\end{equation}
One could worry that this map is ill-defined because it involves evaluating functions of continuous energy at specific discrete values. We will illustrate with a toy model in Section \ref{sec:freeparticle} that this can be justified.

So far, we have a map $V$ on states.
This induces a pullback on operators 
\begin{equation}
    V^*: \mathrm{L}(\mathcal{H}) \to \mathrm{L}(\mathcal{H}_0)~, 
\end{equation}
defined by $\braket{\psi|V^*(O) |\psi}_0 = \braket{\psi|V^\dagger O V |\psi}_0$ for all $\ket{\psi}_0 \in \mathcal{H}_0$, which implies $V^*(O) = V^\dagger O V$.
However, $V$ is not an isometry, and thus it does \emph{not} automatically define a \emph{reconstruction} map
\begin{equation}
    R^*: \mathrm{L}(\mathcal{H}_0) \to \mathrm{L}(\mathcal{H})~.
\end{equation}
Any reconstruction map we define will require extra input.
In \cite{Iliesiu:2024cnh}, it was proposed that we should define
\begin{equation}\label{eq:luca_map}
    R^*(O) = V O V^\dagger~.
\end{equation}
This map has some nice properties.
One is just that it is a natural $R^*$ to write down, if all that's known is $V$.
Moreover, the authors of \cite{Iliesiu:2024cnh} used this $R^*$ to reconstruct a version of the wormhole length operator, and found the desired qualitative behavior: linear growth for a time $\mathrm{e}^S$ followed by a plateau. 

However, this map \eqref{eq:luca_map} also has some undesirable properties (in our view).
One is that it does not reconstruct simple operators in a nice way.
In particular, there are normalization ambiguities that arise even with the identity operator, which we would like to avoid.
Consider the reconstruction of the identity $\mathbb{1}_0 = \int_{-\infty}^\infty \mathrm{d}x \, \ket{x}_0 \bra{x}_0$
\begin{equation}
    R^*(\mathbb{1}_0) = \int\limits_{-\infty}^\infty \mathrm{d}x \, \ket{x} \bra{x},
\end{equation}
where we write $\ket{x} := V\ket{x}_0$. 
But then
\begin{equation}
    \braket{x| R^*(\mathbb{1}_0) | x} = \int\limits_{-\infty}^\infty \mathrm{d}x'\; \lvert \braket{x|x'}\rvert^2 = \infty~, \label{eq:identityproblemreconstruction}
\end{equation}
because $\lvert \braket{x|x'}\rvert$ is generally $O(\mathrm{e}^{-S_0})$. 
This demonstrates $R^*(\mathbb{1}_0)$ is not the properly normalized identity operator because again $\braket{x|x'} \simeq \delta_{x,x'} + O(\mathrm{e}^{-S_0})$.
Similarly, $R^*(x)$ is ill defined, as are the reconstructions of many other simple operators. 
The authors of \cite{Iliesiu:2024cnh} pointed this out, and proposed that we simply ignore the operators with bad reconstructions, or use regulated versions for the ones like $x$ whose physics is important to us.  

Another recent paper \cite{Miyaji:2024ity} describes a regularization for the length operator based on restricting the path integral to a microcanonical window.  Our predictions will be similar but inequivalent.

\subsection{Summary of our results}

Our point of view is that a good reconstruction map $R^*$ should satisfy a certain \emph{compatibility condition} with the holographic map $V$, which we summarize as follows.\footnote{We focus here on \emph{linear} reconstruction maps, even though in general holographic systems the reconstruction maps can be \emph{non}-linear because the holographic map is non-isometric \cite{Akers:2021fut, Akers:2022qdl}. This restriction is acceptable in this setting of pure JT, without matter, as long as we only demand the reconstructions work on a restricted set of states that we call \emph{semiclassical}. We return to JT with matter in Section \ref{sec:addmatter}. 
}
Let $S$ be some subset of states in $\mathcal{H}_0$, which are morally the ``good, semiclassical'' states that the effective description should be a valid description of.
Let $K_S$ be the analog for operators, i.e. the subset of $\mathrm{B}(\mathcal{H}_0)$ that we wish to regard as ``good and semiclassical."
Presumably $K_S$ should include the operators that preserve $S$, like the identity.

Roughly speaking, we say that $R^*$ satisfies the compatibility condition if for a given $V$ and any $\ket{\psi} \in S$ and $O \in K_S$,
\begin{align}
    \bra{\psi} V^\dagger R^*(O) V \ket{\psi} &\simeq \bra{\psi} O \ket{\psi} \label{eq:exp_val_cond_ex}~.
\end{align}
This morally says that $R^*(O)$ must ``act the same'' as $O$ on semiclassical states.
If $R^*$ does \emph{not} satisfy this condition, that means there are some semiclassical operators that $R^*$ cannot be trusted to reconstruct, raising the question of why we should trust it in the first place.

The reconstruction map \eqref{eq:luca_map} from \cite{Iliesiu:2024cnh} does not satisfy this compatibility condition.
One natural set of semiclassical states $S$ are the Hartle-Hawking state and short time evolutions of it.
The inner product between these states is approximately preserved by $V$ (the corrections from higher topologies are very small), so we would expect $R^*(\mathbb{1}_0) \simeq \mathbb{1}$, but as we saw in \eqref{eq:identityproblemreconstruction} this is not the case.

We will define a reconstruction map $R^*$ that \emph{does} satisfy this compatibility condition with $V$.
As a warmup, in Section \ref{sec:freeparticle} we study a particle on a line as an ``effective description" for a particle on a long interval of length $L$.
The analogs of the holographic and reconstruction maps will be intuitive and related in a particular way, different from \eqref{eq:luca_map}.  In a nutshell, the map $V$ will filter the Fourier modes $k\in \mathbb{R}$ (on the entire real line) down to $k\in \pi L^{-1} \mathbb{Z}^+$ (well defined on an interval), with some mild spreading to ensure normalizability.  The reason why this is a nice way to ``truncate" the continuous spectrum of the line down to an interval is because: (\emph{i}) it is easy to make $V$ nearly-equivariant with time-evolution, and (\emph{ii}) wave packets whose extent is small compared to the interval are clearly well-described by either the effective line theory, or the fundamental interval theory, at early times.  This gives us a way to crisply define operators in the fundamental theory by comparing them, in suitable regimes, to effective theory ones.

This will guide us in Section \ref{sec:pure} when we return to JT and construct an $R^*$ satisfying the compatibility condition for a reasonable set of a states $S$.  The strategy will heuristically mirror the reconstruction map from the line to the interval.  
As inspiration for this, first notice that the $\mathrm{JT}_\infty$ density of states \eqref{eq:JTinftyDOS} is not an absolute density of states (there are formally infinitely many eigenstates of $H$ on the line).   We expect that the \emph{absolute} density of states of the fundamental theory takes the form \begin{equation}
    \rho(E) \approx \mathrm{e}^{S_0}\rho_{0}(E); \label{eq:rhoE}
\end{equation}
 $\int \mathrm{d}E \rho(E)$ counts the \emph{actual} number of energy levels in any energy window.  In what follows we neglect corrections to (\ref{eq:rhoE}) in $\mathrm{e}^{-S_0}$.  Indeed, going back to our analogy of a particle on the line, we similarly have that \begin{equation}
     \rho_{\mathrm{interval}}(E) = L \rho_{\mathrm{line}}(E) = L \sqrt{\frac{m}{2\pi^2 E}}. 
 \end{equation}
 where the last step assumes $H = p^2/2m$ for a particle on the line.  
 It is very tempting to interpret finite $\mathrm{e}^{S_0}$ as analogous to finite $L$.   We will argue that this analogy is indeed the most natural one, in that it allows us to make quantitative and unambiguous predictions for the fundamental quantum gravity theory, without detailed knowledge of microscopics. 
More precisely, we will show that this known density of states together with the compatibility condition \eqref{eq:exp_val_cond_ex} largely fixes a reasonable reconstruction map. This reconstruction map will not be completely unique, but what we leave unfixed appears important only for the lowest few energy levels.

\begin{figure}[t]
\centering
\subfigure[The probability distribution $p(x)$ of wormhole lengths at different times, indicating a slow spread of wavepackets.]{\includegraphics[width=0.45\textwidth]{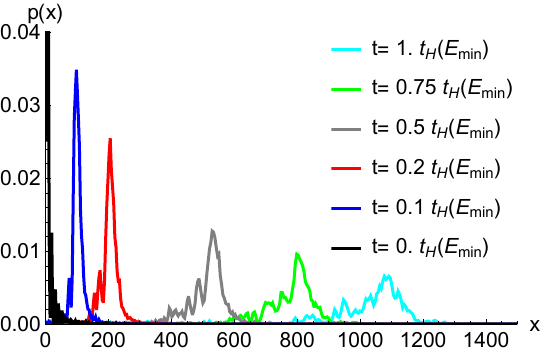}}
\label{fig:microcanonicalspread}
\qquad
\subfigure[Non-monotonicity in the expectation value $\langle x(t)\rangle$ of the wormhole length.]{\includegraphics[width=0.45\textwidth]{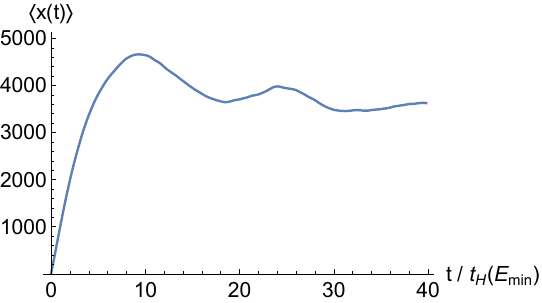}}
\label{fig:microcanonicalbump}
\caption{Numerical illustration of the dynamics of the wormhole length $x$ as a function of the time $t$ in a microcanonical state with minimum energy $E_{\min}$, associated with a Heisenberg time $t_{\htime}(E_{\min}) \sim \exp(S(E_{\min}))$. The slow spread of wavepackets up to late times and non-monotonicity in the expectation value of length are clear hallmarks of ergodic dynamics due to random matrix statistics in the fundamental theory. See Sec.~\ref{sec:wormholedynamics_numerics} for numerical details.}
\label{fig:microcanonicalnumerics}
\end{figure}

In Section \ref{sec:worm} we use this reconstruction map to study the length of the wormhole at long times. An unexpected analogy between quantum JT gravity and quantum ergodicity theory \cite{dynamicalqergodicity} is useful, and allows us to make new quantitative predictions about non-perturbative quantum gravity.  Firstly, we find that the average wormhole length $\langle x(t)\rangle$ is \emph{not monotonically increasing in time}: at times $t\sim t_{\mathrm{H}} \sim \mathrm{e}^{S(E)}$, $\langle x(t)\rangle$ grows above its late-time value before decreasing back.  This is in tantalizing agreement with an independent recent calculation \cite{Balasubramanian:2024lqk}, to which we will compare our results in Section \ref{sec:complexityconjecture}.  An explicit numerical calculation of the wormhole dynamics, describing both the quantum wave function and the averaged wormhole length, is depicted in Fig.~\ref{fig:microcanonicalnumerics}.    Secondly, we predict the magnitude of quantum fluctuations in wormhole length, finding that \begin{equation}
    \Delta x \sim \max\left(\mathrm{\Delta}x(0), x_{\mathrm{sat}} \exp\left[ - \left(\frac{t_{\mathrm{H}}}{t}\right)^2\right]\right)
\end{equation} at short times; here $\Delta x(0)$ is the initial uncertainty in position and $x_{\mathrm{sat}}$ is the late time expectation value of length. This is a clear prediction of our formalism, and a consequence of the random matrix statistics of the fundamental JT Hamiltonian \cite{Saad:2019lba}.  Thirdly, we predict the late time asymptotics of the time-evolved Hartle-Hawking state \eqref{eq:HHstate}: crudely speaking, the probability that the length is measured to lie in the interval $[x,x+\mathrm{d}x]$ is \begin{equation}
   \mathcal{P}(x)\mathrm{d}x \sim \frac{1}{\log (x\mathrm{e}^{-S_0})}\exp\left[-\frac{\beta}{8\pi^2}\left(\log \frac{x}{\mathrm{e}^{S_0}}\right)^2\right] \mathrm{d}x.
\end{equation}   
This wave function is quite heavy-tailed, implying substantial  fluctuations in the wormhole length at late times. 

In Section \ref{sec:velocity} we study the reconstruction of the velocity operator, which measures the growth rate of the wormhole length.
In Section \ref{sec:outlook} we discuss the implications of these results.


\section{``Reconstruction map'' of a free particle}\label{sec:freeparticle}

In this section we introduce, in a simple example, a ``correct'' reconstruction map for a toy model.
There are three takeaways:
\begin{enumerate}
    \item It is \emph{familiar} that a quantum theory with a continuous spectrum can be an effective description of a fundamental theory with a discrete spectrum.
    \item The relationship between these theories can be codified by two maps,
    \begin{equation}
    \begin{split}
        V &: \mathcal{H}_\mathrm{eff} \to \mathcal{H}_\mathrm{fund} \\
        R^* &: \mathrm{L}(\mathcal{H}_\mathrm{eff}) \to \mathrm{L}(\mathcal{H}_\mathrm{fund})
    \end{split}
    \end{equation}
    which respectively identify states and operators in the two theories.
    \item The ``reconstruction'' map $R^*$ is \emph{not} generally given by $R^*(O) = V O V^\dagger$.
\end{enumerate}
These lessons are important for us because in Section \ref{sec:pure} we work to construct an appropriate $R^*$ for JT gravity, which will not be given by simply conjugating by $V$.

\subsection{The line as an effective description}\label{sec:int-line-ex}

Imagine we have a free particle of mass $m$ on a finite interval, with Hilbert space
\begin{equation}
    \mathcal{H}_\mathrm{int} = \mathrm{L}^2\left([0,L]\right)~,
\end{equation}
and infinite potential walls at $0$ and $L$.
Say we are mainly interested in a state supported near $x = L/2$, dying off very quickly relative to $L$, such as (schematically)
\begin{equation}\label{fig:particle_on_interval}
    \includegraphics[width=0.8\textwidth]{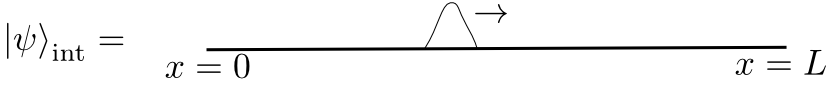}
\end{equation}
To a good approximation, for short enough times, the particle won't notice the finite length, and we can describe the physics of this state using an effective theory as a state of a free particle on an infinite line, with Hilbert space
\begin{equation}
    \mathcal{H}_\mathrm{line} = \mathrm{L}^2\left(\mathbb{R}\right)~,
\end{equation}
mapping the interval state \eqref{fig:particle_on_interval} to the line state
\begin{equation}\label{fig:particle_on_line}
   \includegraphics[width=0.75\textwidth]{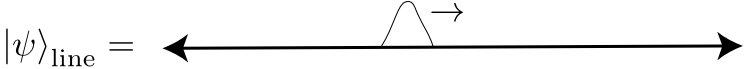}
\end{equation}
Intuitively, this is a good effective description because we can identify the $\ket{x}$ states and the $\hat{x}$ operators in the two systems, and then the expectation values match (for a while): $\braket{\hat{x}}_\mathrm{line} \approx \braket{\hat{x}}_\mathrm{int}$.
Note that the effective description has a continuous spectrum while that of the fundamental theory is discrete, analogous to the relationship between semiclassical JT and the dual quantum theory.

While intuitive, this effective theory functions differently than many other effective descriptions.
The regime of validity is not ``low energies'', but instead related to the complexity of the operation in question.
For example, the description becomes invalid under operations that probe energy differences of order $1/L$ or smaller.

\subsection{Codifying the relationship with $V$ and $R^*$}

We can formally describe the relationship between these two theories by defining two maps,
\begin{equation}
    \begin{split}
        V &: \mathcal{H}_\mathrm{line} \to \mathcal{H}_\mathrm{int} \\
        R^* &: \mathrm{L}(\mathcal{H}_\mathrm{line}) \to \mathrm{L}(\mathcal{H}_\mathrm{int})
    \end{split}
\end{equation}
which identify their states and operators respectively.
Properly defined, these maps will make it clear how to use the effective description, including determining its regime of validity.
In analogy to JT, we will refer to $V$ as the \emph{holographic} map and $R^*$ as the \emph{reconstruction} map.

The starting point is to notice that there is a natural map 
\begin{equation}
\begin{split}
    R&: \mathcal{H}_\mathrm{int} \to \mathcal{H}_\mathrm{line} \\
    &: \ket{n}_\mathrm{int} \to \sqrt{\frac{2}{L}} \int\limits_0^L \mathrm{d}x\, \sin\left(\frac{n \pi x}{L}\right) \ket{x}_\mathrm{line}
\end{split}
\end{equation}
which takes a wave function on the interval and maps it to the same on the line, vanishing outside $x \in [0,L]$.

The reconstruction map $R^*$ is naturally defined as the pullback of this $R$,
\begin{equation}
    R^*(O) = R^\dagger O R
\end{equation}
for $O \in \mathrm{L}(\mathcal{H}_\mathrm{line})$. 
This gives satisfying reconstructions, for example for the identity, Hamiltonian, position, and momentum,
\begin{equation}
\begin{split}
    R^*(\mathbb{1}_\mathrm{line}) &= R^\dagger R = \mathbb{1}_\mathrm{int}~,\\
    R^*(H_\mathrm{line}) &= R^\dagger H_\mathrm{line} R = H_\mathrm{int}~,\\
    R^*(x_\mathrm{line}) &= R^\dagger x_\mathrm{line} R = x_\mathrm{int}~,\\
    R^*(p_\mathrm{line}) &= R^\dagger p_\mathrm{line} R = p_\mathrm{int}~.
\end{split}
\end{equation}

The map $V$ can be constructed as follows.
$V$ will identify states in the effective and fundamental theories, so it is reasonable that it should invert the natural identification given by $R$.
Hence we demand
\begin{equation}\label{eq:V_inverts_R}
    V R = \mathbb{1}_\mathrm{int}~.
\end{equation}
However, this does not fully specify $V$, because the image of $R$ does not include all of $\mathcal{H}_\mathrm{line}$.
We still have to decide how $V$ acts on wave functions supported outside $x \in [0,L]$.
One option is to simply annihilate the part of the state outside that range, but this has the undesirable property that $V$ would not commute with time evolution.
A more useful choice is to define $V$ to satisfy
\begin{equation}\label{eq:warmup_equivariance}
    V H_\mathrm{line} = H_\mathrm{int} V~.
\end{equation}
This is nice for two reasons.
Primarily, it allows the effective description to be a useful description of the fundamental dynamics; we don't lose anything by time evolving in the effective description.
Secondarily, this is a key property satisfied by the holographic map \eqref{eq:V_JT} in JT gravity, so \eqref{eq:warmup_equivariance} ensures we are studying an effective theory analogous to semiclassical JT.

The condition \eqref{eq:warmup_equivariance} ensures $V$ acts diagonally in the energy eigenbasis.
This creates a subtlety: energy eigenstates are true elements of $\mathcal{H}_\mathrm{int}$, but technically not of $\mathcal{H}_\mathrm{line}$.
To construct a well-defined $V$, we will proceed by using $\varepsilon$-approximate eigenstates, and demanding only that \eqref{eq:V_inverts_R} and \eqref{eq:warmup_equivariance} hold in the $\varepsilon \to 0$ limit.

Consider the states
\begin{equation}
    \ket{s_{E,\varepsilon}}_\mathrm{line} = \frac{\sqrt{2 \varepsilon}}{(2 \pi)^{\frac{1}{4}}} \int_{-\infty}^\infty \mathrm{d}x\, \mathrm{e}^{-x^2 \varepsilon^2/4} \sin(\sqrt{2mE}x) \ket{x}_\mathrm{line}~.
\end{equation}
These are approximately normalized states of approximately definite energy, with both approximations improving quickly as $\varepsilon \to 0$.
We could analogously define states $\ket{c_{E,\varepsilon}}_\mathrm{line}$ which replace the sine with cosine, and together these states for $E > 0$ would span $\mathcal{H}_\mathrm{line}$.

We define $V_\varepsilon: \mathcal{H}_\mathrm{line} \to \mathcal{H}_\mathrm{int}$ as
\begin{equation}
    V_\varepsilon = \frac{(2 \pi)^{\frac{1}{4}}}{\sqrt{\varepsilon L}} \sum_n \ket{n}_\mathrm{int} \bra{s_{E_n,\varepsilon}}_\mathrm{line}~,
\end{equation}
where $E_n$ are the eigenvalues of $H_\mathrm{int}$. 
It is straightforward to show this satisfies conditions \eqref{eq:V_inverts_R} and \eqref{eq:warmup_equivariance} with error $O(\varepsilon)$. 
Moreover, $V_\varepsilon$ preserves the normalization of wave functions supported entirely in $x \in [0,L]$, up to terms that vanish as $\varepsilon \to 0$.
We can make $\varepsilon$ as small as we want, so we will drop $\varepsilon$ going forward, imagining that we are working in the $\varepsilon \to 0$ limit.\footnote{For some states, $\lim_{\varepsilon \to 0} V_\varepsilon \ket{\psi} $ will not converge. We regard those $\ket{\psi}$ as outside the regime of validity of the effective theory.}

With these maps $V$ and $R^*$ in hand, we have formalized the sense in which the particle on a line is an effective description of the particle on the interval.
Given some operator and state on the line, these maps tell us what we are supposed to compare to on the interval.
They also allow us to determine the regime of validity of the effective description.
We say it is valid for a given state $\ket{\psi} \in \mathcal{H}_\mathrm{line}$ and observable $O \in \mathrm{L}(\mathcal{H}_\mathrm{line})$ if
\begin{equation}\label{eq:line_int_approx}
    \braket{\psi | O | \psi} \approx \braket{\psi | V_\varepsilon^\dagger R^*(O) V_\varepsilon | \psi}
\end{equation}
to within some desired accuracy. 
For relatively localized line wave functions $\psi(x)$ and simple operators like $x_\mathrm{line}$, \eqref{eq:line_int_approx} will hold.
However, \eqref{eq:line_int_approx} will be badly violated for more complicated operators, like the time evolved position operator $x(t)_\mathrm{line}$ for very large $t$.
For such cases, the correct answer is computed by the right hand side of \eqref{eq:line_int_approx}; the left hand side is no longer a reliable approximation.
This is analogous to JT, in which semiclassical JT fails to compute the correct wormhole length at long times, but we can still compute the correct length by correctly reconstructing the state with $V$ and the operator with $R^*$.

\subsection{Lessons for JT}
The point of this interval/line example was to have a system that we understood completely, whose $V$ and $R^*$ were intuitive.
That way we can study how these $V$ and $R^*$ are related, and extract lessons for what $R^*$ might be in JT.

Recall that one idea, from \cite{Iliesiu:2024cnh}, is 
\begin{equation}\label{eq:iliesiu_rec_ex}
    R_I^*(O) := V_\varepsilon O V_\varepsilon^\dagger~.
\end{equation}
This is very different from the $R^*$ we constructed in Section \ref{sec:int-line-ex}.
For example,
\begin{equation}
    R_I^*(x_\mathrm{line}) = \sum_{n,n'} \ket{n}\bra{n'}_\mathrm{int} \frac{2}{L} \int\limits_{-\infty}^\infty \mathrm{d}x \, x \sin\left( \frac{n \pi}{L} x \right) \sin\left(\frac{n' \pi}{L} x\right) \mathrm{e}^{-x^2 \varepsilon^2/4}
\end{equation}
which is \emph{not} equal to $x_\mathrm{int}$, and in fact has vanishing matrix elements.
This is a general problem with $R_I^*$: it tends to output vanishing or divergent matrix elements, unless the operator is finely tuned.
At any rate, it doesn't give the intuitively correct reconstruction in this line/interval example.
We conclude that conjugation by the holographic map $V$ is not in general the right way to reconstruct operators.

Starting in the next section, we apply this lesson to JT.
We will regard the difficulties with \eqref{eq:iliesiu_rec_ex} in JT as a sign that it is not the correct reconstruction map and we will look for another.

Note a crucial difference between JT and this line/interval example.
In JT, we are starting by knowing $V$ and would like to deduce $R^*$.
This is trickier than the converse, because $V$ can be straightforwardly determined from $R$ via \eqref{eq:V_inverts_R} and \eqref{eq:warmup_equivariance}. 
We'll have to figure out the appropriate $R^*$ by different means.

\section{Finding $R^*$ for JT gravity}\label{sec:pure}
Returning to JT gravity,  we have a holographic map $V: \mathcal{H}_0 \to \mathcal{H}$ given by \cite{Iliesiu:2024cnh}
\begin{equation}\label{eq:V_sec3}
    V \ket{x}_0 = \mathrm{e}^{-S_0/2} \sum_{i = 0}^\infty \phi_{E_i}(x) \ket{E_i}~,
\end{equation}
which tells us how to identify states in the effective and fundamental theories.
The reconstruction map $R^*: \mathrm{L}(\mathcal{H}_0) \to \mathrm{L}(\mathcal{H})$ will tell us how to identify operators.
In Section \ref{sec:freeparticle} we argued that in fact we should \emph{not} in general take $R^*(O) = V O V^\dagger$, even though such a map seems natural.

In this section we will build another reconstruction map.  
This problem is more challenging than the warmup from Section \ref{sec:freeparticle}, because here we start with $V$ and try to build $R^*$, rather than the other way around.
We do this following one guiding principle: 
\begin{center}
\emph{Given a semiclassical state and semiclassical operator, the reconstruction of that operator should match the semiclassical answer to as good an approximation as possible, at short times.}
\end{center}
As a consequence, we will also find that the reconstruction closely matches the semiclassical answer all the way to time scales $\mathrm{e}^S/\sqrt{S}$, nearly the Heisenberg time.\footnote{We expect (but do not prove) that this guiding principle implies that, in generic cases, the dynamics of reconstructed operator remains close to its semiclassical dynamics for the longest possible time, relative to other possible reconstructions of that operator.}

Let us start to make this more precise.
By definition, the reconstruction of an operator $O$ on a state $\ket{\psi}$ has error $\varepsilon$ if
\begin{equation}\label{eq:rec_def}
    \lVert R^*(O) V \ket{\psi} - V O \ket{\psi} \rVert < \varepsilon~.
\end{equation}
Because $V$ is diagonal in the energy basis \cite{Iliesiu:2024cnh}, it follows that the effective theory Hamiltonian $H_0$ is perfectly reconstructed by the fundamental theory Hamiltonian $H$ on all states, which encourages us to define $R^*$ to satisfy
\begin{equation}\label{eq:H0_rec}
    R^*(H_0) = H~.
\end{equation}
This does not yet fully specify $R^*$.
To further constrain it we would like to decide how it acts on the conjugate to $H_0$, called the ``time-shift'' operator $\delta$, which satisfies
\begin{equation}
    \mathrm{e}^{\mathrm{i} H_0 t} \delta \mathrm{e}^{-\mathrm{i} H_0 t}  = \delta + t~.
\end{equation}
Unfortunately, this runs into two complications. One~\cite{Harlow:2018tqv} is that $\delta$ is technically not a well-defined self-adjoint operator because $H_0$ is bounded below, but let us initially proceed as though it were, to illustrate the idea. Another is that in the fundamental theory, $H$ has an irregular random matrix spectrum, and therefore cannot generate perfect shifts of an orthonormal basis of $\delta$-eigenstates. We will have to address these issues by forming an \textit{approximate} conjugate relationship between $H$ and $R^*(\delta)$ that upholds our guiding principle, using tools from quantum ergodicity~\cite{dynamicalqergodicity}.
The rest of this section will explain how to do this more carefully.

We want to decide what operator in the fundamental theory should be $R^*(\delta)$.
A priori there are many options, but our guiding principle instructs us to look for an operator that minimizes the reconstruction error on semiclassical states.
For definiteness, let us consider $\ket{\psi} \in \mathcal{H}_0$ to be the Hartle-Hawking state projected onto a microcanonical window of width $\Delta E \sim O(1)$, and let $\ket{\psi(t)} := \mathrm{e}^{-\mathrm{i}H_0 t} \ket{\psi}$.
We will regard these $\ket{\psi(t)}$ as our semiclassical states, at least for $t$ small compared to the Heisenberg time. 
We would like to find an operator $R^*(\delta)$ in the fundamental theory that satisfies
\begin{equation}\label{eq:delta_rec}
    \lVert R^*(\delta) V \ket{\psi(t)} - V \delta \ket{\psi(t)} \rVert < \varepsilon
\end{equation}
for as small an $\varepsilon$ as possible at short times $t$.
It is helpful to phrase this purely in terms of properties that must be satisfied by the operator $R^*(\delta)$.
This comes out to be
\begin{equation}
    \mathrm{e}^{\mathrm{i} H t} R_\mathrm{MC}^*(\delta) \mathrm{e}^{-\mathrm{i} H t} \simeq R_\mathrm{MC}^*(\delta) + t~,
\end{equation}
where $R_\mathrm{MC}^*(\delta)$ is the operator $R^*(\delta)$ projected into the microcanonical energy window, and with the approximation as good as possible for small $t$.
We will argue this essentially fixes $R_\mathrm{MC}^*(\delta)$, and furthermore that we can effectively stitch these together for each microcanonical window to construct the full $R^*(\delta)$.
The action of $R^*$ on a general operator $\hat{A}$ we then define roughly by writing $\hat{A}$ in terms of $H_0$ and $\delta$ and then replacing each $H_0$ and $\delta$ with their reconstructions.

To make this more rigorous, we first need to discuss an operator pair other than $(H_0, \delta)$ that is more easily defined.
However, we would still like them to play the same roles, with one operator related to time evolution, so that we can apply our guiding principle as above.
A straightforward way to do this is to introduce \emph{action-angle} variables $(N,\theta)$ and then define $R^*$ by its action on those.

We begin in Section \ref{sec:reform_eff_JT} with classical JT gravity, showing how to define suitable action-angle variables that when quantized lead to action and angle operators in the effective theory.
In Section \ref{sec:reform_fund_JT}, we construct action and angle operators in the fundamental theory.
In Section \ref{sec:reconstructionmap} we explain that our guiding principle is best upheld by a particular identification of the action and angle operators in the effective and fundamental theories, and we also explain how this defines the reconstruction map. 
This $R^*$ will not be fully unique, but we will argue that any $R^*$ satisfying the guiding principle will be close to this one.
In Section \ref{sec:length_rec} we apply this to the length operator $x$ to obtain an explicit reconstructed operator, which we study in later sections.

\subsection{Action-angle variables in the effective theory}\label{sec:reform_eff_JT}
We start with classical JT gravity, to define ``action'' and ``angle'' variables that will be useful for defining a reconstruction map following our guiding principle. 

As reviewed in the introduction and derived in \cite{Harlow:2018tqv}, the phase space of classical JT gravity is $\mathbb{R}^2$ with canonical symplectic form \begin{equation}
    \omega = \mathrm{d}p\wedge \mathrm{d}x
\end{equation}
and Hamiltonian function
\begin{equation}
     H = \frac{p^2}{2}+2\mathrm{e}^{-x}~.
     \label{eq:jtHamiltonianclassical}
 \end{equation}
  Classically, trajectories corresponding to an incoming particle from $x=\infty$ with $p<0$, bouncing off the ``soft wall" exponential at $x_{\mathrm{min}}=\log (2/E)$, and then traveling to the right: as $t\rightarrow \infty$, for some constant offset $t_0$, \begin{equation}
    x(t) = \sqrt{2E}(t-t_0).  \label{eq:xtclassical}
\end{equation}

All time-independent classical Hamiltonian systems on a two-dimensional phase space are (locally) integrable.  We can then look for ``action-angle" variables where the integrability is manifest, and $H$ only depends on the momentum (action) variable.  This has been achieved in \cite{Harlow:2018tqv}.  Define the canonical variables
\begin{subequations}\label{eq:PXintermsofpx}\begin{align}
    \widetilde{P} &:= \sqrt{p^2+4\mathrm{e}^{-x}}, \\
    \widetilde{X} &:= -x -2\log \frac{\sqrt{p^2 +4 \mathrm{e}^{-x}} - p}{2} , \label{eq:Xsec2}
\end{align}\end{subequations}
where we note that $\widetilde{X}>0$ when $p>0$.
This $\widetilde{P}$ is (up to a constant) just $\phi_h$, the dilaton value at the black hole horizon. Similarly, $\tilde{X}$ should be identified (up to a constant) with $\phi_h \delta$, where $\delta$ is the time-shift operator across the two boundaries of the wormhole.
Now the phase space is $\mathbb{R}_+\times \mathbb{R}$ with coordinates $\widetilde{P}>0$, symplectic form \begin{equation}
    \omega = \mathrm{d}\widetilde{P}\wedge \mathrm{d}\widetilde{X},
\end{equation}
and Hamiltonian \begin{equation}
    H = \frac{\widetilde{P}^2}{2}.
\end{equation}
The inverse transformation is \begin{subequations}
    \begin{align}
        p &= \widetilde{P} \tanh \frac{\widetilde{X}}{2}, \\
        x &= 2\log \frac{2\cosh \frac{\widetilde{X}}{2}}{\widetilde{P}}. \label{eq:xXP}
    \end{align}
\end{subequations}
Classical trajectories that are incoming from $x=+\infty$ with $p<0$ have $\widetilde{X} \rightarrow -\infty$, while the outgoing trajectory with $p>0$ has $\widetilde{X} \rightarrow+\infty$.

In a system with closed orbits, the action-angle variables $(\widetilde{P}, \widetilde{X})$ would be uniquely defined by the periodicity requirement $\widetilde{X}\sim \widetilde{X}+2\pi$.  In classical JT gravity, we do not have closed orbits.   This means that there is an ambiguity in the action-angle variables.   An alternative choice 
\begin{subequations} \label{eq:bothtildetransforms}
\begin{align}
    \widetilde{P} &:= F(P)~, \label{eq:tildetranformsP}\\
    \widetilde{X} &:= \frac{X}{F'(P)}~.\label{eq:tildetransforms}
\end{align}
\end{subequations}
also has the same symplectic form and phase space, so long as $F(P)$ is monotonically increasing (all canonical transformations must be invertible).  The Hamiltonian is now
 \begin{equation}
     H = \frac{1}{2}F\left(P\right)^2~.
     \label{eq:JTactionenergy_classical}
 \end{equation}

The effective theory $\mathrm{JT}_\infty$ comes from imposing the canonical commutation relations $[\hat{x},\hat{p}]= \mathrm{i}$.
We will define action and angle operators $\hat{P}$ and $\hat{X}$ by their classical relationship to $x,p$ and promoting $x,p$ to operators:
\begin{equation}
\begin{split}
    \hat{P} &= P(\hat{x},\hat{p}) \\
    \hat{X} &= X(\hat{x},\hat{p})
\end{split}
\end{equation}
There are operator ordering ambiguities to worry about, but we will argue later that they lead to only small differences in the questions we study.
Note here we temporarily wrote hats for clarity, but from now on will go back to not using them; \edit{unless explicitly indicated otherwise, $(P,X)$ will henceforth refer to quantum operators acting on the effective Hilbert space $\mathcal{H}_0$.}

\edit{At the level of the effective theory}, we stress that there is no reason to prefer one $(X,P)$ action-angle pair over any other.  But, when we build a reconstruction map in Section \ref{sec:reconstructionmap}, we will explain why there is a preferred choice of $F$ in \eqref{eq:JTactionenergy_classical} that correctly reproduces the spectrum of the fundamental theory.


\subsection{Action-angle operators in the fundamental theory}\label{sec:reform_fund_JT}

Our goal is to define the reconstructions $R^*(P)$ and $R^*(X)$ of the action and angle operators.
Before doing this, it will help to introduce a kind of action $N$ and angle $\theta$ operators in the fundamental theory.
We will see in Section \ref{sec:reconstructionmap} that our guiding principle is best upheld by identifying a particular pair of these as the reconstructions of $P$ and $X$.

Crucially, these fundamental theory action and angle operators must have different properties than their effective theory counterparts.
$P$ and $X$ both had continuous, unbounded spectra (except $P \ge 0$).
In the fundamental theory, the action operator $N$ will have a discrete spectrum because the Hamiltonian does.
Notably though, we will explicitly construct it to have a regular integer spectrum like the harmonic oscillator in place of a random matrix spectrum.
This will allow us to introduce a periodic variable $\theta$ with which $N$ has a natural conjugate relationship, and in fact it will be better defined to deal with $\mathrm{e}^{\pm \mathrm{i} \theta}$ instead of $\theta$ directly.
This reconstruction of unbounded $X$ with bounded $\theta$ will underlie the late time features that show up in the fundamental theory.

To build the operators, consider the fundamental theory's countably infinite Hilbert space
\begin{equation}
    \mathcal{H} = \mathrm{L}^2\left(\left\lbrace |0\rangle, |1\rangle, |2\rangle, \ldots \right\rbrace\right), \label{eq:fundamentalhilbert}
\end{equation}
where these basis states are arbitrary, so far \emph{not} related in any particular way to the eigenbasis of the Hamiltonian $H$.
We define action and angle operators respectively as
\begin{subequations}\label{eq:definefundamentaloperators}\begin{align}
    N|n\rangle &= n |n\rangle, \label{eq:Neigenstates}\\
    \mathrm{e}^{\pm \mathrm{i}\theta} |n\rangle &= |n\pm 1\rangle \label{eq:theta_Ntranslationgenerators}
\end{align}\end{subequations}
We will also call $N$ the ``number operator'' and $\mathrm{e}^{\pm \mathrm{i}\theta}$ the raising/lowering operators.
Here $\ket{-1} = 0$ is a null vector. 
Note this implies
\begin{equation}
    \left[N,\mathrm{e}^{\mathrm{-i}\theta}\right] = -\mathrm{e}^{-\mathrm{i}\theta},
    \label{eq:actionangle_canonicalcommutator}
\end{equation}
which is similar to the harmonic oscillator algebra, however here $\mathrm{e}^{\mathrm{i}\theta}\mathrm{e}^{\mathrm{-i}\theta} \neq N$ (where $\mathrm{e}^{\mathrm{i}\theta} = (\mathrm{e}^{\mathrm{-i}\theta})^\dagger)$.
Also note that since $N$ is not directly given in terms of $\mathrm{e}^{\pm \mathrm{i} \theta}$, the formula for the raising and lowering operators are modified from the textbook form for the harmonic oscillator.  
Note too that $[\mathrm{e}^{\mathrm{i}\theta}, \mathrm{e}^{-\mathrm{i}\theta}]\ne 0$ because  $\mathrm{e}^{\mathrm{i}\theta}\mathrm{e}^{-\mathrm{i}\theta}|0\rangle = 0$.

We emphasize again that we have not yet related the $\ket{n}$ basis to the energy eigenbasis $\ket{E_n}$.
Different relationships define different pairs of operators $N,\mathrm{e}^{\pm \mathrm{i}\theta}$.
We have left this unfixed so that we can determine which choice best upholds our guiding principle.
That is, in the next section we will consider reconstructing $P$ and $X$ roughly as $N$ and $\theta$, and then ask which particular $N$ and $\theta$ make the reconstruction error the smallest at early times.
We will claim (and argue in Appendix \ref{app:ergodic}) that the answer is to line up the bases of $N$ and $H$, identifying $\ket{n} = \ket{E_n}$.


Let us comment on why we have defined $\mathrm{e}^{\pm \mathrm{i}\theta}$ instead of $\theta$ directly.
While naively we would have liked to define a conjugate ``angle'' operator $\theta$ satisfying $\theta \in (-\pi, \pi]$ and $[N,\theta] = -\mathrm{i}$, unfortunately that periodicity and $N \ge 0$ imply such a $\theta$ would not be a well-defined self-adjoint operator on $\mathcal{H}$.
Defining $\mathrm{e}^{\pm\mathrm{i}\theta}$ is a suitable workaround which would imply the canonical commutator if $\theta$ were well-defined, and does correctly account for the periodicity \cite{SusskindGlogowerPhaseOperator, CarruthersNietoPhaseAngle}.

In the absence of a well-defined $\theta$ operator, it is not possible to define eigenstates of $\theta$ in any direct sense.
For example, it was noted in \cite{SusskindGlogowerPhaseOperator, CarruthersNietoPhaseAngle} that $\cos \theta$ and $\sin \theta$ are well-defined \emph{non-commuting} operators, when given by the conventional linear combinations of $\mathrm{e}^{\pm\mathrm{i}\theta}$, so they cannot share an eigenbasis of $\theta$-states.
So instead, we treat the angle variable with the following strategy.
As with the harmonic oscillator, the lowering operator $\mathrm{e}^{\mathrm{-i}\theta}$ admits (non-normalizable) eigenstates $\theta$ with respective eigenvalues $\mathrm{e}^{\mathrm{-i}\theta}$, $\theta \in (-\pi,\pi]$, which form an overcomplete set of coherent states:
\begin{equation}
    \lvert \theta\rangle = \sum_{n=0}^{\infty} \mathrm{e}^{-\mathrm{i}n\theta}\lvert n\rangle.
    \label{eq:fundamental_angle_coherentstates}
\end{equation}
For this family of states, $N$ acts like a generator of translations, where the addition of $\theta$-values is modulo $2 \pi$:
\begin{equation}
    \mathrm{e}^{-\mathrm{i}N \theta_0} \lvert \theta\rangle = \lvert \theta + \theta_0\rangle.
    \label{eq:N_thetaTranslationGenerator}
\end{equation}

\subsection{Defining the reconstruction map}
\label{sec:reconstructionmap}


We now propose a reconstruction map, which we argue best upholds our guiding principle. (The full technical argument can be found in Appendix \ref{app:ergodic}.)
Loosely speaking, the idea is to define $R^*(P) = N$ and $R^*(\mathrm{e}^{\mathrm{i}X}) = \mathrm{e}^{\mathrm{i} \theta}$. \edit{We emphasize that these are maps between operators: $(P,X)$ act on the effective Hilbert space $\mathcal{H}_0$, while $(N,\theta)$ act on the fundamental Hilbert space $\mathcal{H}$.}
This itself is somewhat too quick though, because it does not fully specify how a general function $f(X)$ is to be reconstructed.

We really define $R^*$ as follows. Consider a general effective theory operator written as a function $A(X,P)$. \edit{Let $\clas{A}(\clas{X},\clas{P})$ be the classical variable corresponding to this operator (here, we introduce subscripts $\clas{\cdot}$ to differentiate classical variables from effective and fundamental operators).}
Define \edit{the classical function $\clas{\widetilde{A}}(\clas{k},\clas{P})$ via the Fourier transform
\begin{equation}
    \clas{\widetilde{A}}(\clas{k},\clas{P}) = \frac{1}{2 \pi} \int\limits_{-\pi}^\pi \mathrm{d}\clas{X} \, \mathrm{e}^{-\mathrm{i}\clas{k}\clas{X}} \clas{A}(\clas{X},\clas{P})~,
\end{equation}
}and then, \edit{after promoting $\clas{P}$ to the fundamental theory operator $N$ but retaining $\clas{k}$ as a classical variable,} define the reconstruction map \edit{on \emph{quantum operators} $A(X,P)$ in the effective theory as
\begin{equation}
    R^*\left(A(X,P)\right) = \sum_{\clas{k} = -\infty}^\infty \frac{1}{2}\left\lbrace \mathrm{e}^{\mathrm{i}\clas{k}\theta},\widetilde{A}(\clas{k},N)\right\rbrace~. 
    \label{eq:globalreconstructionmap_def}
\end{equation}
Here, $\widetilde{A}(\clas{k},N)$ is a well-defined operator in the fundamental theory via the replacement $\clas{P} \to N$ if, e.g., $\clas{\widetilde{A}}(\clas{k},\clas{P})$ is defined through a power series in $\clas{P}$, as it only involves the single operator $N$ and therefore no commutation ambiguities.}
What we have done is to define the reconstruction map via the Fourier transform, truncating the range of $X$ from $(-\infty,\infty)$ to $[-\pi,\pi)$.
Note that this reconstruction map is linear, because the Fourier transform is linear.  While this map is not an algebra homomorphism, it is approximately one on very high-energy states: see Appendix \ref{app:actionanglereconstruction}.  

Two comments on why we choose to consider this map.
One, it allows us to define reconstruction with this specification that $R^*(\mathrm{e}^{\mathrm{i}X}) = \mathrm{e}^{\mathrm{i}\theta}$.
Two, we truncate the range of $X$ in preparation for replacing $X$ with $\theta$, otherwise it would over count.

To finish defining the reconstruction map, we need to specify exactly which $N, \mathrm{e}^{\mathrm{i} \theta}$ to use, i.e. specify how the basis $\ket{n}$ relates to the energy eigenbasis $\ket{E_n}$.
We argue in Appendix \ref{app:ergodic} that there is a ``best choice": align the bases $\ket{n} = \ket{E_n}$.
We show that with this choice, our guiding principle is best upheld.
This brings along a number of nice features, such as $R^*(H_0) = H$.\footnote{\label{footnote:suboptimalactions}One could make other choices, e.g. $N' = U_{\Delta n} N U_{\Delta n}^\dagger$, where $U_{\Delta n}$ is a block diagonal permutation matrix mixing $|n\rangle$ states within a window of size $\Delta n$.  At short times, the predictions of this theory would be close to semiclassical predictions, but not \emph{as close} as with our reconstruction.  This permuted theory would have very different dynamics at late times, however.}

\begin{figure}
    \centering
    \includegraphics[width=0.5\linewidth]{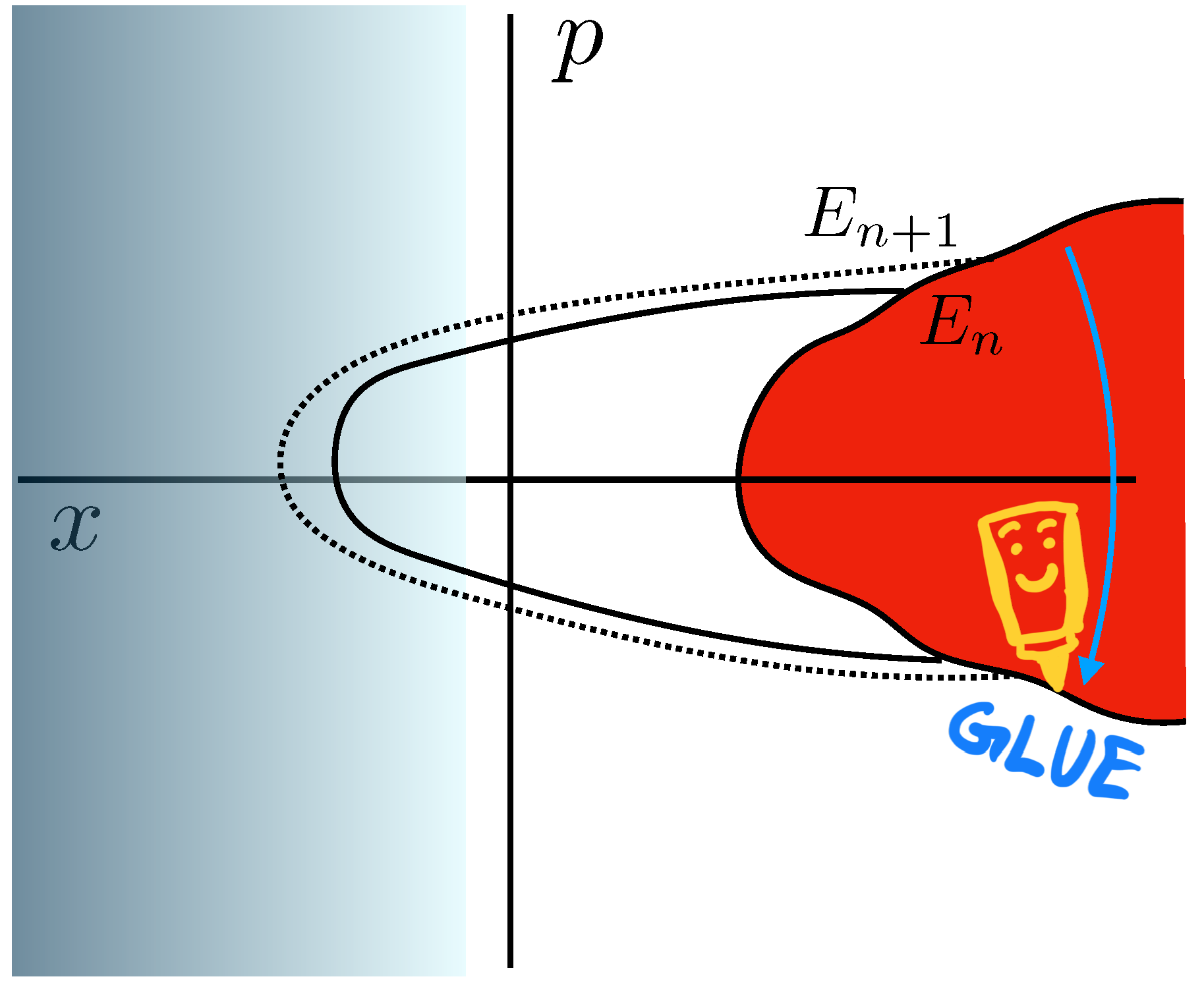}
    \caption{A cartoon of the modification of phase space suggested by our proposed reconstruction map, which truncates the range of $X$ from $(-\infty,\infty)$ to $[-\pi,\pi)$. In $(x,p)$ coordinates, this deletes the red region, which consists of wormholes long compared to $\mathrm{e}^{S(E)}$. The boundary conditions we place at these new walls can be understood as gluing the remaining phase space in the way depicted, making the paths cyclic. }
    \label{fig:Vwall}
\end{figure}

Let us comment further on the physical picture that comes from this reconstruction.
We might think of the truncation of $X$ as a particular modification of phase space, where the region of large $X$ is excised and the remaining portion glued together, which makes the paths in phase space cyclic, as in Figure \ref{fig:Vwall}.

This cut-and-glued picture of phase space illustrates an important idea underlying this reconstruction map, why it makes sense for JT gravity but not all single particle quantum systems.
The original paths in phase space were not closed, hence there were no ``ordinary'' action-angle variables; we had the freedom to redefine the action-angle variables, as in \eqref{eq:bothtildetransforms}.
This allowed us to find action-angle variables $P,X$ whose dynamics closely match those of $N,\theta$ in the fundamental theory (at short times).

For comparison, the same strategy would not have been available if the effective theory had been the harmonic oscillator.
Even though we could have defined action-angle variables, their dynamics would not have matched any action-angle operators in the fundamental theory.
And there would have been no freedom to look for alternative action-angle variables that would have matched more closely, in part because the density of states grows so much faster than that of the harmonic oscillator.

As an exercise, we can build an approximate time shift operator $\delta$ in the fundamental Hilbert space. 
Using \eqref{eq:semiclassicalThetaVsDelta} in Appendix~\ref{app:ergodic}, we can show that the fundamental theory variables $\delta$ and $\theta$ should be related by:
\begin{equation}
    \delta := \frac{1}{2}\lbrace \rho(H), \theta\rbrace = \frac{\mathrm{i}}{2} \sum_{\substack{n,m = 0:\\ n \neq m}}^{\infty} \frac{(-1)^{n-m} [\rho(E_n) + \rho(E_m)]}{n-m} \lvert n\rangle \langle m\rvert~.
    \label{eq:timeshiftglobalreconstruction}
\end{equation}
Notice that $\delta$ essentially reduces to a rescaled version of $\theta$ within narrow energy windows as required.  The results of \cite{dynamicalqergodicity} (see Appendix \ref{app:ergodic}) then imply $\delta$ is as close as possible to an observable whose shifts are generated by $H$ at early times.

\subsection{Explicit reconstruction of the length operator}\label{sec:length_rec}

We now use this reconstruction map on a specific example, the wormhole length operator $x$.
First, we must carry out the classical canonical transformations to deduce $x$ in terms of $P,X$. 
Then we replace $P$ with $N$ and $\mathrm{e}^{\mathrm{i} X}$ with $\mathrm{e}^{\mathrm{i} \theta}$. 
Using \eqref{eq:xXP}, \eqref{eq:bothtildetransforms} we find
\begin{equation}
    x(N,\theta) = 2\log \frac{2\cosh \frac{\theta}{2F^\prime(N)}}{F(N)}  \label{eq:xNtheta}
\end{equation}
To get a usable approximation for $x$, we need to plug in a specific form for $F$ and then write this operator in terms of its matrix elements in the energy basis.
For this reason it will be convenient to write the operator in terms of the Hamiltonian $H$.

To obtain an explicit form for $F$, we proceed as follows.
Using \eqref{eq:JTactionenergy_classical} and $R^*(P) = N$, we can write the fundamental Hamiltonian as
\begin{equation}\label{eq:fund_ham}
    H = \frac{1}{2} F(N)^2~,
\end{equation}
where the discrete spectrum of $N$ enforces
\begin{equation}
    F(n) = \sqrt{2 E_n}~,
    \label{eq:fund_Fdef}
\end{equation}
with $E_n$ the fundamental theory spectrum $\{E_n\}$.
In theory this determines a lot about $F$ (though not completely).
In practice, we will want to be more explicit.
At first this seems difficult, because this $F$ would need to be some function fitting infinitely many random values of $E_n$.
However, for the calculations we want to do, it will suffice to define $F$ coarsely.
While fluctuations in $E_n$ vs. $E_{n+1}$ are drastic from one draw of the ensemble to the next, the overall shape of the distribution hardly fluctuates, and is what's responsible for the qualitative features we study.
A sufficient coarse definition can be built using the leading order average density of states \eqref{eq:JTinftyDOS}.  
Let $N(E)$ denote the number of eigenstates $\le E$; then 
\begin{align}
    \mathbb{E}[N(E)] &=  \mathrm{e}^{S_0} \int\limits_{0}^E\mathrm{d}E^\prime \; \rho_0(E^\prime) = \mathrm{e}^{S_0} \frac{\sqrt{2E}\cosh(2\pi \sqrt{2E})}{(2\pi)^3} -\mathrm{e}^{S_0}  \frac{\sinh(2\pi\sqrt{2E})}{(2\pi)^4} \notag \\
    &\approx \mathrm{e}^{S_0} \frac{\sqrt{2E}\mathrm{e}^{2\pi \sqrt{2E}}}{2(2\pi)^3} \approx \sqrt{\frac{E}{2\pi^2}}\rho(E)  \label{eq:NE}
\end{align}
with $\rho(E)$ given by \eqref{eq:rhoE}. For a random matrix \cite{zdbai} \begin{equation}
    \frac{\sqrt{\mathrm{Var}[N(E)]}}{\mathbb{E}[N(E)]} \lesssim N(E)^{-1/4},
\end{equation}
implying that for $N(E)\gg 1$ (as we will always have for the semiclassical states of interest in this paper), the function $F$ is universal up to extremely small ($\lesssim \mathrm{e}^{-S/4}$) fluctuations. 
Because we need\begin{equation}
    N(E)= F^{-1}\left(\sqrt{2E}\right), \label{eq:Finverse}
\end{equation}
comparing (\ref{eq:NE}) with (\ref{eq:Finverse}) we deduce that 
\begin{align}
    F(N) &\approx \frac{1}{2\pi}\left[\log \frac{N}{\mathrm{e}^{S_0}} - \log\left(\frac{1}{2\pi}\log \frac{N}{\mathrm{e}^{S_0}}\right) + \cdots\right] \approx \frac{1}{2\pi} \log \frac{N}{\mathrm{e}^{S_0}}. \label{eq:FPavg}
\end{align}
For future reference, we also remark that the choice of $F$ is (up to the statistical fluctuations in $E_n$) related to the density of states $\rho(E)$ via \begin{equation}
    \mathrm{e}^{S_0}\rho_0(E) = \mathrm{e}^{S(E)} \approx \frac{1}{F(N(E))F^\prime(N(E))} = \frac{1}{\sqrt{2E} F^\prime(N)}. \label{eq:rho0Frelation}
\end{equation}
Therefore, we can write the fundamental theory Hamiltonian as 
\begin{equation}
    H = \frac{1}{2}F(N)^2 \approx \frac{1}{8\pi^2}\left(\log \frac{N}{\mathrm{e}^{S_0}}\right)^2. \label{eq:JTH_quantum}
\end{equation}

Returning to the length operator, we use \eqref{eq:FPavg} to get
\begin{equation}
    x(N,\theta) \approx \sqrt{2H} \rho(H) |\theta| - \log (2H) +\cdots
    \label{eq:lengthoperatorabstheta}
\end{equation}
To find the matrix elements, it is useful to rewrite $|\theta|$, defined assuming that $\theta\in (-\pi,\pi]$,  as the operator $\vartheta$:
\begin{equation}
    \vartheta := \lvert \theta\rvert = \frac{\pi}{2} \sum_{n=0}^{\infty} \lvert n\rangle\langle n\rvert  - \sum_{\substack{n,m = 0: \\ n \neq m}}^{\infty}\frac{[1-(-1)^{n-m}]}{\pi (n-m)^2}\lvert n\rangle\langle m\rvert. \label{eq:vartheta}
\end{equation}
This gives\footnote{As an aside, we remark that this length operator $R^*(x)$ does not satisfy the off-diagonal eigenstate thermalization hypothesis~\cite{DAlessio2016}, conventionally expected for ``physical'' observables in a theory with a random matrix spectrum. In particular, off-diagonal matrix elements are not suppressed by $e^{-S(E)/2}$ relative to the diagonal elements in the energy eigenbasis, especially near the diagonal. This is partly due to the simple action-angle dynamics of the effective theory, and implies that the expectation value of length does not settle down to a saturation value in special initial states supported on a small number of energy levels (unlike the ``larger'' initial states of Sec.~\ref{sec:worm}, in which the length does saturate).}
\begin{equation}
    \langle n|R^*(x)|m\rangle \approx \left\lbrace\begin{array}{ll} \displaystyle \dfrac{\pi}{2} \sqrt{2E_n} \rho(E_n)-\log(2E_n) &\ n=m \\ \displaystyle \left((-1)^{m-n}-1\right) \dfrac{\sqrt{2E_n}\rho(E_n)+\sqrt{2E_m} \rho(E_m)}{2\pi(n-m)^2} &\ n\ne m \end{array}\right.. \label{eq:Xmatrixelements}
\end{equation}

Let us comment on how this is related to the length operators in previous work.
The authors of \cite{Iliesiu:2021ari,Iliesiu:2024cnh} made explicit predictions for the matrix elements of the length operator, at least for $n\ne m$: 
\begin{align}
    \langle n|R^*(x)|m\rangle &= \frac{-\mathrm{e}^{-S_0}(2\pi)^2}{(E_n-E_m)\sinh(\pi[\sqrt{2E_n} - \sqrt{2E_m}])\sinh(\pi[\sqrt{2E_n} + \sqrt{2E_m}])} \notag \\
    &\approx \frac{-\mathrm{e}^{-S_0}(2\pi)^2 \sqrt{2E_n}}{2\pi(E_n-E_m)^2\sinh(2\pi\sqrt{2E_n})} \approx -\frac{\sqrt{2E_n}\rho(E_n)}{2\pi(n-m)^2}. \label{eq:oldXmatrixelements}
\end{align}
In the second line, we have approximated that $1 \lesssim |m-n| \ll n$, we have further neglected the tiny fluctuations in $E_n$ that encode random matrix statistics, and we have lastly invoked \eqref{eq:rho0Frelation}.  Comparing to  \eqref{eq:Xmatrixelements}, we see that there is a quantitative similarity in the scaling of the matrix elements.  A technical difference is that \eqref{eq:Xmatrixelements} vanishes between inequivalent energy eigenstates with $m-n$ an even number, while \eqref{eq:oldXmatrixelements} does not.  

Note that this reconstruction $R^*(x)$ is not unique.
For one thing, we have used an approximation for $F$, which ignores many subtle features of the spectrum.
Many equally valid approximations could be made.
However, these differences are small and do not affect the qualitative behavior of $\braket{R^*(x)(t)}$.

Another non-uniqueness comes from the choice of operator ordering between the non-commuting $N$ and $\theta$.  
But this difference is also small, at least when evaluated in high energy states.
Given two different choices of operator orderings in the reconstruction, $R^*(x)$ vs. $R^*_{\mathrm{new}}(x)$, notice that 
\begin{equation}
    \frac{\langle R^*(x)\rangle - \langle R^*_{\mathrm{new}}(x)\rangle}{\langle R^*(x)\rangle} \lesssim \frac{|\langle [N,\theta]\rangle |}{\langle N\rangle } \sim \frac{1}{\langle N\rangle}.
\end{equation}
In this equation, we have used the explicit form of \eqref{eq:xNtheta} to estimate the importance of operator ordering ambiguities. 
Notice that for a wave packet of dominant energy $E$, \eqref{eq:NE} implies that \begin{equation}
    \langle N\rangle \sim \sqrt{E} \mathrm{e}^{S(E)}.
\end{equation} Therefore, the operator ordering ambiguities when going from $(N,\theta)$ to $(x,p)$ are not of physical interest to us.

\edit{We also comment on a conceptual point regarding the relationship between $R^*$ and $V$. Our guiding principle ensures that states of the form \eqref{eq:V_sec3} are fairly narrow wavepackets in the eigenbasis of $R^*(x)$ for semiclassical values of $x$, and therefore have expectation values and fluctuations consistent with an eigenstate of the latter to leading order. The crucial requirement for this is that $\phi_{E_n}(x)$ has a slowly varying phase factor as a function of $n$, which holds for $x \ll \mathrm{e}^{S}$. Such slowly varying phase factors are also seen for time-evolved semiclassical states well before the Heisenberg time scale, which indeed remain fairly localized in $R^*(x)$ due to the slow spreading of wavepackets discussed in Sec.~\ref{sec:microcanonical}. However, we do not expect or require that $V\lvert x\rangle_0$ is precisely an eigenstate of $R^*(x)$, only that it is localized to near the $x$-eigenvalue of the latter if $x \ll \mathrm{e}^{S}$. In essence, we are trading off a precise eigenvalue-eigenstate relationship between $R^*$ and $V$ in favor of a guiding principle that allows us to carefully compute nonperturbative quantum dynamics in the fundamental theory.}

Lastly, for brevity we will often write $ x$ instead of $R^*(x)$  when it is clear that the operator acts on the fundamental Hilbert space (e.g. in Figure \ref{fig:microcanonicalnumerics}).


\section{Wormhole dynamics}\label{sec:worm}

Having built a reconstruction map $R^*$ and written down an explicit form for the reconstructed wormhole length $R^*(x)$, we are ready to study the dynamics of wormhole length in the fundamental theory.

\subsection{Microcanonical state}\label{sec:microcanonical}
We begin by considering wormhole dynamics in the following microcanonical state:
\begin{equation}
|\psi\rangle := \frac{1}{\sqrt{d}} \sum_{n=N_0}^{N_0+d-1}|n\rangle, \label{eq:microcanonicalstate}  
\end{equation}  
where $\ket{n}$ are the energy eigenstates of the fundamental Hilbert space \eqref{eq:fundamentalhilbert}. 
We assume that $d\ll N_0$, such that $E_{N_0+d}\approx E_{N_0}$, and this state is supported in a narrow energy window of size \begin{equation}
    \Delta E \sim \frac{d}{\rho_0(E_{N_0})} \sim d \mathrm{e}^{-S}.
\end{equation}
This state appears identical to the states used to develop the theory of quantum ergodicity \cite{dynamicalqergodicity}, which we will use below.

In what follows, it will often be easiest to discuss the ``rescaled wormhole length" operator $\vartheta$, defined in \eqref{eq:vartheta}, and the reader should keep in mind that, following \eqref{eq:xNtheta}, we can approximate
\begin{equation}
    x\approx \mathrm{e}^{S(H)}\sqrt{2H}\cdot \vartheta -\log (2H) + \cdots. \label{eq:sec4vartheta}
\end{equation}
The discussion that follows is accurate provided that $N_0$ is large, and the density of states is approximately constant in this energy window:
\begin{equation}
    \Delta E \cdot \frac{\partial \log \rho(E) }{\partial E} \sim \frac{\Delta E}{\sqrt{E}} \ll 1, \label{eq:microcanonicalassumption}
\end{equation}
which ensures that $x$ and $\vartheta$ are essentially proportional.

We now can generalize \cite{dynamicalqergodicity} to deduce critical features of the probability distribution of $x$ in the time-evolved microcanonical state 
\begin{equation}
\lvert \psi(t)\rangle = \frac{1}{\sqrt{d}} \sum_{n=N_00}^{N_0+d-1} \mathrm{e}^{-\mathrm{i} E_n t} \lvert n\rangle.
\end{equation}
 The details rely heavily on the level statistics of the fundamental theory, and we defer the derivations of the following results to Appendix~\ref{app:ergodic}.  At $t=0$, $|\psi(0)\rangle$ is an approximate eigenvector of $\vartheta$ with $\langle \psi(0)|\vartheta|\psi(0)\rangle \sim d^{-1}$.  If $|\psi(t)\rangle$ is an approximate eigenstate of $\vartheta$ at later times, it should be close to a state of the form 
 \begin{equation}
     |\tilde\theta\rangle := \frac{1}{\sqrt{d}}\sum_{l=0}^{d-1}\mathrm{e}^{-\mathrm{i}l\tilde\theta}|N_0+l\rangle.
 \end{equation}
 Notice that within the microcanonical window of interest, the discrete set $\tilde\theta = 2\pi m/d$ with $m\in\lbrace0,1,\ldots,d-1\rbrace$ form a complete basis.  To be even more quantitative, since we have found approximate quantum action-angle variables, we expect that $|\psi(t)\rangle$ should be close to the state $|\qtheta(t)=2\pi t/t_{\mathrm{H}}\rangle$, where 
\begin{equation}
t_{\htime} := 2\pi \rho(E) = 2\pi \mathrm{e}^{S(E)} \label{eq:heisenbergtime}
\end{equation}
is the Heisenberg time, which is the period of a classical angle variable with (quantized) density of states $\rho(E)$.  We consider the ``persistence amplitudes'' 
\begin{equation}
z(t) := \lvert \langle \tilde\theta(t)\vert \psi(t)\rangle\rvert = \left\lvert \frac{1}{d}\sum_{l=0}^{d-1} \exp\left[-2\pi \mathrm{i} \Delta_l \frac{t}{t_{\htime}}\right] \right\rvert,
\label{eq:persistencedef}
\end{equation}
which measure the closeness of the actual quantum wave function to a sharply peaked wave function following the classical trajectory.  We have defined $\Delta_l$, which measures spectral fluctuations:
\begin{equation}
\Delta_l := \left(\frac{E_l-E_0}{\Delta E}d -  l\right), \label{eq:def_deltan}
\end{equation}
with \begin{equation}
    \Delta E = \frac{d}{\rho(E)}
\end{equation} denoting the width of the energy window.


A short calculation using CUE random matrix statistics (see Appendix \ref{app:persistence}) reveals that \begin{equation}
    z(t) \approx \exp\left[-\log d \frac{t^2}{t_{\mathrm{H}}^2}\right]. \label{eq:main_persistence}
\end{equation}
It follows that the wavefunction remains in a \textit{microcanonical} ``coherent" state $\lvert \qtheta(t)\rangle$ with overwhelming probability for $t \ll t_{\htime}/\Delta$ (for RMT statistics, this time scale is $t_{\htime}/\sqrt{\ln d}$). This means that the wavepacket remains localized to within $\Delta \vartheta \sim 1/d$ of the classical trajectory $\qtheta(t)$ until this time, but spreads sharply thereafter, to $\Delta \vartheta \sim 1$ by $t \sim t_{\htime}$.   The results discussed thus far are a mild extension of earlier work \cite{dynamicalqergodicity}.


We can now extend these results and be even more quantitative about the actual \emph{shape} of the wave function as it begins to sharply spread near $t \sim t_{\htime}$.  In Appendix \ref{sec:wavepacketspreading}, we find (using arguments based on the self-similarity of random matrix statistics) that the width of the wave packet is \begin{equation}
    \mathrm{\Delta}\vartheta(t) \sim \left\lbrace\begin{array}{ll} d^{-1} &\ t \lesssim t_{\mathrm{H}}/\sqrt{\log d} \\ c_1 \exp[-(t_{\mathrm{H}}/c_2 t)^2] &\ t_{\mathrm{H}}/\sqrt{\log d} \lesssim t \lesssim t_{\mathrm{H}} \\ 1 &\ t\gtrsim t_{\mathrm{H}}\end{array}\right.. \label{eq:wormholexfluctuations}
\end{equation}
where $c_{1,2}$ are O(1) constants.  Remembering \eqref{eq:sec4vartheta}, we notice that the semiclassical wave packet can grow all the way to wormhole length $x\sim \mathrm{e}^S\sqrt{2E/S}$ before spreading due to quantum fluctuations!

The quantitative way that the wave function spreads into an incoherent sum over many different wormhole lengths is also very intriguing.  In Appendix \ref{sec:latetimebump}, we predict that (for $ \mathrm{e}^S \ll T \ll \exp[\mathrm{e}^S] $, at which time we encounter quantum recurrences \cite{QuantumRecurrences, BrownSusskind2}) \begin{equation}
    \int\limits_0^T \frac{\mathrm{d}t}{t_{\htime}}\left[\langle \vartheta(t)\rangle - \frac{\pi}{2}\right] \approx 0.
\end{equation}
This means that on average, the wormhole has a specific length at very late times, but the wave function explores longer wormholes than average for a noticeable amount of time.
What actually happens is that for $t\sim t_{\mathrm{H}}$, the wormhole actually \emph{overshoots} its final length, before bouncing back: see Figure \ref{fig:microcanonicalnumerics}.  This is a non-trivial prediction of our formalism.  We note an intriguing similarity with predictions of \cite{Balasubramanian:2024lqk}, which we will discuss further in the concluding section.

\edit{Let us also briefly comment on the difference between our approach and a recent work \cite{Bak:2025eul}, which proposed to account for the discrete spectrum of JT gravity by simply modifying the exponential potential of \eqref{eq:jtHamiltonianclassical} with a wall at large $x$, whose growth is chosen to match the spectrum.  In this approach, the wormhole growth will begin to deviate from its classical value at times $t\sim \mathrm{e}^{S_0}$ -- rather than $t\sim \mathrm{e}^{S(E)}/\sqrt{S(E)}$ -- which is incompatible with our fundamental postulate.
} 

\subsection{Hartle-Hawking state}
\label{sec:HHstate}

Now we discuss the Hartle-Hawking state in \eqref{eq:HHstate}. 
Its probability distribution as a function of energy is given by
\begin{equation}
\mathcal{P}(E) = \frac{1}{Z(\beta)}\rho(E)\mathrm{e}^{-\beta E},
\label{eq:HHdistribution}
\end{equation}
which is normalized to $1$. For the JT density of states, we recall that $\rho(E) = \rho_0(E)\mathrm{e}^{S_0}$ with $\rho_0(E) \propto \sinh(2\pi\sqrt{2E})$ given in \eqref{eq:JTinftyDOS}.
In the high temperature regime $\beta \ll 1$, the distribution $\mathcal{P}(E)$ is approximately a Gaussian
\begin{equation}
\mathcal{P}(E(\beta)+\delta E) = \frac{1}{Z(\beta)} \exp\left[S(\beta)-\frac{\pi \delta E^2}{4} \sqrt{\frac{2}{E(\beta)^3}}+\frac{\pi \delta E^3}{8}\sqrt{\frac{2}{E(\beta)^5}}+\ldots\right]
\label{eq:seriesexpansion}
\end{equation}
with a peak at \begin{equation}
    E(\beta) \approx 2\pi^2 T^2 \label{eq:HHEvsT}
\end{equation}where $T=1/\beta$,  and standard deviation \begin{equation}
    \sigma_E(\beta) \approx 2\pi T^{3/2} \ll E(\beta). \label{eq:HHsigmavsT}
\end{equation}
Here,
\begin{equation}
S_{\mathrm{th}}(\beta) = \log [Z(\beta)\mathcal{P}(E(\beta))] \approx S_0 + 2\pi^2 T-\log\left(8\pi^2\right).
\end{equation}
We emphasize the distinction between the microcanonical entropy $S(E) = \log \rho(E)$, which depends solely on the density of states, and the thermal entropy $S_{\mathrm{th}}(\beta)$.  This discrepancy has interesting consequences we will shortly discuss.

Based on intuition from statistical mechanics, one might expect that for $T\gg 1$, the Hartle-Hawking state behaves very similarly to a microcanonical state.   However, this is not true.  In the discussion of Section \ref{sec:microcanonical}, we had assumed \eqref{eq:microcanonicalassumption}, i.e. $\Delta E \ll E^{1/2}$.  Comparing \eqref{eq:HHEvsT} and \eqref{eq:HHsigmavsT}, we see that for the Hartle-Hawking state, $\Delta E \sim E^{3/4}$.  Therefore we must consider the Hartle-Hawking state to be a superposition of many ($\gtrsim E^{1/4}\sim T^{1/2}$) different microcanonical windows.

A rather tedious (but direct) computation in Appendix \ref{app:microcanonical} confirms that  the length operator $x$, whose matrix elements are given in \eqref{eq:Xmatrixelements}, is ``close enough" to diagonal that one can indeed split apart the wave function into many different microcanonical windows, each of which can essentially by treated ``independently".  More explicitly, \begin{align}
  x&\approx  \sum_{\text{energy window}} \mathbb{P}_{[E,E+\delta E]} x\mathbb{P}_{[E,E+\delta E]} \notag \\
  &\approx \sum_{\text{energy window}} \mathbb{P}_{[E,E+\delta E]} \sqrt{2E} \mathrm{e}^{S(E(\beta)) + \beta \cdot \delta E} \vartheta\mathbb{P}_{[E,E+\delta E]} . \label{eq:xwindow}
\end{align}
where $\mathbb{P}_{[E,E+\delta E]}$ is a projector into a microcanonical state with width $\delta E \ll \sqrt{E}$ obeying \eqref{eq:microcanonicalassumption}.  This also implies that we can borrow many of the technical results from Section \ref{sec:microcanonical} to deduce the late-time dynamics of the Hartle-Hawking state.

Importantly, however, the Hartle-Hawking state is a superposition of many distinct microcanonical windows with different $E$.  This means that some of the breakdown of semiclassical dynamics in this state is simply due to the fact that there is a very broad distribution of Heisenberg times: \begin{equation}
t_{\htime}(E(\beta)+\delta E) = 2\pi \rho(E(\beta) + \delta E) \approx 2\pi \mathrm{e}^{S(E(\beta))+\beta \delta E}. \label{eq:tHdistributed}
\end{equation}
Recalling that $\delta E \sim \sigma_E\sim \beta^{-3/2}$, we see a large \emph{relative} spread in $t_{\htime}$, of order $\mathrm{e}^{\sqrt{T}}$, between different typical microcanonical windows, even though the value of the energy $E$ itself hardly varies.  

Here, let us focus on a few drastic consequences of this large spread, leaving the derivation to Appendix \ref{app:HH}.   First, for very late times \begin{equation}
    t\gg t_{\mathrm{H}}(E) \mathrm{e}^{\mathrm{O}(S(E))}, \label{eq:tggtH}
\end{equation}
after nearly all microcanonical windows have reached their Heisenberg time, the probability to measure a length $x$ is approximately given by \begin{equation}
\mathcal{P}(x) \approx \frac{32\pi^{7/2}}{\beta Z(\beta) \log\left(64\pi^2\mathrm{e}^{-S_0} x\right)}\exp\left[-\frac{\beta}{8\pi^2}\left\lbrace \log\left(64\pi^2\mathrm{e}^{-S_0} x\right)\right\rbrace^2\right]. \label{eq:longtailed_x_dist}
\end{equation}
This is an extremely heavy-tailed distribution, which has drastic implications.  For example, if we compute $\langle x^q\rangle$, for each different value of $q$, the expectation value is dominated by a different part of the distribution $\mathcal{P}(x)$, namely a different energy window in the Hartle-Hawking state near \begin{equation}
    E_q(\beta)\approx (q+1)^2E(\beta). \label{eq:HHEq}
\end{equation}
Taking $q=1$, we see that $\langle x\rangle$ is actually dominated by a very small fraction of the wave function, and therefore will not saturate to its final value until well after the Heisenberg time $t_{\htime}(E(\beta))$ associated to the typical energy of the Hartle-Hawking state.  Indeed the rather stringent inequality \eqref{eq:tggtH} comes from this observation: to truly capture the full distribution \eqref{eq:longtailed_x_dist} requires evolving the wave function for an extraordinarily long time, such that even the microcanonical windows containing an exponentially small fraction of the high-energy states have had time to saturate.  The atypical microcanonical windows are thus qualitatively important in capturing wormhole dynamics in the Hartle-Hawking state.  In Appendix \ref{app:HH}, we further discuss the nature of the ``bump and spread" in the Hartle-Hawking state, which is still present, but is less pronounced than in a microcanonical window.

\subsection{Numerical simulations}
\label{sec:wormholedynamics_numerics}

For numerical simulations, we generate a spectrum with the JT density of states given by \eqref{eq:JTinftyDOS} and CUE level correlations, by using \eqref{eq:NE} and \eqref{eq:Finverse} to ``fold'' a CUE spectrum (which has a uniform density of states on $[0,2\pi)$) to have the JT density of states. First, we define a smooth folding function $\overline{F}$ for an ``averaged'' spectrum via a variant of \eqref{eq:Finverse}:
\begin{equation}
    \mathbb{E}[N(E)] = \overline{F}^{-1}(\sqrt{2E}).
\end{equation}
For a CUE spectrum of $D$ levels with eigenphases $\varphi_n \in [0,2\pi)$, we have the approximate relation:
\begin{equation}
    n \sim \frac{D}{2\pi}\varphi_n +O(\sqrt{\ln D}).
\end{equation}
This suggests that a reasonable strategy to introduce fluctuations into $\overline{F}(n)$ is to directly replace $n$ in the above expression with $D \varphi_n/2\pi$. Owing to the smoothness of $\overline{F}(n)$, the local CUE correlations between the $\varphi_n$ are preserved by this transformation. This gives the energy levels:
\begin{equation}
    E_n = \frac{1}{2}F(n)^2 \equiv \frac{1}{2} \overline{F}\left(\frac{D \varphi_n}{2 \pi}\right)^2.
\end{equation}
We note that a similar strategy may be used to generate a Poisson spectrum with the JT density of states, which we illustrate in Appendix~\ref{app:JTPoisson_numerics} to highlight the lack of a bump and a much faster spread of wavepackets compared to random matrix statistics.

Due to numerical limitations, our simulations necessarily truncate the spectrum to a finite number of levels $D$($=5000$). For the $x$ operator, we use a close approximation to \eqref{eq:Xmatrixelements}:
\begin{equation}
    x_{\text{numerics}} = \frac{1}{2}\left\lbrace \sqrt{2E} \rho(E), \vartheta\right\rbrace - \log(2E),
\end{equation}
in which $\vartheta$ is defined by the matrix elements \eqref{eq:vartheta}, but truncate this expression to the subspace spanned by the $D$ lowest energy levels in the spectrum.

To observe the dynamics of the spread and bump in an initial state $\lvert \psi(0)\rangle$, we respectively plot the distribution of the state in the $x$-eigenbasis $\lbrace x_n\rbrace_{n=0}^{D-1}$ (where $x_n$ are the eigenvalues of the $x$ operator) at different times and the mean value $\langle x(t)\rangle$. For the former, to more accurately represent the continuum limit, we directly plot a discretized probability density function in $x$,
\begin{equation}
p(x_n) \equiv \frac{1}{x_{n+1}-x_n} \lvert \langle x_n\vert \psi(t)\rangle\rvert^2,
\end{equation}
against the eigenvalues $x_n$.


Finally, for the initial state itself, we will illustrate two choices: the Hartle-Hawking state in \eqref{eq:HHstate} with a specific value of $\beta$, truncated to $D$ levels, as well as a microcanonical state in the interval $[E(\beta),3E(\beta)]$ supported over $d$ energy levels near the peak of the Hartle-Hawking distribution. Consequently, both states should show similar behavior in the ``most probable'' regions of their respective distributions, but the Hartle-Hawking state should show a larger spread in velocities and positions. Here, it is important to emphasize a limitation of our finite size numerics: due to the small number of energy levels $D$ and the exponentially growing density of states, we cannot simultaneously ensure that $d$ is large for the microcanonical state and that $\rho(E)$ remains approximately constant (i.e., satisfies \eqref{eq:microcanonicalassumption}) over its support. We have prioritized the former, and our ``microcanonical state'' is supported on $d=595$ energy levels, but $\rho(3E(\beta))$ is several times larger than $\rho(E(\beta))$. This is still a much narrower spread than the Hartle-Hawking state (in particular, the microcanonical state lacks a long tail), and our numerical results still effectively illustrate the qualitative differences in wormhole dynamics between the two initial states that originate in the significantly larger energy spread of the Hartle-Hawking state.

\begin{figure}[t]
\centering
\subfigure[Spread of wavepackets in terms of the probability distribution $p(x)$.]{\includegraphics[width=0.45\textwidth]{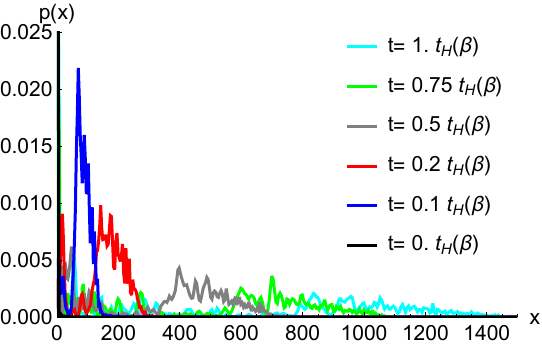}}
\label{fig:HHspread}
\qquad
\subfigure[Non-monotonicity of the expectation value $\langle x(t)\rangle$ of wormhole lengths.]{\includegraphics[width=0.45\textwidth]{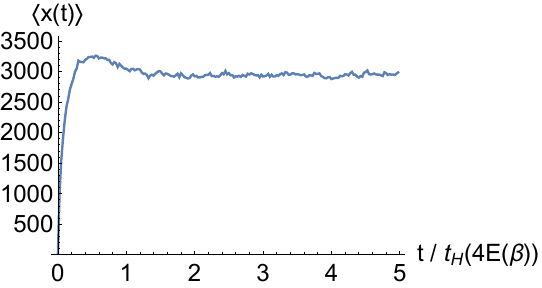}}
\label{fig:HHbump}
\caption{Wormhole dynamics as a function of time in the Hartle-Hawking state with $\beta = 9$. Compared to the microcanonical state in Fig.~\ref{fig:microcanonicalnumerics}, $p(x)$ shows the emergence of a long-tailed distribution due to the thermal spread of velocities, as well as a more spread out bump (see Appendix~\ref{app:HH}).}
\label{fig:HHnumerics}
\end{figure}

A related issue, again due to the exponentially growing density of states, is that the spectrum is truncated to rather low energies $E < E_D$, and we should choose high values of $\beta$ so that $E(\beta) \ll E_D$. In particular, we have chosen $\beta = 9$. This means that our numerics cannot be used to directly verify some of our specific quantitative predictions based on the high-temperature $\beta \to 0$ expansion in \eqref{eq:seriesexpansion}, such as the saturation value of $\langle x(t)\rangle$. The Hartle-Hawking state, in particular, also sees significant contributions from low energies $E \approx 0$, and has the portion of its exponential tail that would stretch to $E > E_D$ in the full JT spectrum cut off. However these effects are not significant in the microcanonical state, due to not being supported on the $E < E(\beta)$ and $E > 3 E(\beta)$ parts of the spectrum. Nevertheless, all the important qualitative features described above, such as the wide distribution of velocities in the Hartle-Hawking state, slow spread due to CUE statistics, and the bump in $\langle x(t)\rangle$, are all clearly visible for both initial states even with a truncated energy spectrum, as seen in Figs.~\ref{fig:microcanonicalnumerics} and \ref{fig:HHnumerics}.

In contrast to \cite{Miyaji:2024ity}, our reconstruction map suggests that the wormhole length is sharply defined all the way up to $t\sim \mathrm{e}^S/\sqrt{S}$.  Our result is a consequence of our guiding principle that semiclassical dynamics should be as accurate as possible at short times.

\section{Velocity operator}\label{sec:velocity}
Finally, we describe the eigenvalues of the velocity operator \begin{equation}
    v = \mathrm{i}[H,x] \label{eq:veloperator}
\end{equation}
characterizing the growth of the wormhole as a function of time. 

First, notice that so long as $x$ is a \emph{real} Hermitian operator, $v$ is antisymmetric and imaginary (thus Hermitian).   As $v$ is antisymmetric, any eigenvalue $\lambda$ has a partner eigenvalue $-\lambda$; since it is Hermitian, these eigenvalues are real.  We deduce that when measuring the velocity operator $v$ in a typical state, as one would expect to find at times $t\gg \mathrm{e}^S$, we have a 50\% chance of measuring $v<0$ and a 50\% chance of measuring $v>0$.

One can wonder whether a stronger statement can be made.
Perhaps the probability to measure $v\approx \sqrt{2E}$ is itself about 50\%, as is the probability to measure $v\approx -\sqrt{2E}$ (see e.g. discussion in \cite{Iliesiu:2024cnh}).  
A heauristic argument for this is that a general state of approximate energy $E$ can be written as a superposition of Hartle-Hawking states evolved for times approximately within $t \in [-\mathrm{e}^{S(E)}, \mathrm{e}^{S(E)}]$.
Semiclassically these states are all either shrinking with velocity $-\sqrt{2E}$ or growing with velocity $\sqrt{2E}$.
The question is: how well does the reconstructed velocity operator match this semiclassical answer?
This is a quantitative prediction that we can address.  For simplicity, let us focus on the microcanonical state discussed in Section \ref{sec:microcanonical}.

 Since we have seen that these microcanonical wormholes follow a semiclassical trajectory \eqref{eq:xtclassical} extremely closely for times $1\ll t\ll \mathrm{e}^S/\sqrt{S}$, we might expect that the wormhole velocity will be measured to have almost the same value during this time window where $\langle x(t)\rangle$ is steadily increasing.  And indeed from the generalized Heisenberg uncertainty principle we have \begin{equation}
    \Delta H \Delta x \ge \frac{1}{2}\Delta v,
\end{equation}
with uncertainty \begin{equation}
    \Delta x^2 = \langle \psi(t)|x^2|\psi(t)\rangle - \langle \psi(t)|x|\psi(t)\rangle^2,
\end{equation}
e.g., measured in a semiclassical time-evolving state.   Our theory of the wave function $\psi(x,t)$ ensures that $\Delta x \sim 1$ for $t\ll \mathrm{e}^S/\sqrt{S}$, while in a microcanonical window we have \begin{equation}
    \Delta H \sim d \mathrm{e}^{-S}.
\end{equation}
This implies that for early times, the probability to measure $v\approx \sqrt{2E}$ is close to 1.

What happens at later times?  Once the wave function has completely delocalized, the Heisenberg uncertainty principle allows for large velocity fluctuations, as $\Delta x \sim \sqrt{2E}\mathrm{e}^S$.  It is easiest to probe what happens at late times by considering the following simple calculation:
\begin{align}
\lim_{T\rightarrow\infty}\frac{1}{T} \int\limits_0^T\mathrm{d}t \; \left\langle \psi(t)\left|v^2\right|\psi(t)\right\rangle &= \frac{1}{d} \sum_{n=N_0}^{N_0+d-1} \left\langle n\left|v^2\right|n\right\rangle  \notag \\
&=\frac{1}{d} \sum_{n=N_0}^{N_0+d-1}\sum_{m\ne n} (E_m-E_n)^2 |\langle m|x|n\rangle|^2 \notag \\
&\approx 2E \cdot \frac{1}{\pi^2}\sum_{m:m\ne n}((-1)^{m-n}-1)^2 \frac{[m-n+\Delta_{m,n}]^2}{(m-n)^4}
\end{align}
where in the last line we have defined  \begin{equation}
    \Delta_{mn} = (E_m - E_n)\rho(E) - (m-n).
\end{equation}
and used that in the microcanonical window, $\rho(E)$ is effectively constant.  $\Delta_{mn}$ capture the fluctuations in the spectrum away from a perfectly rigid (and semiclassical) spectrum.  In a typical member of the JT ensemble, we can therefore estimate \begin{equation}
    \lim_{T\rightarrow\infty}\frac{1}{T} \int\limits_0^T\mathrm{d}t \; \left\langle \psi(t)\left|v^2\right|\psi(t)\right\rangle \approx 2E \left[1+\sum_{k=1,3,\ldots} \frac{8}{\pi^2}\frac{\mathbb{E}\left[\Delta_{n+k,n}^2\right]}{k^4}\right] > 2E.
\end{equation}
Therefore the long-time fluctuations in $v^2$ are \emph{larger} than $2E$ by an O(1) constant.  To check that this is not coming entirely from a very small probability of measuring a very large velocity, we can similarly estimate \begin{align}
    \lim_{T\rightarrow\infty}&\frac{1}{T} \int\limits_0^T\mathrm{d}t \; \left\langle \psi(t)\left|v^4\right|\psi(t)\right\rangle \notag \\
    &\approx \frac{16(2E)^2}{\pi^4} \mathbb{E}\left[\sum_{k=0,\pm2,\pm 4,\ldots}\left|\sum_{l=\pm 1, \pm3,\ldots}\frac{[k-l+{\Delta_{n+k,n+l}}][l+\Delta_{n+l,n}]}{(k-l)^2l^2} \right|^2\right].
\end{align}
The right hand side of this equation does not diverge with $d$ since $|\Delta_{n+k,n}|\ll k$ for nearly all $(n,m)$ at large $m$, and the sum over $l$ converges absolutely. This finite value of $\langle v^4\rangle$ implies that the distribution of velocities decays at least as fast as a power law for large $v^2$, implying a vanishing probability of measuring velocities that diverge with $d$. Moreover, although we did not obtain analytical results for $\langle v^4\rangle$ in terms of the distributions of $\Delta_{n,m}$, making the crude assumption that $\Delta_{n,m}$ and $\Delta_{n^\prime,m^\prime}$ are uncorrelated if $\lbrace n,m\rbrace \ne \lbrace n^\prime,m^\prime\rbrace$, we still find that $\langle v^4\rangle - \langle v^2\rangle^2$ is not small: \begin{equation}
    \langle v^4\rangle - \langle v^2\rangle^2 \sim \frac{16(2E)^2}{\pi^4}\sum_{l=\pm 1,\pm 3\cdots } \frac{\mathbb{E}[\Delta_{n+l,l}^4] - \mathbb{E}[\Delta_{n+l,l}^2]^2}{l^8}.
    \end{equation}
    These results imply that the probability distribution of $v$ at a typical late time has an O(1) spread around $\pm \sqrt{2E}$.
 We deduce that the distribution of $v^2$ cannot be too sharply peaked at $2E$.

\section{Outlook}\label{sec:outlook}
We have proposed an explicit reconstruction map for pure JT gravity, with the bulk/effective theory given by JT gravity at $S_0 \to \infty$  \cite{Harlow:2018tqv} and the boundary/fundamental theory given by a single draw of the matrix model dual \cite{Saad:2019lba}. 
As a case study, we wrote down an explicit proposal for the wormhole length operator in the dual theory and studied its behavior as a function of time in (initially semiclassical) states of interest in quantum gravity.

\subsection{How unique are the predictions of our reconstruction map?} \label{sec:generic} 
Our reconstruction was proposed based on the postulate that semiclassical gravity is an excellent approximation to the fundamental theory when acting on good semiclassical states, specifically the Hartle-Hawking state (or its microcanonical version) and small time evolutions of it.
It follows that any other reconstruction map that also satisfies this postulate will be close to ours at short times. 

What about long times?
The reason reconstruction maps are interesting is because they allow us to make predictions that \emph{do not} match the semiclassical answer.
To what extent should we trust those predictions from our map, e.g. for the length at times $t\gtrsim t_{\mathrm{H}}$? 
One could worry that tiny differences in the reconstruction map could keep the error small at early times but accumulate to very different answers at late times.
Perhaps different choices of map all satisfy our postulate but give wildly different answers.

 Where our reconstruction map may be most faulty is at large $x\sim \mathrm{e}^S$.  Intuitively, the ``cut-and-glue" prescription of Figure \ref{fig:Vwall} may not be accurate.  Quantitatively, the matrix elements of $R^*(x)$ may differ from one reconstruction map to the next, even if they all  reproduce semiclassical dynamics for the wormhole on times $t\ll t_{\mathrm{H}}$.  Nevertheless, we have some optimism that our key predictions are insensitive to the details of $R^*$.  As an explicit example, suppose that we modify \eqref{eq:Xmatrixelements} by \begin{equation} \label{eq:gvartheta}
     R^*(x) \rightarrow R^*_{\mathrm{alt}}(x) = \frac{1}{2}\left\lbrace \sqrt{2H}\rho(H), g(\vartheta)\right\rbrace
 \end{equation}
 for some function $g(\vartheta)=\vartheta+\cdots $ whose Taylor expansion matches the semiclassical expectations at short times, but differs at long times.  The arguments of Appendix \ref{app:ergodic} generalize to this case, and we still find that $\langle x_{\mathrm{alt}}(t)\rangle$ (using obvious abbreviated notation) overshoots its late time value, and that the wave function exhibits an explosive growth in fluctuations just before the Heisenberg time.  An explicit example replaces \eqref{eq:Xmatrixelements} with a formula much closer to \cite{Iliesiu:2021ari}: \begin{equation}
    \langle n|R^*_{\mathrm{alt}}(x)|m\rangle \approx \left\lbrace\begin{array}{ll} \displaystyle \dfrac{\pi}{3} \sqrt{2E_n} \rho(E_n)-\frac{2}{3}\log(2E_n) &\ n=m \\ \displaystyle -\dfrac{\sqrt{2E_n}\rho(E_n)+\sqrt{2E_m} \rho(E_m)}{2\pi(n-m)^2} &\ n\ne m \end{array}\right..  \label{eq:Xmatrixelements2}
\end{equation}
We have determined these matrix elements according to certain consistency conditions. The off-diagonal elements for large $(n-m)$ must have the same average (over some range of $(n-m)$) as those of $R^*(x)$ to reproduce the correct semiclassical dynamics of the velocity $v(t)$, which is responsible for the $(-1)$ prefactor above. The diagonal elements are chosen to ensure that $x_{\mathrm{alt}}(0) \approx 0$ in e.g. the microcanonical or Hartle-Hawking states. Moreover, to satisfy the guiding principle of Sec.~\ref{sec:pure}, the off-diagonal matrix elements reduce to functions of $(n-m)$ in microcanonical windows, rather than e.g. $(E_n-E_m)$ as in \cite{Iliesiu:2024cnh, Iliesiu:2021ari}.\footnote{This discrepancy is rather small due to the spectral rigidity, once $|m-n| \gg 1$.}

\begin{figure}[t]
\centering
\subfigure[Microcanonical state.]{\includegraphics[width=0.47\textwidth]{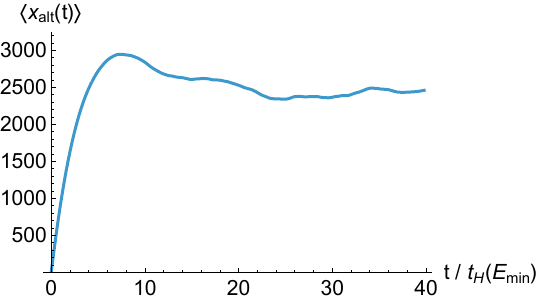}}
\subfigure[Hartle-Hawking state.]{\includegraphics[width=0.47\textwidth]{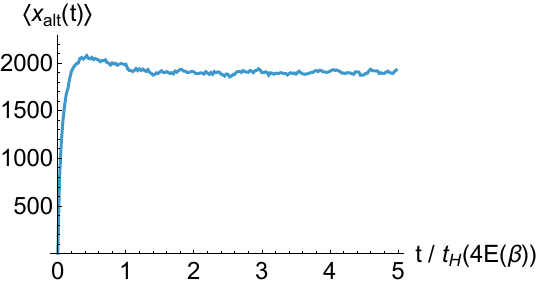}}
\caption{Bump in $\langle x_{\mathrm{alt}}(t)\rangle$, given \eqref{eq:Xmatrixelements2}, which is qualitatively similar to the behavior of $\langle x(t)\rangle$ shown in Figures \ref{fig:microcanonicalnumerics} and \ref{fig:HHnumerics}.}
\label{fig:alternatebumpnumerics}
\end{figure}

The alternate length operator $x_{\mathrm{alt}}(t)$ exhibits extremely similar dynamics to that observed in Figure \ref{fig:microcanonicalnumerics}: see Figure \ref{fig:alternatebumpnumerics}.  This happens because $R^*_{\mathrm{alt}}(x)$ (in microcanonical windows) is still an even function of $(n-m)$; it simply arises from a different choice of $g(\vartheta)$ in \eqref{eq:gvartheta}.

\edit{In this context, we also comment on an interesting open problem. The effective theory, due to the $S_0 \to \infty$ limit, corresponds entirely to the leading cylinder (single wormhole) topology in the gravitational path integral. However, the fundamental theory involves a sum over topologies, but our construction only identifies abstract operators acting on the fundamental Hilbert space without a direct geometric interpretation. In other words, we have argued that $R^*(x)$ corresponds to the wormhole length in the leading topology at short times, but it would be nice to have a more explicit understanding of which geometric quantities it corresponds to in the full nonperturbative path integral.  A full understanding should resolve the ambiguities highlighted here.}


\subsection{Complexity = volume?}
\label{sec:complexityconjecture}

The key outcome of our construction is that the reconstructed length operator could be studied in detail using the theory of quantum ergodicity \cite{dynamicalqergodicity}.  This allowed us to leverage many known results from random matrix theory to make clear predictions for the dynamics of wormholes before and after the Heisenberg time.  The most striking of these predictions are a non-monotonic growth in the averaged wormhole length operator, along with heavy-tailed moments of the length operator in the Hartle-Hawking states.   As we have emphasized above, we believe that this prediction is generic and is not sensitive to particular assumptions in our reconstruction procedure.

What are the implications for the ``complexity = volume'' conjecture~\cite{Susskind:2014rva}?
Often ``complexity'' is assumed to mean circuit complexity, the minimal number of gates needed to prepare the state from some reference state.
It is unclear that this conjecture for \textit{circuit} complexity is consistent with the non-monotonic growth.


That said, it is very intriguing to compare our result to an independent result which has recently appeared in the literature \cite{Balasubramanian:2024lqk} regarding a different notion of complexity.  These authors pointed out that the ``spread complexity" \cite{spreadcomplexity} of a thermofield double state exhibits a similar non-monotonic dependence in time as our $\langle x(t)\rangle$ -- we believe that the origin of this phenomena is analogous to what we have shown: the wormhole shrinks at $t\sim t_{\mathrm{H}}$ because of level repulsion in the spectrum of the fundamental theory.  The authors of \cite{Balasubramanian:2024lqk} have proposed that spread complexity describes the length operator in quantum gravity; see also \cite{SonnerDSSYK}.  We have arguably ``derived" a closely related equivalence between wormhole length and quantum ergodicity in our formalism.  A clear advantage, in our view, of our approach to defining the length operator is that the spread complexity is based on a notion of Krylov complexity \cite{Parker:2018yvk, KrylovReview} which is sensitive to initial states; our approach defines the length operator independently of the initial state.

Interestingly, these results have implications for the \textit{non-gravitational} physics of quantum many-body systems. We have shown, with very reasonable assumptions, that ergodic dynamics in the fundamental (matrix) theory drives the dynamics of observables in JT gravity such as the wormhole length over the Heisenberg timescale $t_{\htime} \sim \exp(S(E))$. In turn, the ``complexity = volume'' conjecture~\cite{Susskind:2014rva} posits a relation between the length operator, together with a possibly larger class of gravitational observables with nontrivial dynamics over this timescale~\cite{ComplexityAnything}, and the growth of the (e.g., spread) complexity of a state in the fundamental theory over the same timescale. This suggests that ergodicity may be the physical mechanism driving the dynamics of non-gravitational observables such as complexity in the fundamental theory over the Heisenberg timescale, which may be of relevance in suitable classes of quantum systems.

In short, for JT gravity, our calculations arguably reduce the ``complexity = volume" conjecture to a \emph{non-gravitational conjecture} that ``complexity = ergodicity". It may be possible to explicitly resolve this conjecture in future work.

\subsection{The linearity of $R^*$ and adding matter}
\label{sec:addmatter}
A peculiar feature of our reconstruction map is that it is linear.
It is well-known that a linear reconstruction map $R^*$ cannot be good if $V$ is non-isometric, i.e. it cannot satisfy $R^*(O)V\ket{\psi} \simeq V O \ket{\psi}$ to a good approximation for all $\ket{\psi}$ and even just simple $O$ \cite{Akers:2022qdl, Antonini:2024yif}.
Here we evade the issue because we do \emph{not} demand the reconstruction is good on all states.
We only demand it on states that we call ``semiclassical''.
For example, we do not require it works on highly entangled states like $\int \mathrm{d}x \, \ket{x}_0 \ket{x}_R$, where $R$ is some auxiliary reference system and perhaps the integral runs over some finite range of $x$. 
This state is intuitively non-semiclassical, because $R$ has decohered the wormhole length into a incoherent mixture of very high (non-semiclassical) precision.
It is these kinds of highly entangled states that easily lead to issues with linear reconstruction maps in non-isometric codes \cite{Akers:2022qdl, Antonini:2024yif}. 

This situation will be harder once matter is included \cite{Iliesiu:2024cnh,Penington:2019kki,Penington:2023dql,Boruch:2024kvv}. Then it will be possible to highly entangle the matter with a reference system while keeping the state ``semiclassical''.
For this reason, once matter is included, we expect a linear reconstruction map will have to be abandoned.

\section*{Acknowledgements}
We thank Tom Faulkner and Luca Iliesiu for useful discussions.  This work was supported by the Heising-Simons Foundation under Grant 2024-4848.

\begin{appendix}

    \section{Quantum dynamical ergodicity}\label{app:ergodic}
This appendix contains a review of technical details of the theory of quantum dynamical ergodicity \cite{dynamicalqergodicity}, which play a crucial role in our choice of reconstruction $R^*$ and ultimate predictions for wormhole dynamics.
\subsection{Microcanonical wavepackets and quantum ergodicity}
Here we provide a more formal discussion of quantum ergodicity \cite{dynamicalqergodicity} within microcanonical windows, and justify how our reconstructing map satisfies the guiding principle in Sec.~\ref{sec:pure}.
\edit{At an intuitive level, the goal of this subsection is to associate the time-shift operator $\delta$ with the basis of discrete Fourier transforms of the energy eigenstates $\lvert E_n\rangle$ in narrow energy windows, owing to the expected conjugate relationship between $(E,\delta)$ that is represented in standard quantum mechanics by Fourier transforms. This will end up implying that the Hamiltonian $H$ acts as close as possible to a translation of $\delta$, and in this way satisfies our guiding principle.}

\edit{Such an association may be formulated exactly for perfectly regular spectra such as in the harmonic oscillator. But in the case of JT gravity,} 
because the $E_n$ have an irregular and bounded spectrum, we cannot expect that any Hermitian operator $\delta$ generates exact time-translations, as emphasized in the main text. Such a conjugate relationship requires precise action-angle variables, such as $(N,\theta)$ defined in the main text, whose analogues in the classical phase space are given by the expressions (from \eqref{eq:fund_ham} and \eqref{eq:fund_Fdef}):
\begin{subequations} \label{eq:semiclassicalNandTheta}
\begin{align}
    \clas{N} &= \int_0^{\clas{H}} \diff E\ \rho(E)~, \label{eq:semiclassicalNvsE}\\
    \clas{\theta} &= \frac{\clas{\delta}}{\rho(H)}~,\label{eq:semiclassicalThetaVsDelta}
\end{align}
\end{subequations}
where $\rho(E)$ is the (averaged) density of states. \edit{Here, we use the notation $\clas{A}$ to indicate that a variable $A$ is classical, rather than in the (quantum) effective or fundamental theories.}

Still, we might hope for $(H,\delta)$ to effectively function as a good action-angle pair at \emph{short times} in \emph{sufficiently narrow energy windows}.  More explicitly, consider some narrow range of energies $[E_{\mc}, E_{\mc}+\Delta E]$ such that the averaged density of states is constant,
\begin{equation}
    \rho(E_{\mc}+\Delta E) \approx \rho(E_{\mc}),
\end{equation}
which we will call a ``microcanonical'' energy window. Here, we might hope that a pair $(H,\delta)$ is ``close"  to the ``ideal'' action-angle pair $(N,\theta)$ up to rescaling:
\begin{subequations} \label{eq:semiclassicalNandTheta2}
\begin{align}
    N &\approx H \rho(E_{\mc}) + \text{const.}~, \label{eq:semiclassicalNvsE2}\\
    \theta &\approx \frac{1}{\rho(E_{\mc})}\delta~.\label{eq:semiclassicalThetaVsDelta2}
\end{align}
\end{subequations}
Therefore, $N$ is fully capable of generating time shifts via the relation between $\theta$ and $\delta$.  Clearly, \eqref{eq:semiclassicalNvsE2} cannot be exact on all time scales, due to the irregular spectrum we have just mentioned.

\begin{figure}[!t]
\centering
\includegraphics[width=0.75\textwidth]{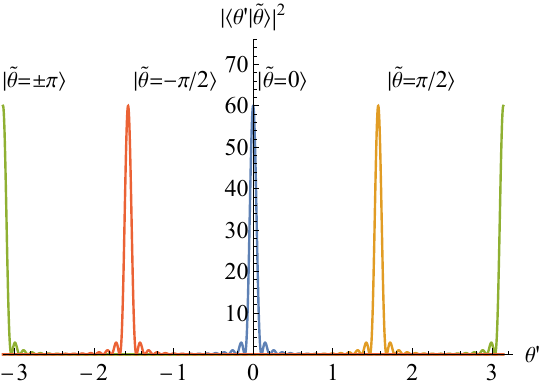}
\caption{Probability disributions $\lvert \langle \theta'\vert \qtheta\rangle\rvert^2$ for the wavepackets in \eqref{eq:quasianglewavepackets} for different values of $-\pi \leq \theta \leq \pi$ in $\theta' \in (-\pi,\pi]$, with $d=60$. These wavepackets become increasingly localized as $d\to\infty$, and always contain a complete orthonormal basis for any finite $d$. The $N_0$-dependent oscillations in the phase of the wavepackets are not visible in this plot of probability density, but will be visible in the real and imaginary parts.}
\label{fig:thetawavepackets}
\end{figure}
\begin{figure}[t]
\centering
\subfigure[CUE statistics, showing ergodicity in $\vartheta$ via the probability of the classical trajectory (dots), a slow spread (light horizontal lines), and indications of a late-time bump (concentration near $\vartheta=\pi$).]{\includegraphics[width=0.43\textwidth]{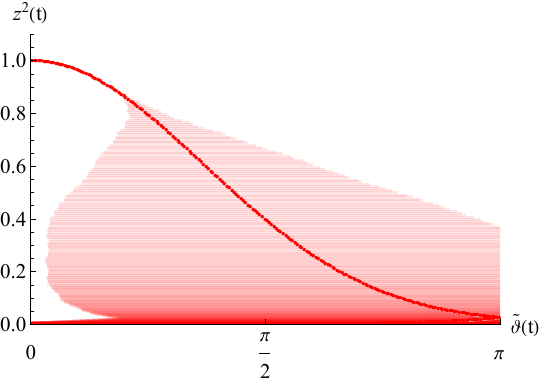}}
\label{fig:cuepersistence} \qquad \qquad
\subfigure[Poisson statistics, showing non-ergodicity in $\vartheta$  via the probablility of the classical trajectory (dots) and a relatively fast spread (light horizontal lines).]{\includegraphics[width=0.43\textwidth]{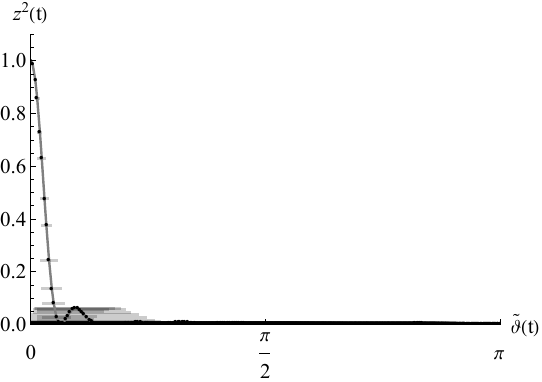}}
\label{fig:poissonpersistence}
\caption{Plot of the persistence probability $z^2(t)$ of the classical trajectory from Eq.~\eqref{eq:persistencedef}, against the classical value of $\qvartheta(t)$ corresponding to $\qtheta(t) = 2t/t_{\htime}$ at different values of time, for CUE and Poisson statistics in a microcanonical window with $d=1024$. Horizontal bars indicate the smallest symmetric region around $\qtheta(t)$ in which there is a $0.9$ probability of measuring $\theta$. The growth of these bars indicates the spreading of the wavepacket at different values of $\qtheta(t)$.}
\label{fig:persistenceinX}
\end{figure}

Nevertheless, let us see if we can make sense of \eqref{eq:semiclassicalNandTheta2} at early times $t\ll 1/\rho(E_{\mathrm{mc}})$.   Explicitly, we will look for  the optimal reconstruction of $(N,\theta)$ in terms of the energy eigenbasis in a microcanonical window of $d$ consecutive eigenstates of $N$, with $n \in [N_0, N_0+d-1]$ and $\Pi_{d}$ denoting the projector onto this subspace: \begin{equation}
    \Pi_d = \sum_{n=N_0}^{N_0+d-1}|n\rangle\langle n|.
\end{equation} We introduce finite-dimensional angle coherent states in this subspace as projections of the fundamental angle states in \eqref{eq:fundamental_angle_coherentstates}:
\begin{equation}
\lvert \qtheta\rangle := \frac{1}{\sqrt{d}} \sum_{n=N_0}^{N_0+d-1} \mathrm{e}^{-\mathrm{i} n \qtheta} \lvert n\rangle,
\label{eq:DFTtheta}
\end{equation}
which satisfy the overcompleteness relation:
\begin{equation}
\int\limits_0^{2\pi}\frac{\diff \qtheta}{2\pi d^{-1}}\ \lvert \qtheta\rangle\langle \qtheta\rvert = \Pi_d.
\end{equation}
In contrast to the fundamental angle coherent states $\lvert \theta\rangle$, these states are normalizable. Among these coherent states, any regularly spaced subset of $d$ states forms an orthonormal basis:\footnote{The numerics in Figs.~\ref{fig:persistenceinX}, \ref{fig:WavepacketSpreading} and \ref{fig:bumpfeature} implicitly use such a finite basis in discrete steps of time rather than the continuum of coherent states.}
\begin{equation}
\lvert C_k(\qtheta_1)\rangle \equiv \lvert \qtheta = \qtheta_1 + 2\pi k /d\rangle = \frac{1}{\sqrt{d}} \sum_{n=0}^{d-1} \mathrm{e}^{-\mathrm{i} n \qtheta_1} \mathrm{e}^{-2\pi \mathrm{i} n k /d} \lvert n\rangle.
\label{eq:DFT}
\end{equation}
These basis states can be regarded as forming an ``eigenbasis'' of the angle variable $\qtheta$ within the microcanonical window\footnote{A generic observable $A(\theta)$ reconstructed as in \eqref{eq:operator_reconstruction_def}, when projected onto the microcanonical window, has eigenstates highly localized in $\theta$, and we expect that there is a sufficiently close operator for which the $\lvert C_k(\theta_1)\rangle$ are exact eigenstates. For example, $\lvert C_0(\theta_1)\rangle$ is an exact eigenstate with eigenvalue $1$ for the microcanonical projection of the operator $A(\theta)$ for which $a(\theta) = \delta(\theta-\theta_1)$.}, although such a notion cannot be extrapolated to the full infinite dimensional Hilbert space $\mathcal{H}$ in \eqref{eq:fundamentalhilbert}. In the full Hilbert space, the microcanonical angle states have the wavefunctions:
\begin{equation}
\langle \theta'\vert \qtheta\rangle = \mathrm{e}^{\mathrm{i} N_0 (\theta'-\theta)} \delta_d\left(\frac{\theta' - \qtheta}{2\pi}\right),
\label{eq:quasianglewavepackets}
\end{equation}
where 
\begin{equation}
\delta_d(\xi) = \frac{1}{\sqrt{d}}\sum_{n=0}^{d-1} \mathrm{e}^{2\pi \mathrm{i} n \xi} = \frac{1}{\sqrt{d}}\frac{\sin(\pi \xi d)}{\sin (\pi \xi)} \mathrm{e}^{\mathrm{i} \pi \xi (d-1)},
\end{equation}
which has period $1$ in $\xi$. Thus, for large $d$, the wavepackets in \eqref{eq:quasianglewavepackets} are localized to within $\Delta \theta \sim 1/d$ of $\theta' = \qtheta$. Such wavepackets are illustrated in Fig.~\ref{fig:thetawavepackets}. 

Given that the spectrum of $N$ is fixed in the fundamental theory, our freedom to reconstruct $(N,\theta)$ is represented by unitary transformations $\lvert \qtheta\rangle \to U_{d}\lvert \qtheta\rangle$ and $N \to U_{d} N U_{d}^\dagger$, where $U_{d}$ acts nontrivially only within the microcanonical window, reducing to identity for all states outside this subspace. Then for all $\lvert \qtheta\rangle$ and any small $\theta_0$ (typically, smaller than $2\pi /d$, including the $\theta_0 \to 0$ limit for finite $d$), our criterion is that the optimal choice of $(N,\theta)$ must satisfy: 
\begin{equation}
    \left\lvert \langle \qtheta + \theta_0 \rvert \exp\left(-i H \rho(E_{\mc}) \theta_0\right)\lvert \qtheta \rangle\right\rvert \text{ is maximized over all } \lvert \qtheta\rangle \to U_{d} \lvert \qtheta\rangle.
    \label{eq:actionoptimizationproblem}
\end{equation}

Imposing \eqref{eq:actionoptimizationproblem} for small $\theta_0$ implies~\cite{dynamicalqergodicity} the following constraints:
\begin{enumerate}
    \item $N$ is an exactly conserved quantity within the microcanonical window:
    \begin{equation}
          [N_d, H_d]  = 0,
         \label{eq:actioncommuteswithN}
    \end{equation}
    where $N_d = \proj_d N \proj_d$ and $H_d = \proj_d H \proj_d$.
    \item The eigenvalues of $H$ within the microcanonical window are arranged in ascending order of $n$:
    \begin{equation}
     H_d = \sum_{n=N_0}^{N_0+d-1} E_n \lvert n\rangle \langle n\rvert,\ \text{ with } E_{n+1} \geq E_n.
    \label{eq:actioneigenvalsareascending}
\end{equation}
\end{enumerate}

Strictly within the microcanonical window, this implies that any set of the $\lvert C_k(\theta_1)\rangle$ form the best possible eigenbasis for the time-shift operator $\delta$, being the orthonormal basis on which $H$ acts closest to a shift operation. For example, we could consider an optimal reconstruction $\Pi_d\delta_{\text{opt}}\Pi_d$ of the time shift operator in our microcanonical window,\footnote{The operator $\delta_{\mathrm{opt}}$ formally depends on the choices $N_0$ and $d$, but we suppress this to simplify the notation. } whose eigenbasis is precisely $\lvert C_k(-\pi)\rangle$: 
\begin{equation}
   \Pi_d \delta_{\text{opt}} \Pi_d = \sum_{k=0}^{d-1} \frac{\pi(2 k-d)}{d} \rho(E_{\mc}) \lvert C_k(-\pi)\rangle \langle C_k(-\pi)\rvert
\end{equation}
 However, our primary task here is to reconstruct a \textit{global} operator $\delta$ that is as close as possible to these optimal microcanonical operators $\delta_{\text{opt}}$ in \emph{every} microcanonical window, without knowledge of $N_0$ or $d$ directly. To do so precisely is a technically challenging problem that we will not claim to solve. Instead, we will be content with observing that for a reconstruction of the form \eqref{eq:timeshiftglobalreconstruction},
 \begin{equation}
    \Pi_d \delta \Pi_d \approx \Pi_d \delta_{\mathrm{opt}} \Pi_d
    \label{eq:reconstruction_error_tolerance}
\end{equation}in all high-energy microcanonical windows, where this approximation should improve as $d\rightarrow \infty$.  Indeed, notice that \begin{align}
    \langle m|\Pi_d\delta_{\mathrm{opt}}\Pi_d|n\rangle &= \frac{2\pi \rho(E_{\mathrm{mc}})}{d^2}\sum_{k=0}^{n-1}\mathrm{e}^{2\pi\mathrm{i}k(m-n)/d + \pi\mathrm{i}(m-n)}k=\frac{(-1)^{m-n}\rho(E_{\mathrm{mc}})}{\frac{d}{2\pi}(\mathrm{e}^{2\pi \mathrm{i}(m-n)/d}-1)}.
    \end{align}
In the above expression, both $m$ and $n$ are taken within the microcanonical window.
Taylor expanding in the limit of large $d$, we observe that the largest matrix elements of $\delta$ (which have $|m-n|\ll\delta$) are quantitatively in agreement with \eqref{eq:timeshiftglobalreconstruction}.

For other (suboptimal) choices $(N', \theta')$, the long-time dynamics of $\delta$ would generally be drastically different even to leading order. 
However, any global reconstruction of the above form continues to be sensitive to the optimal microcanonical dynamics of $\theta$ identified by our physical principle \textit{at all times}, to within some resolution $\Delta \theta \ll 1$ determined by the approximation in \eqref{eq:reconstruction_error_tolerance}. \edit{We note that in an entirely analogous manner, our guiding principle ensures that the length operator essentially reduces to the absolute value of the angle variable $\lvert \theta\rvert$ in microcanonical windows, via the expression in \eqref{eq:lengthoperatorabstheta}.}

\subsection{Comparing quantum and classical angle dynamics}\label{app:persistence}

Here, we will review some properties of the persistence amplitude $z(t)$ in \eqref{eq:persistencedef}, which measures the ``closeness'' of the quantum dynamics of the angle coherent states to classical action-angle dynamics, from \cite{dynamicalqergodicity}. \edit{Its behavior may be succinctly summarized as follows. In the case of random matrix statistics, the angle coherent states follow classical dynamics extremely closely up to a time that is almost as large as the Heisenberg time, being smaller only by a low power-law in the entropy, $\exp(S(E))/
\sqrt{S(E)}$; this will lead to the ``slow wavepacket spread'' phenomenon in Sec.~\ref{sec:wavepacketspreading}. These states also retain some noticeable memory of classical dynamics for at least half the period of the angle variable both forward and backward in time, which is similar (though not identical) to the bump phenomenon to be discussed in Sec.~\ref{sec:latetimebump}. However, with Poisson statistics, strong deviations from classical dynamics are observed at the \textit{exponentially} smaller time scale $\exp(S(E)/2)$, thereby being quite far from either of these features.}

In generic spectra, the $\Delta_n$ defined in \eqref{eq:def_deltan} follow a Gaussian distribution. This is particularly true to a high degree of accuracy for RMT statistics~\cite{GRMGaussian}, and is also a reasonable approximation for Poisson statistics. In this case, the persistence amplitude behaves as a Gaussian in time:
\begin{equation}
z(t) = \exp\left[-2\pi^2 \Delta^2\frac{t^2}{t_{\htime}^2}\right] + \mathrm{O}(d^{-1/2}),
\label{eq:Gaussian_persistence}
\end{equation}
with \begin{equation}
    \Delta^2 := d^{-1}\sum_n \Delta_n^2 \label{eq:spectralrigidity}
\end{equation}being a measure of \textit{spectral rigidity}~\cite{Mehta, DeltaStar2} within the microcanonical window. 

We note that $\lvert \qtheta(t)\rangle$ retains some imprint of the classical angle variable as long as $z(t) >c d^{-1/2}$ for any large constant $c$ (as $d\rightarrow \infty$). 
To trace out a full classical cycle (i.e., a half-cycle each way in time) before losing this imprint, we need $z(\pm t_{\htime}/2) > cd^{-1/2}$, and to avoid periodicity, we need $z(\pm t_{\htime}) \leq c d^{-1/2}$ (possibly up to logarithmic factors). These two conditions suggest
\begin{equation}
\Delta^2 \in \frac{[1,4)}{4\pi^2}\log d,
\end{equation}
which is the precise range of spectral rigidity for RMT statistics~\cite{Mehta, DeltaStar2, ShenkerThouless}. Specifically, for the circular unitary ensemble (CUE), whose eigenvalue statistics is expected for JT gravity, and for Poisson statistics representing a generic non-ergodic system, we have
\begin{subequations}\begin{align}
\Delta^2_{\text{CUE}} &\approx \frac{1}{2\pi^2} \log d, \label{eq:CUE_SR}\\
\Delta^2_{\text{Poisson}} &\sim d. \label{eq:Poisson_SR}
\end{align}
\label{eq:SR_functions}
\end{subequations}
 We quote these results to later contrast the behavior of CUE (ergodic and aperiodic) and Poisson (non-ergodic, and still aperiodic) to isolate the role of spectral statistics in wormhole dynamics.  The case of CUE statistics was quoted in \eqref{eq:main_persistence}.

\subsection{Early time dynamics from spectral self-similarity}
\label{sec:wavepacketspreading}

Both RMT and Poisson spectra show a degree of self-similarity: if one selects any subset of $d_s$ \textit{consecutive} energy levels from such spectra, they continue to show the respective RMT or Poisson behavior in statistical measures such as the distribution of $\Delta_n$ and spectral rigidity (with $d$ replaced by $d_s$). They also retain the same Heisenberg time $t_{\htime}$ as the full spectrum. We can use this property to quantitatively derive how an initial coherent state $\lvert \qtheta(0)\rangle$ spreads in $\qtheta$ as a function of time. \edit{In particular, we will substantiate our claim that the spread of the coherent state is parametrically slower with random matrix statistics (as expected for JT gravity) than for a Poisson spectrum.}

At an intuitive level, our argument proceeds as follows: Within a smaller microcanonical window of $d_s$ levels, each coherent state $\lvert \qtheta\rangle_s$ has an angular resolution of $\Delta \qtheta \sim 2\pi/d_s$, as there are $d_s$ basis states spanning $\qtheta \in [-\pi,\pi)$. By \eqref{eq:Gaussian_persistence}, an initial coherent state $\lvert \qtheta(0)\rangle$ in the full $d$-level window retains a substantial overlap with some $d_s$-level coherent state in each smaller window until a time $t \sim t_{\htime}/\Delta(d_s)$ (where $\Delta(d_s)$ is given by \eqref{eq:SR_functions} with $d \to d_s$), after which it spreads out to multiple states. At a given time $t$, this spreading is seen for a choice of $d_s \sim \Delta^{-1}(t_{\htime}/t)$, which corresponds to the initial state spreading beyond an angular resolution of $\Delta \qtheta \sim 2\pi/\Delta^{-1}(t_{\htime}/t)$. This gives a quantitative estimate of the spreading of the wavepacket in time. Now, we will proceed with the formal derivation of this estimate, eventually arriving at \eqref{eq:ThetaSpread}.

For simplicity in the argument that follows, we suppose that $d_s \ll d$ and that $d/d_s$ is an integer. For coherent states $\lvert \qtheta_s\rangle_s$ in a smaller region of the spectrum, given by the analogue of \eqref{eq:DFTtheta} for $d_s$ consecutive energy levels from $n=n_s$ to $n=n_s+d_s-1$,
\begin{equation}
\lvert \qtheta\rangle_s = \frac{1}{\sqrt{d_s}} \sum_{n=n_s}^{n_s+d_s-1} e^{-i (n-n_s) \qtheta} \lvert E_n\rangle,
\label{eq:DFTthetasmaller}    
\end{equation}
the overlap with the microcanonical coherent states $\lvert \qtheta\rangle$ over $d$ levels is
\begin{equation}
\langle \qtheta\vert \qtheta_s\rangle_s = \frac{1}{\sqrt{d_s}}\sum_{n=n_s}^{n_s+d_s-1} e^{i n \qtheta} e^{-i(n-n_s)\qtheta_s} = e^{i n_s \qtheta}\delta_{d_s}\left(\frac{\qtheta-\tilde{\theta_s}}{2\pi}\right).
\label{eq:thetawavepackets}
\end{equation}
These are wavepackets localized to within a width of $\Delta \qtheta \sim 1/d_s$ around $\qtheta = \qtheta_s$. More concretely, the probability that a measurement of the \textit{microcanonical} angle operator $\qtheta$ (as constructed in terms of one of the eigenbases in \eqref{eq:DFT})\footnote{For the microcanonical $\qtheta$ operator, the existence of a complete orthonormal eigenbasis \eqref{eq:DFT} allows us to reconstruct all functions of $\qtheta$ as traditional commuting operators, including projectors whose expectation values give measurement probabilities, without the complications associated with the global $\theta$ variable in Sec.~\ref{sec:reconstructionmap}.} gives an outcome in the interval $\qtheta \in [\qtheta_0, \qtheta_0+\Delta \qtheta]$ is given by the expectation value of the projector
\begin{equation}
\proj(\qtheta_0, \Delta \qtheta) = \sum_{\substack{k:\\ 0\leq k \leq \frac{(\Delta \qtheta) d}{2\pi}}} \lvert C_k(\qtheta_0)\rangle \langle C_k(\qtheta_0)\rvert.
\label{eq:thetaproj_def}
\end{equation}
Then, the width $\Delta \qtheta = \lambda_\kappa/d_s$ of a symmetric interval around $\qtheta_s$ within which one has a probability $1-\kappa$ of measuring $\qtheta$, can be determined according to
\begin{align}
\mathbb{P}\left(\qtheta \in \left[\qtheta_s - \frac{\lambda_\kappa}{d_s}, \qtheta_s + \frac{\lambda_\kappa}{d_s}\right] \text{ in } \lvert \qtheta_s\rangle_s\right) &= {_s}\left\langle \qtheta_s\left\rvert \proj\left(\qtheta_s - \frac{\lambda_\kappa}{d_s}, \qtheta_s + \frac{\lambda_\kappa}{d_s}\right)\right\lvert \qtheta_s\right\rangle_s \notag \\ &= 1-\kappa. \label{eq:kappa_def1} \end{align}
As $d_s \rightarrow \infty$, the expectation value of this projector can be approximated by an integral that becomes independent of $d_s$ to leading order, using \eqref{eq:thetawavepackets}, giving a relation between $\kappa$ and $\lambda_{\kappa}$: \begin{align}
 \int\limits_0^{\lambda_\kappa}\diff y\ \sinc^2(\pi y) = \frac{\Si(2\pi \lambda_\kappa)}{\pi} &+\frac{\cos(2\pi\lambda_\kappa)-1}{2\pi^2\lambda_\kappa} = \frac{1-\kappa}{2}. \label{eq:kappa_integral}
\end{align}
For small $\kappa$, we get $\lambda_\kappa \sim 1/\kappa$ up to oscillatory (but monotonic) factors.

Conversely, by splitting the full energy window into different consecutive sets of $d_s$ levels, we can write 
\begin{equation}
\lvert \qtheta\rangle \approx \sqrt{\frac{d_s}{d}}\sum_s e^{in_s \qtheta} \lvert \qtheta_s = \qtheta\rangle_s.
\label{eq:thetasum}
\end{equation}

First, let us consider the time-evolving state $\lvert \qtheta_s(t)\rangle_s$ in each such subspace, such that $\qtheta_s(0) = \qtheta(0)$. We have seen that this remains peaked at the classical value for long times: it overlaps with the classical state $\lvert \qtheta_{\cl}(t)\rangle_s$ with probability $z^2_s(t)$, given by \eqref{eq:Gaussian_persistence} with $\Delta \to \Delta(d_s)$ (due to self-similarity), where $\qtheta_\cl(t) = \qtheta_s(0) + 2\pi t/t_{\htime}$.

Combining these observations with the discussion following \eqref{eq:thetawavepackets}, in each $\lvert \qtheta_s(t)\rangle$, the probability of measuring $\qtheta$ in the interval $\qtheta_{\cl}(t)\pm \lambda_\kappa/(2d_s)$ is guaranteed to remain relatively large at early times.  Quantitatively, measuring $\tilde\theta$ in a restricted window of $d_s$ states, we want to determine the range of early times over which the probability $\mathbb{P}(\cdots)$ of measuring a value close to the classical expectation value is
\begin{equation}
\mathbb{P}\left(\qtheta \in \left[\qtheta_\cl(t) - \frac{\lambda_\kappa}{d_s}, \qtheta_\cl(t) + \frac{\lambda_\kappa}{d_s}\right] \text{ in } \lvert \qtheta_s(t)\rangle_s\right) > 1-\epsilon,
\label{eq:probabilitywidth1}
\end{equation}
for some small $\epsilon > 0$.  To calculate the time window $0<t<t_\kappa$ in which this is accurate, notice that\begin{equation}
    |\tilde\theta(t)\rangle_s =  \mathrm{e}^{\mathrm{i}\alpha(t)}\left( z_s(t) |\tilde\theta_{\mathrm{cl}}(t)\rangle_s + \sqrt{1-z_s(t)^2}|\tilde\theta_{\mathrm{cl}}(t)^\perp\rangle_s\right)
\end{equation}
with the overall phase $\alpha(t)$ unimportant and henceforth ignored, and $|\tilde\theta_{\mathrm{cl}}(t)^\perp\rangle_s$ denoting the orthogonal part of the state (normalized).   If $Q$ denotes the projector \eqref{eq:thetaproj_def} corresponding to the desired range of $\tilde\theta$ in \eqref{eq:probabilitywidth1} at the time $t$, we have for real-valued $\alpha$, $\beta$: 
\begin{subequations}
    \begin{align}
        \langle \tilde\theta_{\mathrm{cl}}(t)|Q|\tilde\theta_{\mathrm{cl}}(t)\rangle_s &:= \alpha^2 =  1-\kappa, \\
        \langle \tilde\theta_{\mathrm{cl}}(t)^\perp|Q|\tilde\theta_{\mathrm{cl}}(t)^\perp\rangle_s &:= \beta^2 \le 1, \\
        \left\lvert \langle \tilde\theta_{\mathrm{cl}}(t)|Q|\tilde\theta_{\mathrm{cl}}(t)^\perp\rangle_s\right\rvert &\leq \alpha \beta,
    \end{align}
    \end{subequations}
    where the first line follows from \eqref{eq:kappa_def1}, with the next two lines being general properties of projectors. Therefore,


\begin{align}
    \langle \tilde\theta(t)|Q|\tilde\theta(t)\rangle_s &\ge z_s(t)^2 \alpha^2 - 2\alpha\beta z_s(t)\sqrt{1-z_s(t)^2} + \left(1-z_s(t)^2\right)\beta^2
\end{align}
Here, the left hand side is identical to the left hand side of \eqref{eq:probabilitywidth1}. For sufficiently large persistence amplitude,
\begin{equation}
z_s(t)^2 \geq \frac{1}{1+\alpha^2} = \frac{1}{2-\kappa},   
\label{eq:zs_constraint}
\end{equation}
the above expression attains its minimum among all possible values of $\beta$ at $\beta = 1$. In this regime, we can write:
\begin{align}
\langle \tilde\theta(t)|Q|\tilde\theta(t)\rangle_s  &\geq \left(z_s(t) \sqrt{1-\kappa} - \sqrt{1-z_s(t)^2}\right)^2 \label{eq:tkappa1},
\end{align}
which increases monotonically with $z_s(t)$ as long as \eqref{eq:zs_constraint} is satisfied. 

Now, suppose that $0<\kappa \le 1/2$, and that 
\begin{equation}
    z_s(t) > z_{\kappa} := \frac{1}{\sqrt{1+\kappa}}. \label{eq:zkappabound}
\end{equation}
Choosing
\begin{equation}
    \epsilon = 1-\left(z_\kappa \sqrt{1-\kappa} - \sqrt{1-z_\kappa^2}\right)^2 = \frac{\kappa + 2\sqrt{\kappa-\kappa^2}}{1+\kappa},
    \label{eq:tkappa2}
\end{equation}
which can be inverted to determine the appropriate value of $\kappa \leq 1/2$ for a given value of $0 < \epsilon \leq 1$, we get that \eqref{eq:probabilitywidth1} holds in the regime where \eqref{eq:zkappabound} holds, which by \eqref{eq:Gaussian_persistence} occurs for $t< t_{\kappa}$ with
\begin{equation}
    t_{\kappa}  = \frac{t_{\htime}}{4\pi \Delta(d_s)} \log \frac{1}{1+\kappa}.
\end{equation}
The crucial point is that for a sufficiently small but O(1) choice of $\kappa$, the wave function restricted to $d_s$ states will be sharply peaked in $\theta$ before time $t_{\htime}/\Delta$.


We now want to combine each window of $d_s$ states back together to reproduce the full microcanonical wave function.  We claim that -- for the most part -- each smaller microcanonical window can be studied independently (for a more careful demonstration of such a microcanonical decomposition for the wormhole length operator, see Appendix~\ref{app:microcanonical}).  To see this, let us write an arbitrary wave function on the entire $d$-dimensional microcanonical Hilbert space 
\begin{equation}
|\psi\rangle = \sum_s c_s |\psi_s\rangle    
\end{equation}
as a sum of contributions in each smaller window $s$, with $\langle \psi_s|\psi_s\rangle = 1$.  
For two different windows $s\ne s^\prime$, we have \begin{equation}
    \langle \psi_s | \proj(\qtheta_0, \Delta \qtheta) |\psi_{s^\prime}\rangle = \sum_{\substack{k:\\ 0\leq k \leq \frac{(\Delta \qtheta) d}{2\pi}}} \langle \psi_s|\qtheta\rangle_s \langle \qtheta|\psi_{s^\prime}\rangle_{s^\prime} \mathrm{e}^{\mathrm{i}(n_{s^\prime}-n_s)\qtheta}.
\end{equation}
Due to the rapidly oscillating factor of $n_{s^\prime}-n_s = md_s$ (an integer multiple of $d_s$), we expect that (approximating the sum by an integral as in \eqref{eq:kappa_integral}) \begin{equation}
    \langle \psi_s | \proj(\qtheta_0, \Delta \qtheta) |\psi_{s^\prime}\rangle \lesssim \mathrm{sinc}\frac{m d_s \cdot \Delta \qtheta}{2} \sim \frac{1}{m d_s \cdot \Delta \qtheta} 
\label{eq:thetainterference}
\end{equation}
where the last scaling expression on the right hand side assumes that $\Delta \qtheta \gg d_s^{-1}$.  Hence, if we are only interested in an angular resolution large compared to $d_s^{-1}$, it makes sense to decompose a larger microcanonical wave function into a sum over smaller microcanonical wave functions, each over $d_s$ levels.  

Of course, this decomposition clearly fails to be sensitive to the width of $\lvert \psi\rangle$ for any choice of $d_s < d$ at time $t=0$, where we know that the spread in $\tilde\theta$ scales as $d^{-1}$, which is smaller than $d_s^{-1}$.   So we should choose $d_s$ ``dynamically" as a function of $t$, making $d_s$ as small as possible, while still ensuring that \eqref{eq:thetainterference} is small, thus ensuring that interference effects between different windows are unimportant for an angular resolution $\Delta \qtheta$ comparable to the actual width of $\lvert \psi\rangle$.  Suppose that if we measured $\tilde\theta$ on the full microcanonical wave function, we saw that
\begin{equation}
\mathbb{P}\left(\qtheta \in \left[\qtheta_\cl(t) - \frac{\Delta \qtheta(t)}{2}, \qtheta_\cl(t) + \frac{\Delta \qtheta(t)}{2}\right] \text{ in } \lvert \qtheta(t)\rangle\right) > 1-\epsilon. \label{eq:DeltathetaP}
\end{equation} for $\epsilon$ related to $\kappa$ as in \eqref{eq:tkappa2}.   From \eqref{eq:probabilitywidth1}, so long as $t<t_\kappa$, \begin{equation}
    d_s \cdot \Delta\qtheta < \lambda_\kappa. \label{eq:dstinequality}
\end{equation}
We should choose $d_s(t)$ such that \eqref{eq:dstinequality} is approximately saturated.  This occurs when $t_\kappa \sim t$.  From \eqref{eq:tkappa2} we see that we should choose
\begin{equation}
    d_s  = \mathrm{\Delta}^{-1}\left[ \frac{t_{\htime}}{4\pi t} \log \frac{1}{1+\kappa}\right],
\end{equation}
where $\Delta^{-1}$ is the inverse of the spectral rigidity function \eqref{eq:spectralrigidity}.

\begin{figure}[t]
\centering
\includegraphics[width=0.9\textwidth]{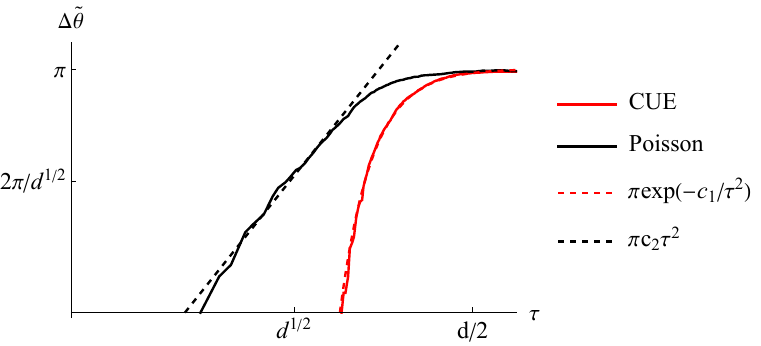}
\caption{Log-Log plot of the size $\Delta \qtheta(t)$ of the smallest symmetric region around $\qtheta_{\cl}(t)$ in which there is a probability of greater than $1-\epsilon = 0.95$ of measuring $\qtheta$, for CUE and Poisson statistics, as a function of the normalized time $\tau = t d/t_{\htime}$. These are respectively compared with the analytical estimates above by plotting functions of the form Eq.~\eqref{eq:CUEspread} and Eq.~\eqref{eq:Poissonspread} with constants $c_1$ and $c_2$ chosen by hand. These are for a microcanonical window with $d=1024$; see also Fig.~\ref{fig:persistenceinX} for another illustration of the spread.}
\label{fig:WavepacketSpreading}
\end{figure}

For CUE and Poisson statistics, \eqref{eq:CUE_SR} and \eqref{eq:Poisson_SR} imply that
\begin{subequations}\begin{align}
d_s(t)^{-1}\sim \left.\Delta\qtheta(t)\right\rvert_{\text{CUE}}  &\lesssim \lambda \exp\left[-\frac{t_{\htime}^2}{\gamma t^2}\right], \label{eq:CUEspread}\\
d_s(t)^{-1} \sim \left.\Delta\qtheta(t)\right\rvert_{\text{Poisson}}  &\lesssim \mu \frac{t^2}{t_{\htime}^2}. \label{eq:Poissonspread}
\end{align}
\label{eq:ThetaSpread}
\end{subequations}
where $\lambda,\mu,\gamma$ are O(1) constants. 
The drastic distinction between the two cases is illustrated in Fig.~\ref{fig:WavepacketSpreading}.  Therefore, we obtain \eqref{eq:wormholexfluctuations} from \eqref{eq:CUEspread}.  

\subsection{A bump at late times}
\label{sec:latetimebump}

There is one final feature of the dynamics of the angle variable that deserves attention.  Notice that in Fig.~\ref{fig:persistenceinX}, there is a brief period near $t\sim t_{\htime}$ where it appears that the system is \emph{more likely} to be found close to $\qtheta=\pi$, relative to $\qtheta=0$.  
For $t\gg t_{\htime}$ the dynamics is completely out of phase and non-classical, with $\langle |\qtheta|\rangle = \pi/4$.  We now show that there is a ``bump" --- or non-monotonic $t$-dependence --- in expectation values of observables of the form $\langle A(\qtheta(t))\rangle$, which is a universal consequence of level repulsion in RMT spectra indicating an ``ergodic'' exploration of the entire domain of the angle variable $\qtheta \in [-\pi, \pi)$.

\edit{Intuitively, this is because the time integral of the deviation of any such expectation value from its long-time saturation value is determined by the density of level spacings in the spectrum near $0$, due to the standard Fourier relationship between energy and time. Consequently, if level repulsion (vanishing density of spacings near $0$) is present in the spectrum, these deviations must ``equally'' explore either side of the saturation value so that they cancel each other out to make the time integral vanish. This is directly responsible for the bump feature. In the absence of level repulsion, these deviations need not exactly cancel each other out over time, leading to the absence of a prominent bump.}


To detect this feature analytically, it is useful to look at a coarse-grained measure of ergodicity (see Appendix B of \cite{qergthesis}) by taking the Fourier components of a state $\lvert \qtheta_0(t)\rangle$ along $\qtheta$.  This metric is sensitive to level statistics: for integer $r > 0$, 
\begin{align}
\int\frac{\diff \qtheta}{2\pi d^{-1}}\ \mathrm{e}^{\mathrm{i}r(\qtheta-\qtheta_0)} \left\lvert \langle \qtheta\vert \qtheta_0(t)\rangle\right\rvert^2 &= \frac{1}{d}\sum_{n=0}^{d-1-r} \mathrm{e}^{\mathrm{i}(E_{n+r}-E_n)t} \nonumber \\
&= \left(1-\frac{r}{d}\right)\int\diff S_r P^{(r)}(S_r) \mathrm{e}^{\mathrm{i} S_r t}, \label{eq:levelspacingfourier1}
\end{align}
where $S_r = E_{n+r}-E_n$ denotes the $r$-th nearest neighbor level spacings (numbering $d-r$ in a set of $d$ consecutive levels) and $P^{(r)}(S_r)$ is their probability distribution~\cite{Mehta}.   An analogous expression is obtained for $r<0$. In the large $d$ limit, and upon taking $\tilde\theta_0\rightarrow 0$ as is appropriate for our application, this implies
\begin{equation}
\left\langle \cos\left[r\qtheta(t)\right]\right\rangle_{\qtheta(0)=0} \approx \int\diff S_r P^{(r)}(S_r) \cos(S_r t).
\label{eq:cosine_levelspacing}
\end{equation}

 The observable $\cos[r\qtheta(t)]$ partitions the circle $\qtheta\in [0, 2\pi)$ into $r$ regions where it is positive and another $r$ regions where it is negative, such that both kinds of regions have the same length. For ``coarse-grained'' ergodicity, we want $\qtheta(t)$ to ``equally'' explore both types of regions over time, which means the expectation value $\langle \cos[r\qtheta(t)]\rangle_{\qtheta(0)=0}$ should integrate to zero\footnote{It is perhaps more natural to consider a coarse-grained version of ergodicity in terms of Walsh-Hadamard functions (which take the values $+1$ and $-1$ on these same intervals) rather than cosines, but these will involve more complicated combinations of level spacings so we stick with cosines. But our conclusions will turn out to be similar in either case.}. Note that we eventually expect $\qtheta$ to become uniformly distributed as $t\to\infty$ in all generic spectra without degeneracies, so a time \textit{average} of such observables (as in conventional ergodicity) will always decay to zero; a time integral is therefore a more appropriate quantity to probe ``coarse-grained ergodicity''.


These time integrals are given by
\begin{equation}
\int\limits_0^{\infty}\frac{\diff t}{t_{\htime}}\ \left\langle \cos\left[r\qtheta(t)\right]\right\rangle_{\qtheta(0)=0} \approx \frac{\pi}{2t_{\htime}}  P^{(r)}(S_r \to 0/t_{\htime}).
\label{eq:coslevelspacingintegral}
\end{equation}
Here, $t_{\htime}$ is introduced to fix the time scale of interest and make the integral dimensionless, which is important when taking limits such as $d\to\infty$. We write $S_r \to 0/t_{\htime}$ rather than $S_r = 0$ to suggest that we want to take $S_r \to 0$ over the scale of $t_{\htime}^{-1}$, rather than precisely zero (for which the right hand side will be zero for all nondegenerate spectra)\footnote{To make this more precise, we really mean the result one would get by integrating the right hand side of Eq.~\eqref{eq:cosine_levelspacing} from $t = 0$ to $t = \lambda t_{\htime}$, and then taking the limit $\lambda \to \infty$ \textit{after} $d\to\infty$, while keeping $t_{\htime}$ fixed.}; in what follows, we will simply call this $P^{(r)}(0)$.

In parallel with \eqref{eq:levelspacingfourier1}, it is convenient to write observables that depend only on the angle variable $\qtheta$ in terms of the microcanonical coherent states (note that this form is consistent with our global reconstruction map \eqref{eq:globalreconstructionmap_def}):
\begin{align}
    A(\qtheta) :=& \int_{-\pi}^{\pi}\frac{\diff \qtheta}{2\pi d^{-1}}\  a(\qtheta) \lvert \qtheta\rangle \langle \qtheta \rvert \nonumber \\
    =& \sum_{n,m = 0}^{d-1} \left[\int_{-\pi}^{\pi}\frac{\diff \qtheta}{2\pi}\  a(\qtheta) \mathrm{e}^{-\mathrm{i}(n-m)\qtheta} \right] \lvert n\rangle \langle m\rvert.
    \label{eq:operator_reconstruction_def}
\end{align}
Now we specialize to an observable $A(\qvartheta(t))$ that depends on the angle variable only through $\qvartheta = \lvert \qtheta\rvert$ (i.e., is invariant under $\qtheta \to -\qtheta$) --- such as the reconstruction of the wormhole length operator $x(N,\theta)$ in \eqref{eq:xNtheta} (where the $N$-dependence can be dropped in micrcanonical windows, see also Appendix \ref{app:microcanonical}). If $A(\qvartheta)$ remains sufficiently smooth as $d\to\infty$, it can be written as a Fourier series that retains only even combinations of the terms in \eqref{eq:operator_reconstruction_def}:
\begin{equation}
A(\qvartheta(t)) = a_0 + \sum_{r=1}^{\infty} a_r \cos\left[r\qtheta(t)\right].
\label{eqs:genobservableFourier}
\end{equation}
Here, $a_0$ is the microcanonical expectation value of the observable, which is the value to which it will saturate at long times (if the spectrum is nondegenerate). From \eqref{eq:coslevelspacingintegral}, the time integral of the difference between the observable and this saturation value is:
\begin{equation}
\int_0^{\infty}\frac{\diff t}{t_{\htime}}\ \left[\langle A(\qvartheta(t))\rangle_{\qtheta(0)=0} - a_0\right] = \frac{\pi}{2t_{\htime}}\sum_{r=1}^{\infty} a_r P^{(r)}(0).
\label{eqs:generalobservableerg}
\end{equation}

Over these scales, both CUE and Poisson spectra have $P^{(r>1)}(0) = 0$, but differ~\cite{Mehta} for $r=1$. While a CUE spectrum has complete level repulsion with $P^{(1)}(0) = 0$, a Poisson spectrum lacks nearest-neighbor level repulsion and has $P^{(1)}(0) > 0$. It follows that every observable $\langle A(\qvartheta(t))\rangle$ must oscillate ``equally'' around its saturation value for a CUE spectrum (such as in JT gravity), indicating ``coarse-grained ergodicity'', but this can not happen universally for all such observables given a Poisson spectrum.\footnote{We also note that a largely complementary argument in \cite{WilkieBrumerReturnProbabilities}, originally for suitably averaged return probabilities over shorter timescales, can be applied to our case to show that if all the $P^{(r)}(S)$ attain a maximum away from $S = 0$, irrespective of level repulsion at $S=0$, then $[A(\qtheta(t)) - a_0]$ cannot be everywhere positive or everywhere negative in $t$. However, such an argument does not estimate the magnitude of this effect, e.g., whether $[A(\qtheta(t))-a_0]$ explores both sides equally, which would not be the case in the absence of level repulsion.} This constitutes a coarse-grained form of ergodic quantum dynamics, in which \textit{level repulsion} in the spectrum leads to observables ``equally'' exploring both sides of their long-time saturation value.

For more intuition, let us focus on the observable $\cos[\qtheta(t)]$ (i.e. $a_r = \delta_{r,1}$ in \eqref{eqs:generalobservableerg}), which is $>0$ when $\qtheta(t)$ is closer to the initial state $\qtheta=0$, and $<0$ when it is closer to the ``diametrically opposite'' value $\qtheta \approx \pi$.  If $\langle \cos \tilde\theta\rangle < 0$, it suggests that the system has a larger probability of $\vartheta>\pi/2$ than $\vartheta<\pi/2$ (although mathematically this is not guaranteed).  Using the nearest-neighbor level spacing statistics for CUE and Poisson distributions in the $d\to\infty$ limit:
\begin{subequations}\begin{align}
P^{(1)}_{\text{CUE}}(S) &= \frac{t_{\htime}}{2\pi} \frac{32}{\pi^2}\left(\frac{S t_{\htime}}{2\pi}\right)^2 \exp\left[-\frac{4}{\pi}\left(\frac{S t_{\htime}}{2\pi}\right)^2\right], \label{eq:CUEspacings}\\
P^{(1)}_{\text{Poisson}}(S) &= \frac{t_{\htime}}{2\pi} \mathrm{e}^{-S t_{\htime}/2\pi}, \label{eq:Poissonspacings}
\end{align}\end{subequations}
we find very different quantitative behaviors in $\langle \cos\left[\vartheta (t)\right]\rangle$ for CUE vs. Poisson level statistics.



\begin{figure}[t]
\centering
\subfigure[The bump in $\langle \cos(\qtheta)\rangle$.]{\includegraphics[width=0.43\textwidth]{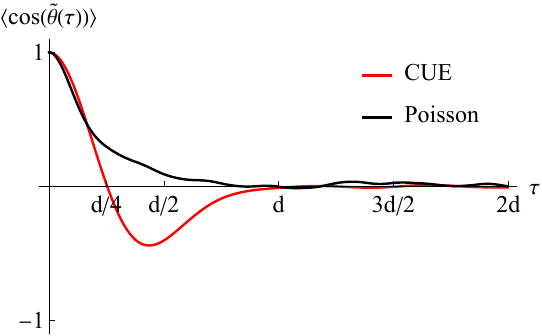}}
\label{fig:cosThetabump} \qquad \qquad
\subfigure[The bump in $\langle \qvartheta\rangle$.]{\includegraphics[width=0.43\textwidth]{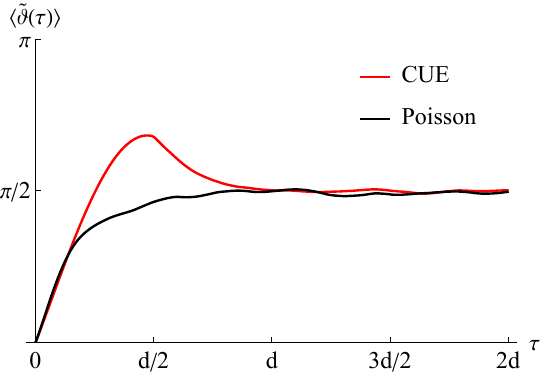}}
\label{fig:Xbump}
\caption{Illustration of ``coarse-grained ergodicity'' via the late-time bump feature for the observables $\cos(\qtheta)$ and $\vartheta$ for microcanonical windows of $d=1024$ levels with CUE statistics, and its absence with Poisson statistics, as a function of the normalized time $\tau = td/t_{\htime}$.}
\label{fig:bumpfeature}
\end{figure}

If the system has CUE statistics, then 
\begin{equation}
\left.\left\langle \cos\left[\qtheta(t)\right]\right\rangle\right\rvert_{\text{CUE}} \approx \left(1-\frac{\pi^3 t^2}{2t_{\htime}^2}\right)e^{-\pi^3 t^2/(4 t_{\htime}^2)}
\end{equation}
At $t=0$, this starts of at $1$, but crosses over to negative values at $t = t_{\htime} \sqrt{(2/\pi^3)} \approx 0.254 t_{\htime}$ --- just a little higher than the time $t = t_{\htime}/4$ when the classical trajectory would cross $\qtheta=\pi/4$. Subsequently, it reaches a maximum negative value of $-2 \mathrm{e}^{-3/2} \approx -0.446$ at $t \approx 0.440 t_{\htime}$, after which it gradually decays to $0$. Thus, $\cos(\qtheta)$ captures a coarse grained variant of ergodicity in which the initial wavepacket at $\qtheta=0$ manages to explore the ``opposite side'' near $\qtheta = \pi$ before truly randomizing over the full range of angles.  This is illustrated in Fig.~\ref{fig:bumpfeature}.

In contrast, if we have Poisson level statistics:
\begin{equation}
\left.\left\langle \cos\left[\qtheta(t)\right]\right\rangle\right\rvert_{\text{Poisson}} \approx \frac{t_{\htime}^2}{t_{\htime}^2+4\pi^2 t^2}.
\end{equation}
While this again starts off at $1$ at $t=0$, it never decays to negative values, and always remains positive. Thus, even in this coarse grained sense, the dynamics of the wavepacket under a Poisson spectrum remains close to the initial state at $\qtheta=0$ and does not shift towards the opposite side with $\qtheta=\pi$.


As discussed in Sec.~\ref{sec:worm}, \eqref{eqs:generalobservableerg} leads to a robust bump in the expectation value of the wormhole length operator $\langle x(t)\rangle$, which equally explores either side of its late-time saturation value. Our calculations show that this bump is a clear and universal signature of level repulsion, a key property of random matrix spectra, in quantum ergodicity.


\section{Non-uniqueness of $R^*$ in JT gravity}
\label{app:actionanglereconstruction}



Recall that $\mathrm{e}^{\mathrm{i}\theta}$ and $\mathrm{e}^{-\mathrm{i}\theta}$ do not commute due to their action on the ground state $|0\rangle$.  As such,  there is a slight ambiguity in how to reconstruct operators, which we can highlight by showing that the reconstruction map is not an exact algebra homomorphism:  $R^*(AB)\ne R^*(A)R^*(B)$.  Nevertheless, we will see that it is a very good \emph{approximate} homomorphism on high-energy states.  

Suppose that we have already obtained the classical canonical transformation from $(x,p)$ to $(N,\theta)$, and suppose that we are given two observables $A(N, \theta)$ and $B(N,\theta)$ which do not depend on $N$.  Given the Fourier transforms of each function, we see from \eqref{eq:globalreconstructionmap_def} that \begin{subequations}
    \begin{align}
        R^*(A) &= \sum_{m=-\infty}^\infty \frac{1}{2}\left\lbrace A_m(N), \mathrm{e}^{\mathrm{i}m\theta}\right\rbrace, \\ 
        R^*(B) &= \sum_{m=-\infty}^\infty \frac{1}{2}\left\lbrace B_m(N), \mathrm{e}^{\mathrm{i}m\theta}\right\rbrace, \\
        R^*(AB) &= \sum_{m=-\infty}^\infty \frac{1}{2}\left\lbrace C_m(N), \mathrm{e}^{\mathrm{i}m\theta}\right\rbrace, 
    \end{align}
\end{subequations}
where the Fourier coefficients \begin{equation}
    C_m(N) = \sum_{k=-\infty}^\infty A_{m-k}(N)B_k(N).
\end{equation}
Let's now compare the operators $R^*(A)R^*(B)$ and $R^*(AB)$ by acting with both operators on the state $|n\rangle$: \begin{subequations}
    \begin{align}
         R^*(A)R^*(B)|n\rangle &= \frac{1}{4}\sum_{m_1,m_2=-\infty}^\infty [A_{m_1}(n+m_1+m_2) + A_{m_1}(n+m_2)] \notag \\
         &\cdot [B_{m_2}(n)+B_{m_2}(n+m_2)]\delta_{n+m_2\ge 0}\delta_{n+m_1+m_2\ge 0} |n+m_1+m_2\rangle,  \\
         R^*(AB)|n\rangle &= \frac{1}{2}\sum_{m_1,m_2=-\infty}^\infty [A_{m_1}(n+m_1+m_2)B_{m_2}(n+m_1+m_2)\notag \\
         &\;\;\;\;\;+A_{m_1}(n)B_{m_2}(n)]\delta_{n+m_1+m_2\ge 0}|n+m_1+m_2\rangle.
    \end{align}
    \label{eq:ABcomparison_n}
\end{subequations}
Clearly there are two simple differences between the two expressions: (\emph{1}) the first sum has the Fourier coefficients $A_{m_1}$ and $B_{m_2}$ evaluated at distinct values of $n$, and (\emph{2}) the fact that $|0\rangle$ is the ground state puts an extra constraint on the first sum.


Happily, if the Fourier coefficients $A_m$ and $B_m$ decay relatively quickly with $m$, i.e. the functions of $\theta$ are reasonably smooth, and $n\gg 1$, these discrepancies will be negligible in practice \edit{when acting on \textit{high-energy} states, including semiclassical states. We remind the reader that these states have the form: 
\begin{equation}
    \lvert \psi\rangle = \sum_n c_n \lvert n\rangle,
\end{equation}
where the distribution of $\lvert c_n\rvert^2$ is peaked (and has most of its weight) at $n \sim \mathrm{e}^S\gg 1$. Then, as long as $A_m(n)$ and $B_m(n)$ are slowly varying with $n$ (which we expect to be the case at large $n$ for most reconstructed observables of physical interest), we have $A_{m_1}(n+\delta n) B_{m_2}(n+\delta n') \approx A_{m_1}(n) B_{m_2}(n)$, due to which
\begin{equation}
    R^*(A)R^*(B) \lvert \psi\rangle \approx R^*(AB) \lvert \psi\rangle,
\end{equation}
by \eqref{eq:ABcomparison_n}. However, for low-energy states, the ground state cutoffs such as $\delta_{n+m_2 \geq 0}$ in \eqref{eq:ABcomparison_n} play a more important role, and we are no longer guaranteed that e.g. $A_m(n)$ is slowly varying with $n$, making the two actions different in general. As long as we can neglect the ground state truncation of the Hilbert space and have smooth coefficients $A_m(n)$, the algebra homomorphism is essentially exact up to operator ordering ambiguities, which are $\sim e^{-\mathrm{O}(S)}$.}

\edit{For example, } the length operator \edit{has coefficients $A_m(n)$ that are smooth in $n$ but fluctuate only in $m$, and its} matrix elements \eqref{eq:Xmatrixelements} decay such that $\langle 0|x|n\rangle \sim n^{-2}$, meaning that discrepancies in operator reconstruction will be suppressed by factors of $\mathrm{e}^{-\mathrm{O}(S)}$ \edit{relative to the expectation value of the operator.}

\edit{Finally, the approximate algebra homomorphism also holds at late times; the late-time deviations from semiclassical dynamics are entirely due to the discrete fundamental Hilbert space and its implications for the expectation values of observables, including $A$ and $B$ separately or either reconstruction of the product $AB$.}  \edit{It is not clear to us that this approximate homomorphism would hold for the prescription chosen in \cite{Iliesiu:2021ari,Iliesiu:2024cnh}.}

\section{Microcanonical approximations to length operators}
\label{app:microcanonical}

Here, we will show that for physically relevant states such as the Hartle-Hawking state, it is safe to replace the full length operator, approximated here by
\begin{equation}
    x = \frac{1}{2}\left\lbrace \sqrt{2E} \rho(E), \vartheta\right\rbrace,
\end{equation}
with a ``microcanonical'' length operator $x_{\mc}$~\eqref{eqs:xmcdef} that is block diagonal with respect to some choice of microcanonical energy windows, and directly expressed in terms of the angle variable in each such window.

Let $\proj_r$ project onto a Hilbert space $\mathcal{H}_r$ spanned by $d$ consecutive energy levels from $E_{r}$ to $E_{r+d-1}$, each such set of levels constituting a microcanonical window, such that $\sum_r \proj_r = 1$. The index $r = (m-1)d$ represents the ``base index'' of the $m$-th microcanonical window in the energy spectrum. In practice, we will be interested in a subset of these microcanonical windows spanning some interval $[E_{\min}, E_{\max}]$; let $R$ be the set of values of the base index $r$ corresponding to this interval.

While we are interested in the regime where $d$ is very large, we also assume that the density of states scales suitably so as to satisfy
\begin{equation}
    \sqrt{2E_{r+k}} \rho(E_{r+k})-\sqrt{2E_r} \rho(E_r) < \epsilon \sqrt{2E_r} \rho(E_r),\ \text{ for all } 0 < k < d,
\end{equation}
for some small constant $\epsilon$, as long as $E_r \geq E_{\min}$.
In other words, the energy-dependent prefactor $\sqrt{2E} \rho(E)$ in $x$, which increases monotonically with energy, doesn't increase significantly relative to its ``base value'' within each microcanonical window with a minimum energy of at least $E_{\min}$. It is correspondingly convenient to introduce the operator 
\begin{equation}
    \eta_r := \sum_{n=r}^{r+d-1} \eta_r(E_n) \lvert E_n\rangle\langle E_n\rvert = \sum_{n=r}^{r+d-1} \frac{\sqrt{2E_{n}} \rho(E_{n})-\sqrt{2E_r} \rho(E_r)}{\sqrt{2E_r} \rho(E_r)} \lvert E_n\rangle \langle E_n\rvert \label{eq:etardef}
\end{equation}
that measures the  relative variation of the prefactor within the window.

Now, we define the microcanonical length operator,
\begin{equation}
    x_{\mc} :=  \sum_{r\in R}\sqrt{2 E_r} \rho(E_r) \vartheta_r,
    \label{eqs:xmcdef}
\end{equation}
where $\vartheta_r$ represents the absolute value of the angle variable within each microcanonical window, whose eigenstates are the $\lvert C_k(0)\rangle$ in each microcanonical window with eigenvalues $\lvert \qtheta\rvert$ (where $\qtheta \in (-\pi, \pi]$), i.e.
\begin{equation}
    \vartheta_r \equiv \sum_{k=0}^{d-1} \min\left\lbrace \frac{2\pi k}{d}, \frac{2\pi (d-k)}{d} \right\rbrace \lvert C_{k,r}(0)\rangle \langle C_{k,r}(0)\rvert,
    \label{eqs:varthetadef}
\end{equation}
where
\begin{equation}
    \lvert C_{k,r}(0)\rangle \equiv \frac{1}{\sqrt{d}}\sum_{n=0}^{d-1}\mathrm{e}^{-2\pi \mathrm{i} n k/d}\lvert E_{r+n}\rangle.
\end{equation}

Let $\mathcal{H}_{\mc}(\nu_{\mc})$ be the set of states that are supported on $\nu_{\mc}$ consecutive microcanonical windows spanning $[E_{\min}, E_{\max}]$ and are unbiased in each window: 
\begin{equation}
    \lvert \psi\rangle = \sum_{r\in R} \sqrt{\frac{p_r}{d}} \sum_{n=0}^{d-1} \mathrm{e}^{\mathrm{i}\varphi_{r+n}} \lvert E_{r+n}\rangle,
    \label{eqs:microcanonical_unbiased}
\end{equation}
where $p_r \equiv \langle \psi\rvert \proj_r \lvert \psi\rangle$. Then, we will show that $x$ can be replaced by $x_{\mc}$ for the Hartle-Hawking state in 3 stages:
\begin{enumerate}
\item The microcanonical length operator $x_{\mc}$ and the operator
\begin{equation}
    x_{\text{block}} :=  \sum_{r\in R}\sqrt{2 E_r} \rho(E_r) \proj_r \vartheta \proj_r,
\end{equation}
which is the direct block diagonal restriction of $x$ to microcanonical windows, approach each other as $d\to\infty$ for any state in $\mathcal{H}_{\mc}(\nu_{\mc})$: 
\begin{equation}
    \max_{\lvert \psi\rangle \in \mathcal{H}_{\mc}(\nu_{\mathrm{mc}})} \lvert \langle \psi\rvert (x_{\mc}-x_{\text{block}})\lvert \psi \rangle\rvert \leq \sqrt{2 E_{\max}} \rho(E_{\max})  \frac{c_0 \log d}{d}.
    \label{eqs:xmcminusblockbound}
\end{equation}
Here, $c_0$ is an $O(1)$ constant as $d\to\infty$.
    \item $x$ and $x_{\text{block}}$ have almost the same expectation values for states in $\mathcal{H}_{\mc}(\nu_{\mc})$:
    \begin{equation}
        \max_{\lvert \psi\rangle \in \mathcal{H}_{\mc}(\nu_{\mathrm{mc}})} \lvert \langle \psi\rvert (x-x_{\text{block}})\lvert \psi\rangle \rvert \leq \sqrt{2 E_{\max}} \rho(E_{\max}) \left[c_2 \epsilon + \frac{\nu_{\mc} c(1+\epsilon)\log d}{d}\right],
        \label{eqs:xminusxblockbound}
    \end{equation}
    where $c$, $c_2$ are $O(1)$ constants as $d\to\infty$.
    \item The Hartle-Hawking state has a large overlap with a state in $\mathcal{H}_{\mc}(\nu_{\mathrm{mc}})$.
\end{enumerate}
Combining \eqref{eqs:xmcminusblockbound} and \eqref{eqs:xminusxblockbound} we obtain \begin{equation}
    \max_{\lvert \psi\rangle \in \mathcal{H}_{\mc}(\nu_{\mathrm{mc}})} |\langle \psi|\left(x-x_{\mathrm{mc}}\right)|\psi\rangle| \le \sqrt{2 E_{\max}} \rho(E_{\max}) \left[c_3 \epsilon + \frac{ c_4 \nu_{\mathrm{mc}}\log d}{d}\right],
\end{equation}
for appropriate constants $c_3$ and $c_4$.

For the first statement, let us consider the matrix elements of $x_{\mc}$ and $x_{\text{block}}$ within a microcanonical Hilbert space $\mathcal{H}_r$. We have, for $n,m \in [r, r+d)$:
\begin{equation}
\langle E_n\rvert x_{\mc}\lvert E_m\rangle = \sqrt{2E_r} \rho(E_r) \langle E_n\rvert \vartheta_r\lvert E_m\rangle.
\end{equation}
From \eqref{eqs:varthetadef}, we have the matrix elements:
\begin{align}
\langle E_n\rvert \vartheta_r\lvert E_m\rangle &= \frac{1}{d}\sum_{k=0}^{d-1} \min\left\lbrace \frac{2\pi k}{d}, \frac{2\pi (d-k)}{d} \right\rbrace e^{-2\pi i (n-m) k/d} \nonumber \\
&= \begin{dcases}
    \frac{\pi}{2},\ &\text{ for } n=m,\\
    -\frac{\pi (1+(-1)^{m-n})}{d^2 \sin^2\left(\frac{m-n}{d}\pi\right)},\ &\text{ for } n \neq m.
\end{dcases}
\label{eqs:varthetamatrixelements}
\end{align}
For $x_{\text{block}}$, we recall that $\vartheta$ has the matrix elements [from \eqref{eq:Xmatrixelements}]:
\begin{equation}
\langle E_n\rvert \vartheta\lvert E_m\rangle = \begin{dcases}
\dfrac{\pi}{2},\ &\text{ for } n=m \\ \displaystyle -\frac{\left(1-(-1)^{m-n}\right)}{\pi(m-n)^2}, &\text{ for } n \neq m.
\end{dcases}
\label{eqs:Xmatrixelementsagain}
\end{equation}
Intuitively, we note that the difference between the matrix elements of $\vartheta_r$ and $\vartheta$ within $\mathcal{H}_r$ must generally be small, as the off-diagonal elements in \eqref{eqs:Xmatrixelementsagain} are comparable to \eqref{eqs:varthetamatrixelements} for large $d$ (when $\lvert m-n\rvert \ll d$), or both are negligibly small (when $d \sim \lvert m-n\rvert < d(1-\epsilon)$). The difference between the matrix elements is only significant for $\lvert m-n\rvert \approx d$, which occurs when $m \approx r+d$ and $n \approx r$ or vice-versa; however, these are a negligibly small fraction of matrix elements, and should hardly contribute  to expectation values in the unbiased states in $\mathcal{H}_{\mc}(\nu_{\mathrm{mc}})$. Formally, we have for $n\neq m$:
\begin{align}
    \left\lvert \langle E_n\rvert \vartheta_r\lvert E_m\rangle - \langle E_n\rvert \vartheta\lvert E_m\rangle \right\rvert =  \frac{\left(1-(-1)^{m-n}\right)}{\pi(m-n)^2}\left[\frac{1}{\sinc^2\left(\frac{m-n}{d}\pi\right)}-1\right],
\end{align}
while this difference vanishes for $n=m$. For an unbiased state $\lvert \psi\rangle \in \mathcal{H}_{\mc}(\nu_{\mc})$ \eqref{eqs:microcanonical_unbiased}, this gives (using the triangle inequality for the expression in terms of matrix elements, and that $1-(-1)^{m-n} \leq 2$):
\begin{align}
    \left\lvert \langle \psi\rvert (\vartheta_r - \proj_r\vartheta\proj_r)\lvert \psi\rangle\right\rvert &\leq \frac{2 p_r}{\pi d}\sum_{\substack{n,m = r \\ n \neq m}}^{r+d-1}\frac{1}{(m-n)^2}\left[\frac{1}{\sinc^2\left(\frac{m-n}{d}\pi\right)}-1\right] \nonumber \\
    &= \frac{4 p_r}{\pi d} \sum_{q=0}^{d-1} \frac{d-\lvert q\rvert}{q^2} \left[\frac{1}{\sinc^2(\pi q/d)}-1\right].
\end{align}
In the second line, we have set $q = m-n$, used the evenness of the sum in $q$, and noted that $\sinc^2(0) = 1$ by definition to allow $q=0$. Introducing the variable $\xi = q/d$, we can write the sum in the above expression as:
\begin{equation}
    \sum_{q=0}^{d-1} \frac{1}{d}\frac{1-\lvert q/d\rvert}{(q/d)^2} \left[\frac{1}{\sinc^2(\pi q/d)}-1\right]\approx
\int_{0}^{1-(1/d)} \diff \xi\ \frac{1-\lvert \xi\rvert}{\xi^2}\left[\frac{1}{\sinc^2(\pi \xi)}-1\right].
\end{equation}
This integral diverges as $\log d$ near $\xi \sim 1$, so we conclude that for some O(1) constant $c_0$,
\begin{equation}
    \sum_{q=0}^{d-1} \frac{1}{d}\frac{1-\lvert q/d\rvert}{(q/d)^2} \left[\frac{1}{\sinc^2(\pi q/d)}-1\right] < \frac{\pi c_0}{4} \log d,
\end{equation}
which gives
\begin{equation}
    \left\lvert \langle \psi\rvert (\vartheta_r - \proj_r\vartheta\proj_r)\lvert \psi\rangle\right\rvert \leq \frac{p_r c_0 \log d}{d}.
    \label{eqs:varthetar_difference_1block}
\end{equation}
For the difference between the operators themselves, the triangle inequality gives:
\begin{equation}
    \lvert \langle \psi\rvert (x_{\mc}-x_{\text{block}})\lvert \psi \rangle\rvert \leq \sum_r \sqrt{2 E_r} \rho(E_r)\left\lvert \langle \psi\rvert (\vartheta_r - \proj_r\vartheta\proj_r)\lvert \psi\rangle\right\rvert.
\end{equation}
Using \eqref{eqs:varthetar_difference_1block} with $\sqrt{2E_r}\rho(E_r) \leq \sqrt{2 E_{\max}} \rho(E_{\max})$ and $\sum_r p_r = 1$, we get \eqref{eqs:xmcminusblockbound}.

To show the second statement and \eqref{eqs:xminusxblockbound}, we note that $x$ can be massaged into the following form [recall $\eta_r$ was defined in \eqref{eq:etardef}]:
\begin{align}
    x &= \frac{1}{2}\sum_r \left[\sqrt{2E} \rho(E) \proj_r \vartheta + \vartheta \sqrt{2E} \rho(E) \proj_r \right] \nonumber \\
    &= \frac{1}{2}\sum_r \sqrt{2E_r}\rho(E_r) \left[ (1+\eta_r)\proj_r \vartheta + \vartheta \proj_r (1+\eta_r)\right] \nonumber \\
    &= \frac{1}{2}\sum_{r,r'} \sqrt{2E_r} \rho(E_r) \left[ (1+\eta_r)\proj_r \vartheta\proj_{r'} + \proj_{r'}\vartheta \proj_r (1+\eta_r)\right] \nonumber \\
    &= x_{\text{block}} + \frac{1}{2}\sum_r \sqrt{2E_r} \rho(E_r) \lbrace \eta_r, \proj_r \vartheta \proj_r\rbrace \nonumber \\ &+ \frac{1}{2}\sum_{r} \sum_{r' \neq r} \sqrt{2E_r} \rho(E_r) \left[ (1+\eta_r)\proj_r \vartheta\proj_{r'} + \proj_{r'} \vartheta \proj_r (1+\eta_r)\right].
\end{align}
For a given state $\lvert \psi\rangle$, the triangle inequality then gives
\begin{align}
    \lvert \langle \psi\rvert (x-x_{\text{block}})\lvert \psi\rangle\rvert \leq  &\left\lbrace\left\lvert \sum_r \sqrt{2E_r} \rho(E_r) \langle \psi\rvert \eta_r \proj_r \vartheta \proj_r\lvert \psi\rangle \right\rvert \right. \nonumber \\
    &+\left. \left\lvert \sum_r \sum_{r' \neq r} \sqrt{2E_r} \rho(E_r) \langle \psi\rvert (1+\eta_r) \proj_r \vartheta \proj_{r'}\lvert \psi\rangle \right\rvert \right\rbrace. 
    \label{eqs:xmctwoterms}
\end{align}
The first term measures the error in replacing the energy dependent prefactor with the base value $\sqrt{2E_r} \rho(E_r)$ in each microcanonical block, while the second term largely measures interference between the different blocks.

We note that both of these contributions consist of terms of the form:
\begin{equation}
    M_{r,r'} = \langle \psi\rvert (\zeta+\eta_r)\proj_r \vartheta \proj_{r'}\lvert \psi\rangle,
\end{equation}
where $\zeta \in \lbrace 0,1\rbrace$; in addition, $\zeta = 0$ when $r=r'$. When $\lvert \psi\rangle$ is given by an unbiased state as in \eqref{eqs:microcanonical_unbiased}, this becomes
\begin{equation}
    M_{r,r'} = \frac{\sqrt{p_r p_{r'}}}{d}\sum_{n=r}^{r+d-1} \sum_{m=r'}^{r'+d-1} (\zeta+\eta_r(E_n)) \mathrm{e}^{\mathrm{i}(\varphi_m-\varphi_n)} \langle E_n\rvert \vartheta\lvert E_m\rangle.
\end{equation}

When $r\neq r'$ in $M_{r,r'}$, i.e. for the interference contributions to \eqref{eqs:xmctwoterms}, only the off-diagonal matrix elements of $X$ \eqref{eqs:Xmatrixelementsagain} are relevant. In this case, we get
\begin{equation}
    M_{r\neq r'} = -\frac{\sqrt{p_r p_{r'}}}{d}\sum_{n=r}^{r+d-1} \sum_{m=r'}^{r'+d-1} (\zeta+\eta_r(E_n)) \mathrm{e}^{\mathrm{i}(\varphi_m-\varphi_n)} \frac{(1-(-1)^{m-n})}{\pi(m-n)^2}.
\end{equation}
Using the triangle inequality to replace each term with its magnitude, together with $\eta_r(E_n) < \epsilon$, $(m+n) \geq (m-n)$ and $2 \geq 1-(-1)^{m-n}$, we get the bound:
\begin{equation}
    \lvert M_{r\neq r'}\rvert \leq \frac{2(\zeta+\epsilon)\sqrt{p_r p_{r'}}}{d^3}\sum_{n=r}^{r+d-1} \sum_{m=r'}^{r'+d-1} \frac{1}{\pi(\frac{m-n}{d})^2}.
\end{equation}
On the right hand side, we have a double sum of positive terms over $d$ values of each of $m$ and $n$, which approaches a double integral as $d\to\infty$. For sufficiently large $d$, it follows that we can bound the sum by a multiple $c_m > 1$ of this integral (where $x$ corresponds to $n/d$ and $y$ to $m/d$):
\begin{align}
    \frac{1}{d^2}\sum_{n=r}^{r+d-1} \sum_{m=r'}^{r'+d-1} \frac{1}{(\frac{m-n}{d})^2} &\leq c_m \int_{r/d}^{(r-1)/d+1}\diff x \int_{r'/d}^{(r'-1)/d+1}\diff y\ \frac{1}{(x-y)^2}.\nonumber \\
    &= c_m \log\frac{(r-r')^2}{(r-r'+d-1)(r-r'-d+1)}.
\end{align}
When $\lvert r-r'\rvert =d$, the second line gives a constant times $\log d$; for $\lvert r-r'\rvert > d$, the second line is at most a constant. In either case, the sum is less than $(\pi c/2) \log d$ for some constant $c$. On the whole, we obtain the bound:
\begin{equation}
    \lvert M_{r\neq r'}\rvert \leq \frac{(\zeta+\epsilon)c \sqrt{p_r p_{r'}} \log d}{d}.
\end{equation}

For $r = r'$, we have
\begin{equation}
    M_{rr} = \frac{p_r}{d}\sum_{n = r}^{r+d-1}(\zeta+\eta_r(E_n))\left\lbrace\frac{\pi}{2} - \sum_{\substack{m = r,\\ m\neq n}}^{r+d-1}\mathrm{e}^{\mathrm{i}(\varphi_m - \varphi_n)}\frac{(1-(-1)^{m-n})}{\pi(m-n)^2}\right\rbrace.
\end{equation}
The inner sum over $m$ converges as each term is $\sim 1/m^2$ (i.e. the magnitude of each term is bounded by the terms of a convergent series), and its magnitude cannot exceed some constant $c_s > 0$ (say) for any value of $n$. The outer sum over $n$ then adds up $d$ finite terms, contributing an $\mathrm{O}(d)$ factor that cancels out the $1/d$ prefactor. On the whole, again using the triangle inequality, $\eta_r(E_n) < \epsilon$, as well as $\zeta=0$ when $r=r'$ in \eqref{eqs:xmctwoterms}, we get the inequality
\begin{equation}
    \lvert M_{rr}\rvert \leq p_r \epsilon c_2,
\end{equation}
for some constant $c_2$.

Substituting these bounds on $M_{r,r'}$ in \eqref{eqs:xmctwoterms}, and noting that they hold for any unbiased state of the form in \eqref{eqs:microcanonical_unbiased}, we get
\begin{equation}
\max_{\lvert \psi\rangle \in \mathcal{H}_{\mc}(\nu_{\mathrm{mc}})} \lvert \langle \psi\rvert (x-x_{\text{block}})\lvert \psi\rangle \rvert  \leq \sum_r \sqrt{2E_r} \rho(E_r)\left[p_r \epsilon c_2 +\sum_{r' \neq r} \frac{(1+\epsilon)c\sqrt{p_r p_{r'}} \log d}{d}\right]
\label{eqs:xminusxblockbound_part}
\end{equation}
As these states are supported on $\nu_{\mc}$ microcanonical windows with $\sum_r p_r = 1$, we also have $\sum_r \sqrt{p_r} \leq \sqrt{\nu_{\mc}}$; using these inequalities in \eqref{eqs:xminusxblockbound_part}, together with $\sqrt{2 E_r}\rho(E_r) < \sqrt{2 E_{\max}}\rho(E_{\max})$, we get \eqref{eqs:xminusxblockbound}.


Finally, to show that the Hartle-Hawking state~\eqref{eq:HHstate} 
can be approximated by a microcanonical state in $\mathcal{H}_{\mc}(\nu_{\mathrm{mc}})$, we define its (normalized) microcanonical approximation as (recalling that the $r$-th microcanonical window has $d$ levels):
\begin{equation}
    \lvert \beta_{\mc}\rangle := \frac{1}{\sqrt{Z_{\mc}(\beta)}} \sum_{r \in R} d \mathrm{e}^{-\beta E_r/2}.
\end{equation}
Normalization $\langle \beta_{\mc}\vert \beta_{\mc}\rangle = 1$ then gives
\begin{equation}
    Z_{\mc}(\beta) = \sum_{r \in R} d \mathrm{e}^{-\beta E_r}.
    \label{eqs:partitionfunctionmc}
\end{equation}
The overlap between these two states is given by:
\begin{align}
    \lvert \langle \beta_{\mc}\vert \beta\rangle\rvert &= \frac{1}{\sqrt{Z(\beta)Z_{\mc}(\beta)}} \sum_{r\in R} \sum_{k=0}^{d-1} \mathrm{e}^{-\beta (E_r + E_{r+k})/2} \nonumber \\
    &= \sqrt{\frac{Z(\beta)}{Z_{\mc}(\beta)}} \frac{1}{Z(\beta)} \sum_{r\in R} \sum_{k=0}^{d-1} \mathrm{e}^{-\beta E_{r+k}} \mathrm{e}^{\beta(E_{r+k}-E_r)/2}
     \geq \sqrt{\frac{Z(\beta)}{Z_{\mc}(\beta)}}.
\end{align}
The last inequality follows from noting that $E_{r+k} \geq E_r$.

Now focusing on $Z_{\mc}(\beta)$, we can rewrite \eqref{eqs:partitionfunctionmc} as:
\begin{equation}
    Z_{\mc}(\beta) = Z(\beta) + \sum_{r\in R} \sum_{k=0}^{d-1} \mathrm{e}^{-\beta E_{r+k}} (\mathrm{e}^{\beta(E_{r+k}-E_r)}-1).
\end{equation}
We expect that the microcanonical windows are such that
\begin{equation}
    e^{\beta(E_{r+d-1}-E_r)}-1 < \epsilon_{\beta},
\end{equation}
for some small $\epsilon_\beta > 0$ (note that the density of states is monotonically increasing with energy, so if this can be established for the $d$ lowest levels in $[E_{\min}, E_{\max}]$, we should expect it to be the case for [almost] all energy levels throughout the spectrum, especially given spectral rigidity, as successive sets of $d$ levels span increasingly narrower energy ranges). We get
\begin{equation}
    Z_{\mc}(\beta) \leq Z(\beta) (1+\epsilon_\beta),
\end{equation}
which implies
\begin{equation}
  \lvert \langle \beta_{mc}\vert \beta\rangle\rvert \geq \frac{1}{\sqrt{1+\epsilon_\beta}}.  
\end{equation}
As we discuss in Sec.~\ref{sec:HHstate}, most of the probability of the Hartle-Hawking state can be found in a finite ($\beta$-dependent) energy range, which we choose to be $[E_{\min}, E_{\max}]$, we can truncate the support of $\lvert \beta_{\mc}\rangle$ to this range and normalize this truncated version so that it formally belongs to $\mathcal{H}_{\mc}(\nu_{\mathrm{mc}})$. But it is also the case that to accurately capture expectation values such as $\langle x^q\rangle$, the choice of microcanonical windows that must be accounted for is rather different: see Appendix \ref{app:HH}.

\section{Dynamics of the Hartle-Hawking state}
\label{app:HH}
This appendix contains explicit calculations for the dynamics of the Hartle-Hawking state.
\subsection{Probability distribution of wormhole lengths}
Here we present the derivation of \eqref{eq:longtailed_x_dist} and \eqref{eq:HHEq}, starting with the former.  Recall that the length operator is approximated well by its truncation to microcanonical windows in \eqref{eq:xwindow}.  In each window, $\langle\vartheta\rangle\approx \pi/2$ at late times.  Let \begin{equation}
    x_{\mathrm{sat}}(E) = \sqrt{2E}\rho(E)\cdot \frac{\pi}{2} \label{eq:xsatE}
\end{equation}
denote the average value of $x$ at late times, in a microcanonical window centered around $E$.  Since we know the distribution of energy $E$ in the Hartle-Hawking state is given by \eqref{eq:HHdistribution}, we wish to invert this relation to find a distribution of $x_{\mathrm{sat}}$.
Introducing the function $W_s(z)$ via the following property (for $y\geq 0$):\footnote{$W_s(z)$ reduces to the Lambert product logarithm function $W_0(z)$ for large argument ($y e^{y} = z \iff y = W_0(z)$ for $y\geq 0$).}
\begin{equation}
2 y \sinh(y) = z \iff y = W_s(z),
\end{equation}
we get the expression
\begin{equation}
2\pi \sqrt{2 E(x_{\text{sat}})} = W_s\left(\mathrm{e}^{-\widetilde{S}_0} x_{\text{sat}}\right), \label{eq:WsE}
\end{equation}
where \begin{equation}
    \mathrm{e}^{-\widetilde{S}_0} = 32\pi^2 \mathrm{e}^{-S_0}
\end{equation}
is introduced for future convenience.  We also note the approximation \begin{equation}
    W_s(z)\approx \log z\;\;\;\; (z\gg 1)
\end{equation}
which will prove useful later.  Combining \eqref{eq:HHdistribution} with \eqref{eq:WsE}, we obtain the following distribution of saturation values:  
\begin{align}
\mathcal{P}_{\text{sat}}(x_{\text{sat}}) &= \frac{\diff E}{\diff x_{\text{sat}}} \mathcal{P}(E(x_{\text{sat}})) \nonumber \\
&= 8 \pi \frac{W_s\left(\mathrm{e}^{-\widetilde{S}_0} x_{\text{sat}}\right) \exp\left[-\frac{\beta}{8 \pi^2}W_s^2\left(\mathrm{e}^{-\widetilde{S}_0} x_{\text{sat}}\right)\right]}{Z(\beta) \left[1+W_s\left(\mathrm{e}^{-\widetilde{S}_0} x_{\text{sat}}\right)\sqrt{1+\left\lbrace\frac{2 W_s\left(\mathrm{e}^{-\widetilde{S}_0} x_{\text{sat}}\right)}{\mathrm{e}^{-\widetilde{S}_0} x_{\text{sat}}}\right\rbrace^2}\right]} \notag \\
&\approx \frac{8 \pi}{Z(\beta)} \exp\left[-\frac{\beta}{8\pi^2}\left\lbrace\log\left(\mathrm{e}^{-\widetilde{S}_0} x_{\text{sat}}\right)\right\rbrace^2\right],    
\end{align}
for large $x_{\text{sat}} \gg \mathrm{e}^{S_0}$. The distribution of $x$ itself at extremely long times (after almost all of the distribution saturates) can be modeled by a uniform probability $1/(2 x_{\text{sat}})$ in $x \in [0, 2 x_{\text{sat}}]$ for each microcanonical window with respective saturation length $x_{\text{sat}}$. This gives the following late-time probability distribution of $x\gg \mathrm{e}^{S_0}$:
\begin{align}
\mathcal{P}_x(x; t\to\infty) &= \int_{x/2}^{\infty}\diff x_{\text{sat}}\ \frac{\mathcal{P}_{\text{sat}}(x_{\text{sat}})}{2x_{\text{sat}}} \nonumber \\
&\approx \frac{8 \pi }{Z(\beta)}\int_{\ln(x/2)}^{\infty}\diff (\ln x_{\text{sat}})\ \exp\left[-\frac{\beta}{8\pi^2}\left\lbrace \ln\left(\mathrm{e}^{-\widetilde{S}_0} x_{\text{sat}}\right)\right\rbrace^2\right] \nonumber \\
&= \frac{8\pi^2}{Z(\beta)} \sqrt{\frac{2\pi}{\beta}}\erfc\left[\sqrt{\frac{\beta}{8\pi^2}}\ln\left(2\mathrm{e}^{-\widetilde{S}_0} x\right)\right] 
\end{align}
which leads to \eqref{eq:longtailed_x_dist} in the main text at large $x$.   

We now calculate the moments $\langle x^q\rangle$ with respect to this late time distribution, which we see is approximately Gaussian in $\log x$.  Observe that (dropping algebraic factors in $\log x$) we find \begin{align}
    \left\langle \mathrm{e}^{q\log x}\right\rangle &\propto \int\limits_{-\infty}^\infty \left(x \mathrm{d}\log x\right)\mathrm{e}^{q\log x - \beta (\log (2\mathrm{e}^{-\widetilde{S}_0}x))^2/8\pi^2}\notag \\
    &\sim\int\limits_{-\infty}^\infty  \mathrm{d}\log x\mathrm{e}^{- \beta (\log (2\mathrm{e}^{-\widetilde{S}_0}x)-(q+1)4\pi^2 T)^2/8\pi^2} \mathrm{e}^{(q+1)^22\pi^2T-(q+1)\log(2\mathrm{e}^{-\widetilde{S}_0})} \notag \\
    &\sim \mathrm{e}^{(q+1)^22\pi^2T-(q+1)\log(2\mathrm{e}^{-\widetilde{S}_0})}.
\end{align}
From the middle line above, together with \eqref{eq:xsatE} and \eqref{eq:JTinftyDOS}, we can deduce the typical energy of the microcanonical windows which dominate $\langle x^q\rangle$ to be given by \eqref{eq:HHEq}.  Notice that even the average wormhole length, $q=1$, is dominated by completely \emph{atypical} parts of the Hartle-Hawking state, with $E_{q=1}-E_{q=0} \sim T^2 \gg \sigma_E$.  More generally, each higher moment is dominated by a completely different (and even more atypical) part of the Hartle-Hawking state.

\subsection{Early time dynamics}
Now, let us discuss the evolution of the Hartle-Hawking state at earlier times.  The classical solution of Hamilton's equations,  \eqref{eq:xtclassical}, 
implies that the classical trajectory has approximately the same velocity in nearly all microcanonical windows which make up the Hartle-Hawking state:
\begin{align}
x_{\cl}(t; E(\beta)+\delta E) &\approx  \left(\sqrt{2E(\beta)}\right) t\;\;\; \text{ for } \;\;\; t \lesssim \frac{1}{2} t_{\htime}(E). \label{eq:earlytimexcl}
\end{align}
Focusing on times \begin{equation}
    t < t_{\mathrm{H}}^*= \exp[S(E(\beta))-c\beta \sigma_E(\beta)]
\end{equation}
for some large O(1) constant $c$, \eqref{eq:earlytimexcl} applies to the bulk of the microcanonical windows in the energy distribution $\mathcal{P}(E)$. From \eqref{eq:CUEspread}, we can estimate the spread $\Delta x$ in the wave function as
\begin{align}
\Delta x(t; E(\beta)+\delta E) &\approx \sqrt{2E(\beta)} \mathrm{e}^{S(E(\beta))+\beta \delta E} \Delta\qtheta(t) \nonumber \\
&\sim \sqrt{2E(\beta)} \mathrm{e}^{S(E(\beta))+\beta \delta E} \exp\left[-\frac{\mathrm{e}^{2S(E(\beta))+2\beta \delta E}}{\gamma t^2}\right].
\label{eq:microcanonicalHHspread}
\end{align}
Taking $\delta E \sim -c \sigma_E$ for a large O(1) constant $c$, we deduce that for $t\ll t_{\mathrm{H}}^*$, the vast majority of the Hartle-Hawking state consists of microcanonical windows which form nearly perfect semiclassical wave packets.  Therefore, the overall wave function will be quite tightly concentrated around the semiclassical value.

For $t\gg t_{\mathrm{H}}^*$, typical microcanonical windows in the Hartle-Hawking state reach their Heisenberg time, and so the wave function will begin to explosively spread over many different wormhole lengths.  However, we note that the exponentially small fraction of the Hartle-Hawking state in energy windows with $E\gg E(\beta)$ remain sharply peaked, following their semiclassical trajectory. 
We can use this intuition to estimate the probability distribution $\mathcal{P}(x,t)$ for distances $x \gg \sqrt{2E_{\mathrm{H}}(t)}t$, where $E_{\mathrm{H}}(t)$
is the energy scale at which we have currently reached the Heisenberg time, which we find by solving \eqref{eq:heisenbergtime}: \begin{equation}
    E_{\mathrm{H}}(t) \sim \frac{(\log (2\pi t))^2}{8\pi^2}.
\end{equation}
So for $x\gtrsim \sqrt{2E_{\mathrm{H}}(t)}t$ we expect \begin{equation}
    \mathcal{P}(x,t)\sim \int\mathrm{d}E \delta(x-\sqrt{2E}t) \frac{\rho(E)\mathrm{e}^{-\beta E}}{Z(\beta)}\sim \frac{x}{t^2}\mathrm{e}^{2\pi x/t - \beta x^2/2t^2},
\end{equation}
while for $x\lesssim \sqrt{2E_{\mathrm{H}}(t)}t$ we expect $\mathcal{P}(x,t)$ is approximately given by \eqref{eq:longtailed_x_dist}.


Finally, we expect that
\begin{equation}
\int\limits_0^T \frac{\mathrm{d}t}{t_{\htime}(E_{q=1})}\left(\langle x(t)\rangle - \langle x_{\mathrm{sat}}\rangle\right) = 0. \label{eq:HHavgxt}
\end{equation}
In other words in the Hartle-Hawking state, the average value of $\langle x(t)\rangle$ must ``overshoot" its late time value.   This will occur so long as the large time $T \gg t_{\mathrm{H}}(E_{q=1}+c\sigma_E)$ for a large O(1) constant $c$, to ensure that (as discussed above) all microcanonical windows that dominate the average value of $x$ -- which are \emph{atypical} in the overall Hartle-Hawking state -- have reached their Heisenberg time. For these dominant microcanonical windows, the Heisenberg times are of the same order of magnitude for $O(1)$ values of $\beta$, which means that level repulsion continues to be a dominant effect in the spectral statistics despite the exponentially varying density of states. Indeed, \eqref{eq:HHavgxt} follows from an identical argument to \eqref{eqs:generalobservableerg} applied to each of the microcanonical windows of interest.  However, because each microcanonical window is saturating at a very different Heisenberg time (though with comparable orders of magnitude), as per the discussion around \eqref{eq:tHdistributed}, the bump will not appear as dramatic as in a microcanonical state; rather, $\langle x(t)\rangle$ will asymptote to $\langle x_{\mathrm{sat}}\rangle$ much more slowly.

\section{Wormhole dynamics with Poisson statistics}
\label{app:JTPoisson_numerics}

\begin{figure}[t]
\centering
\subfigure[Rapid spread of wavepackets in terms of $p(x)$.]{\includegraphics[width=0.45\textwidth]{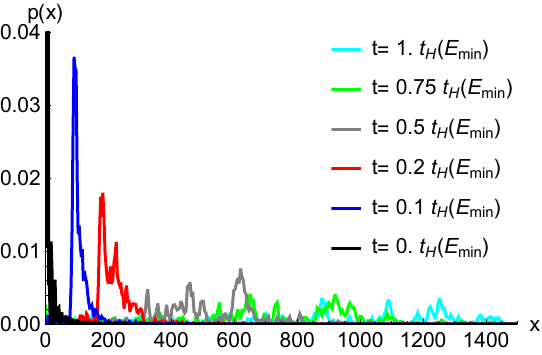}}
\label{fig:microcanonicalspreadpoisson}
\qquad
\subfigure[Lack of a distinct non-monotonic feature in $\langle x(t)\rangle$.]{\includegraphics[width=0.45\textwidth]{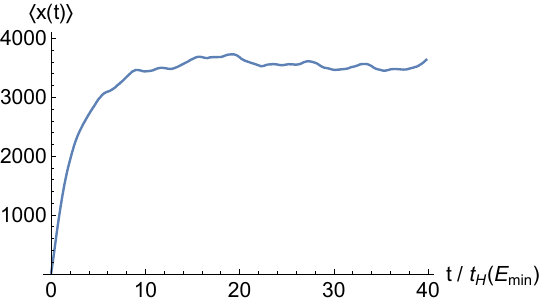}}
\label{fig:microcanonicalbumppoisson}
\caption{Wormhole dynamics with Poisson statistics in the microcanonical state. In contrast to Fig.~\ref{fig:microcanonicalnumerics} with random matrix statistics, the wavepacket spreads rapidly in $x$, and there is no clear bump at late times in $\langle x(t)\rangle$.}
\label{fig:microcanonicalnumericspoisson}
\end{figure}

\begin{figure}[t]
\centering
\subfigure[Extremely fast spread of wavepackets in terms of $p(x)$.]{\includegraphics[width=0.45\textwidth]{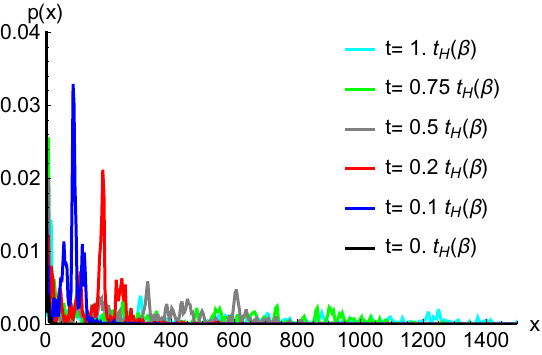}}
\label{fig:HHspreadPoisson}
\qquad
\subfigure[Lack of a distinct non-monotonic feature in $\langle x(t)\rangle$.]{\includegraphics[width=0.45\textwidth]{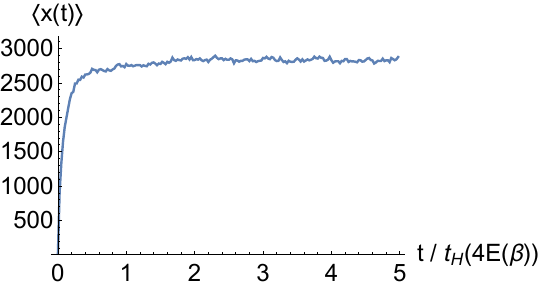}}
\label{fig:HHbumpPoisson}
\caption{Wormhole dynamics with Poisson statistics in the Hartle-Hawking state. Here, the wavepacket spread is extremely rapid, due to a combination of non-ergodicity from Poisson statistics (as in Fig.~\ref{fig:microcanonicalnumericspoisson}), and a thermal distribution of velocities in the initial state (as in Fig.~\ref{fig:HHnumerics}). The lack of a clear non-monotonic bump is once again evident.}
\label{fig:HHnumericspoisson}
\end{figure}

Here, we contrast the numerics in Sec.~\ref{sec:wormholedynamics_numerics}, for JT gravity with the expected RMT spectral statistics, with a quantum system with the same density of states and operators, but with Poisson statistics in the energy spectrum. Our results in this Appendix further demonstrate that the slow wavepacket spread and nonmonotonic wormhole length dynamics in Figs.~\ref{fig:microcanonicalnumerics} and \ref{fig:HHnumerics} are clear consequences of the random matrix spectrum in the fundamental theory. In addition, Poisson statistics also emulates (to an extent) a suboptimal reconstruction of length operators, for example in terms of action-angle variables with randomly permuted energy levels as described in footnote~\ref{footnote:suboptimalactions}, illustrating the key role played by our guiding principle in Sec.~\ref{sec:pure} for the presence of these features.

At a technical level, we can simulate this system by following the procedure of Sec.~\ref{sec:wormholedynamics_numerics} except with the CUE eigenphases $\varphi_n$ replaced by a uniform, random (and sorted) distribution of points in $[0,2\pi)$. With Poisson statistics, we expect the initial state to spread out rapidly to the point of exhibiting no clear imprints of the classical dynamics near the Heisenberg time scale due to a lack of spectral rigidity, and the absence of a bump in $\langle x(t)\rangle$ (more precisely, the absence of an ``equal'' exploration of values below and above the saturation value) due to a lack of nearest-neighbor level repulsion.  This is illustrated in Figs.~\ref{fig:microcanonicalnumericspoisson} and \ref{fig:HHnumericspoisson}, respectively for a microcanonical state and a Hartle-Hawking state.


\end{appendix}

\bibliographystyle{jhep}
\bibliography{mybib2021}

\end{document}